\documentclass[preprint,3p,times,twocolumn]{elsarticle}
\usepackage{amssymb}
\usepackage{epsfig,color}
\usepackage{amsfonts}
\usepackage{amssymb}
\usepackage{amsmath}
\usepackage{graphicx}
\usepackage{color} 
\newcommand{\secref}[1]{Sec.~\ref{#1}}

\journal{Journal of Theoretical Biology}

\begin{document}

\begin{frontmatter}


\author{Robert West and Mauro Mobilia}

\title{Fixation properties of rock-paper-scissors games in fluctuating populations}

 \address{Department of Applied Mathematics, School of Mathematics, University of Leeds, Leeds LS2 9JT, United Kingdom}

\begin{abstract}
Rock-paper-scissors games metaphorically model cyclic dominance in ecology and microbiology. In a static environment, 
these models are characterized by fixation probabilities  obeying two different ``laws'' in large and small well-mixed populations. 
Here, we investigate the evolution of these three-species models subject to a randomly switching  carrying capacity modeling the
endless change between states of resources scarcity 
and abundance. Focusing mainly on the zero-sum rock-paper-scissors game, equivalent to the cyclic Lotka-Volterra model,
we study  how the \emph{coupling} of demographic and environmental noise influences the 
fixation properties. More specifically, we investigate which species is the most likely to prevail in a population of fluctuating size
and how the outcome depends on the environmental variability.
We show that demographic noise coupled with environmental randomness  ``levels the field'' of cyclic 
competition by balancing the effect of selection. In particular, we show that fast switching  effectively reduces the selection 
intensity proportionally to the variance of the carrying capacity. We  
determine the conditions under which  new fixation scenarios arise, where the 
most likely species to prevail changes with the rate of switching and the variance of the carrying capacity.
Random switching has a limited effect on 
the mean fixation time that scales linearly with the average population size. Hence,
environmental randomness makes the cyclic competition more egalitarian, but does not prolong the species coexistence.
We also show how the fixation probabilities of  close-to-zero-sum rock-paper-scissors games  
can be obtained from those of the zero-sum model
by rescaling the selection intensity.

%\begin{itemize}
%\item Coupling of demographic and environmental noise balances the effect of selection
%\item Fast switching  effectively reduces the selection intensity 
%\item New fixation scenarios:  success depends on environmental variability 
%\item EN makes the RPS competition more egalitarian, but does not prolong coexistence
%\end{itemize}

\end{abstract}


\end{frontmatter}






 \section{Introduction}
Studying what affects the extinction and survival  of species in ecosystems 
is of paramount importance~\cite{Pennisi05}.
It is well known that birth and death events cause demographic fluctuations (internal noise, IN)
that can ultimately lead to species extinction and fixation -- when one species takes over the 
entire population~\cite{Kimura,Ewens}. IN being 
stronger in small communities than in large populations, various survival and fixation 
scenarios arise in  populations of different size and structure~\cite{Frean01,Ifti03,Berr09,Frey10,RMF07,cycl-rev}.  
For instance,  experiments on colicinogenic microbial 
communities have demonstrated that cyclic rock-paper-scissors-like competition between three strains
leads to intriguing behavior~\cite{Kerr02}:  the colicin-resistant 
strain  is the only one to survive in a large 
well-mixed population,
whereas all species coexist for a long time on a plate. 
These observations,
and the rock-paper-scissors being the paradigmatic model of cyclic dominance in ecology and microbiology, see, {\it e.g},
Refs.~\cite{Jackson,Sinervo96,Hofbauer,Szabo07,Nowak,Broom,cycl-rev,JPAtopical},
have motivated
the study of the survival/fixation properties of  
 the cyclic Lotka-Volterra model (CLV). This is characterized by 
 a zero-sum rock-paper-scissors competition between three species~
 \cite{Hofbauer,Nowak,Broom,Ifti03,Frean01,RMF06,RMF08b,Berr09,Tainaka89,Tainaka93,Tainaka94,He10,Ni10,Venkat10,Dob12,Mitarai16,Szabo02,Perc07,West18}. 
 Remarkably, it 
 has been shown that, when the population size is {\it constant},
 the  fixation probabilities in the CLV 
 obey two simple laws~\cite{Berr09,Frean01,Tainaka93}: 
 In a large well-mixed population,
 the species receiving the lowest payoff
 is the most likely to survive and fixate,
  a result referred to as the ``law of the weakest'', whereas a different law, called the 
  ``law of stay out'', arises in smaller populations.

  In fact, the fate of a population is influenced by numerous endlessly changing environmental conditions
  (e.g. light, pH, temperature,  nutrient abundance)~\cite{Fux05}. Detailed knowledge about  
  exogenous factors being generally unknown, these are often modeled 
as  environmental (external) noise (EN)~\cite{May73,Kussell05,Acer08,He10,Visco10,Dobramysl13,Assaf13a,Assaf13,Ashcroft14,Hufton16,Hidalgo17,Xue17,West18,Danino18,Hufton19}. 
In many  biological applications
the population size varies in time due to changing external 
factors~\cite{Levins,Roughgarden79}. The EN-caused
 fluctuations in the population size  
in turn affect the demographic fluctuations which results in a coupling of IN and EN
leading to feedback loops that shape the population's long-term 
evolution~\cite{Leibler09,Melbinger2010,Gore13a,Gore13b,Harrington14,Melbinger2015a,KEM1,KEM2,McAvoy2018}.
This is particularly relevant in microbial communities that are subject to sudden and extreme environmental changes
leading, {\it e.g.}, to population bottelnecks or to the collapse of 
biofilms~\cite{Wahl02,Brockhurst07a,Patwas09,Lindsey13}.
While EN and IN are naturally interdependent in many
biological applications, the theoretical understanding of their coupling is still
limited. Recently, progress has been made in
simple  two-species models~\cite{KEM1,KEM2}, but the analysis of  EN and IN coupling in populations consisting of
 many interacting species is a formidable task.

Here, we  study the {\it coupled effect} of environmental and 
internal noise on the fixation properties of three-species rock-paper-scissors games in a population of {\it fluctuating  size},
when the resources continuously vary between states of scarcity and abundance.
Environmental randomness is modeled by assuming that the population is subject to
a carrying capacity, driven by a  dichotomous Markov noise~\cite{HL06,Bena06,PDMP1,Davis84}, 
randomly switching between two values. 
A distinctive feature of this model is the coupling of demographic noise with environmental variability:
Along with the carrying capacity,  the population size can  fluctuate 
 and switch between values 
dominated  by  either the law of the weakest or stay out. It   
is therefore a priori not clear which species will be the most likely to prevail and how the outcome depends 
on the environmental variability.
Here, we show that environmental variability generally balances the effect of selection and can yield
novel fixation scenarios.

The models considered in   this work are introduced in \secref{model}. 
Section 3 is dedicated to the analysis of the long-time dynamics of
the cyclic Lotka-Volterra model (CLV) with a
constant carrying capacity. This paves the way to the detailed study of the survival and fixation properties in 
the CLV subject to a randomly switching  carrying capacity presented in 
\secref{swK}. In Section
5, our results are extended to   close-to-zero-sum rock-paper-scissors games.
Our conclusions are presented in \secref{disc}. Technical details 
and supporting information are provided in a series of appendices.
\section{Rock-paper-scissors games with a carrying capacity}
\label{model}
We consider a well-mixed population (no spatial structure) of  fluctuating size $N(t)$ containing 
three species, denoted by $1$, $2$, and $3$. At time $t$, the population consists of 
$N_i(t)$ individuals of species $i\in\{1,2,3\}$, such that $N(t)=N_1(t)+N_2(t)+N_3(t)$.
As in all rock-paper-scissors (RPS) games~\cite{Hofbauer,Nowak,Szabo07,Broom}, species are engaged in a cyclic competition:
Species $1$ dominates over type $2$, which outcompetes species $3$, which in turn wins against 
species $1$ closing the cycle. In a game-theoretic formulation, the underpinning cyclic competition can be 
generically described in terms of the  payoff matrix~\cite{Maynard,Hofbauer,Nowak,Szabo07,Broom,JPAtopical,Claussen08,Mobilia10,Galla11}:
\begin{eqnarray*}\label{payoff}
\hspace{-5mm}
{\cal P}=
\begin{tabular}{c|c c c}
Species &   $1$   & $2$  & $3$  \\
\hline
$1$   &  $0$ & $r_1$  & $-r_3(1+\epsilon)$\\ 
 $2$   & $-r_1(1+\epsilon)$ & $0$ & $r_2$ \\
 $3$   & $r_3$ & $-r_2(1+\epsilon)$ & $0$ \\
\end{tabular}
\end{eqnarray*}
Here, $0 < r_i= {\cal O}(1)$,  with $\sum_{i}^{3}r_i=1$, and $\epsilon>-1$.
According to ${\cal P}$, an $i$-individual gains a payoff $r_i$  against
an $(i+1)$-individual and gets a  negative payoff $-r_{i-1}(1+\epsilon)$ against 
an $(i-1)$-player (with cyclic ordering, i.e. $1-1 \equiv 3$ and $3+1 \equiv 1$, see below).
  Hereafter, species $i-1$ is therefore referred to as the 
``strong opponent'' of type $i$, whereas  species $i+1$ is its ``weak opponent''.
 Interactions between individuals of same species do not 
provide any payoff.
When  $\epsilon=0$,  ${\cal P}$ underlies 
a zero-sum RPS game, also referred to as ``cyclic Lotka-Volterra model'' 
(CLV)~\cite{Tainaka89,Tainaka94,Tainaka93,Frean01,Ifti03,Maynard,Hofbauer,Nowak,RMF06,RMF08b,Berr09,He10,
Ni10,Venkat10,Knebel13,Szabo02,Perc07,Dob12,West18,JPAtopical}: 
what $i$ gains is exactly what $i+1$ loses.
When $\epsilon\neq 0$,   ${\cal P}$ describes
the general, non-zero-sum, RPS cyclic competition: What an $i$ loses against  $i-1$, $r_{i-1}(1+\epsilon)$,  
differs from the payoff $r_{i-1}$ received by 
 $i-1$ against $i$, see, {\it e.g.}, 
\cite{MayLeonard,Claussen08,Mobilia10,Galla11,JPAtopical,cycl-rev,RMF07,RMF07b,RMF08,He11,SMR13,SMR14,MRS16,Yang17,Rucklidge17}.
In Secs. 3 and 4, we focus on the CLV, and then discuss close-to-zero-sum  RPS games
($|\epsilon|\ll 1$) in Sec.~5.

In terms of the densities $x_i\equiv N_i/N$ of each species in the population, that span the phase space simplex 
$S_3$~\cite{Berr09,West18},
species $i$'s expected payoff is 
\begin{eqnarray}
 \label{eq:Pi}
 \Pi_i&=&({\cal P}\vec{x})_i=r_i x_{i+1} -r_{i-1}(1+\epsilon)x_{i-1}, \\
 \bar{\Pi}&=&\vec{x}\cdot {\cal P}\vec{x}=
-\epsilon \sum_{i=1}^{3} r_ix_i x_{i+1}, \nonumber
\end{eqnarray}
where  $\vec{x}=(x_1,x_2,x_3)$
and   $\bar{\Pi}$ is the population's average  payoff which vanishes when $\epsilon=0$ (zero-sum game).
Here and in the following, the indices are ordered cyclically:  In Eq.~(\ref{eq:Pi}), 
$x_{1-1}\equiv x_3, r_{1-1}
\equiv r_3$ and
$x_{3+1}\equiv x_1, r_{3+1}\equiv r_1$.
In evolutionary game theory, it is common to define the fitness $f_i$ of species
$i$ as a linear function of the expected payoff $\Pi_i$~\cite{Hofbauer,Nowak,Szabo07,Broom}:
\begin{eqnarray}
\hspace{-8mm}
 \label{eq:fi}
 f_i=1+s\Pi_i \quad \text{and} \quad \bar{f}=1+s\bar{\Pi} \quad \text{(average fitness)},
\end{eqnarray}
where $s>0$ is a parameter measuring the contribution to the fitness 
arising from ${\cal P}$, i.e. the ``selection intensity'': species
have close fitness  in the biologically relevant case  $s\ll 1$ (weak selection), whereas 
the fitness fully features the cyclic dominance when $s={\cal O}(1)$ (strong selection). The average
fitness $\bar{f}=\sum_{i=1}^{3}x_if_i =1$ in the CLV ($\epsilon=0$).

\begin{figure}
	\centering
	\centering
	\includegraphics[width=0.90\linewidth]{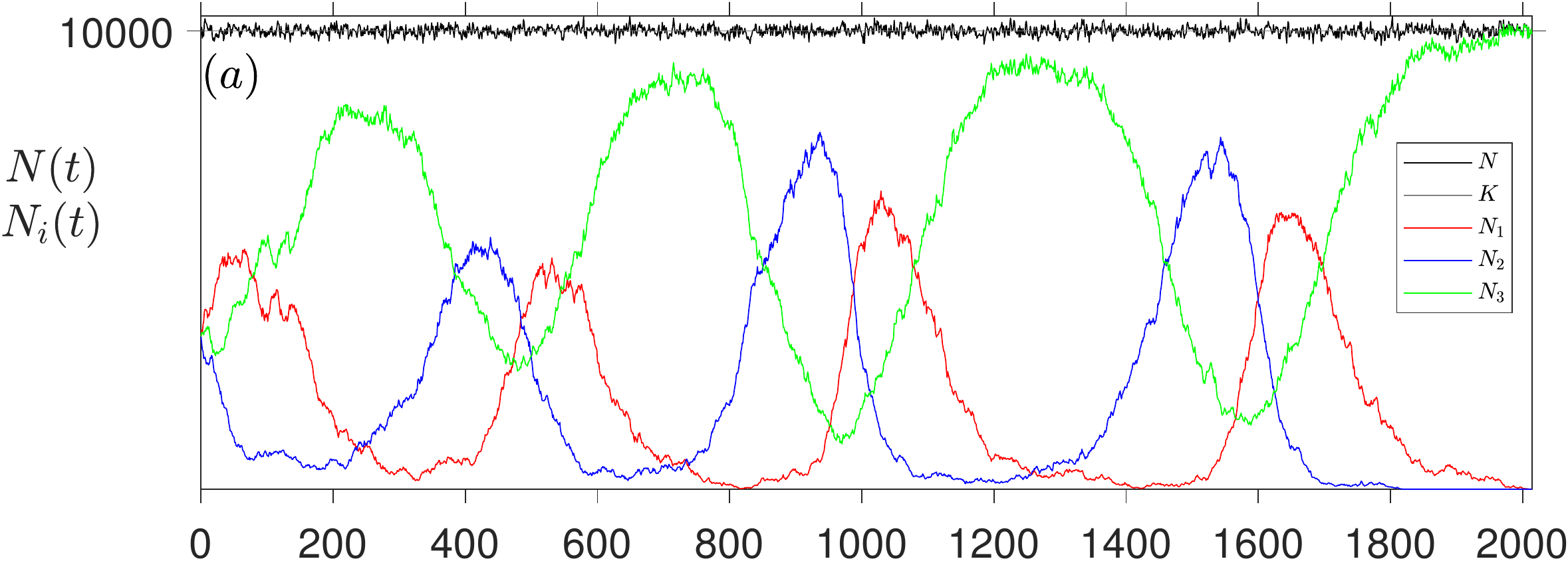}\\
	\includegraphics[width=0.90\linewidth]{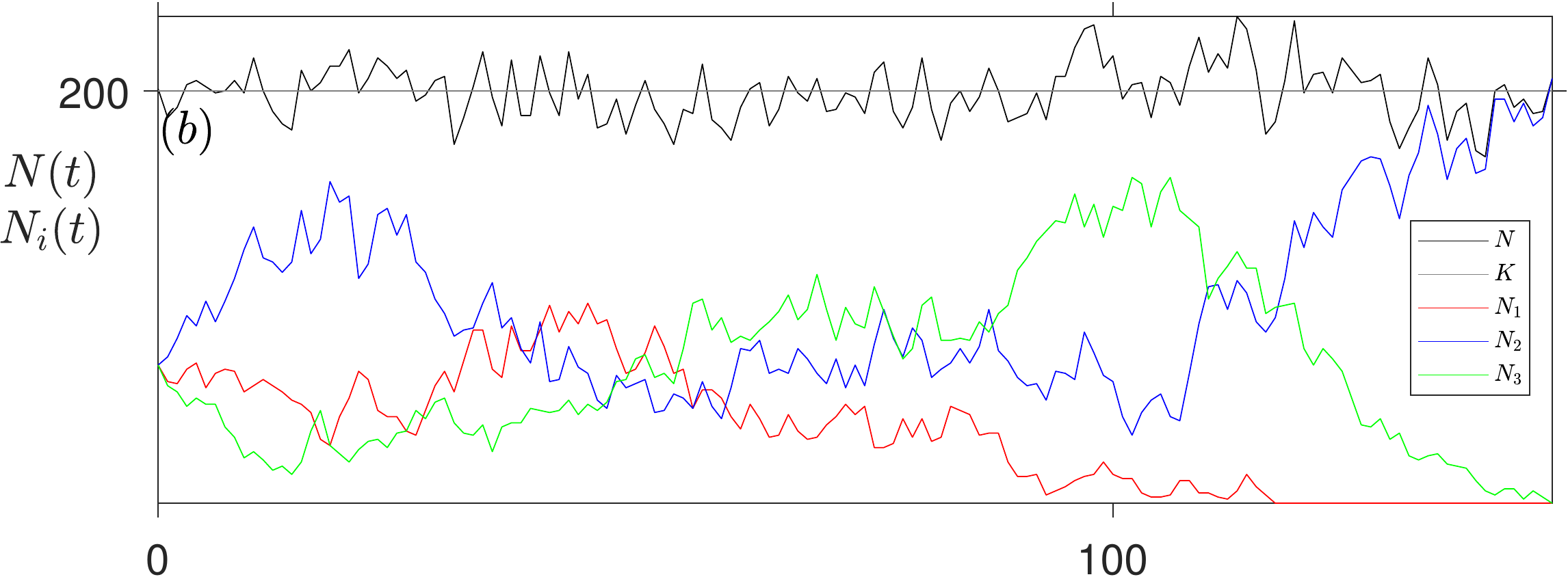}\\
	\includegraphics[width=0.90\linewidth]{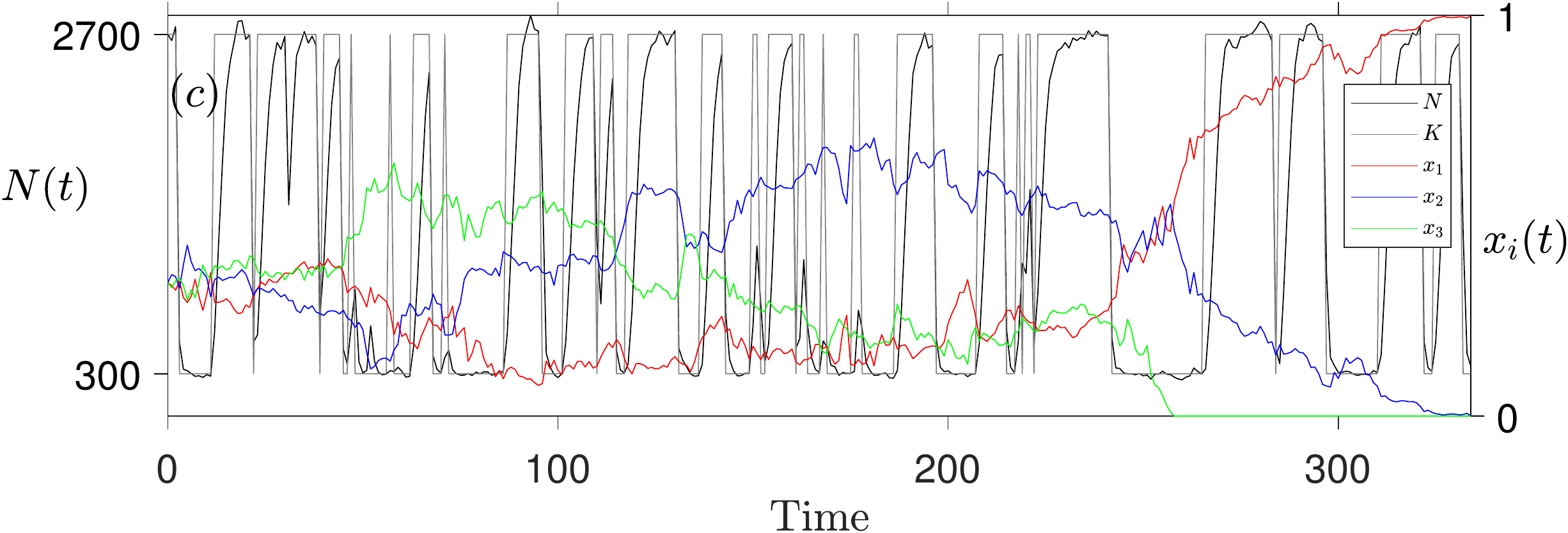}
	\caption{(a,b) Sample paths of  $N(t)$ (black), and
	$N_i(t)$ (colored) with constant  carrying capacity $K=10^4$ in (a) and  $K=200$ in (b);
	 solid gray lines show $N(t)=K$. Parameters are $(s,r_1,r_2,r_3)=(1/10,3/5,1/5,1/5)$.
	$N(t)$ quickly fluctuates about $K$, while $N_i$
	evolve on a much slower  timescale, see text. Fluctuations and extinction  properties vary with $sK$,  see Sec.~3.
(c) Sample paths of  $N(t)$ (black), densities 
	$x_i(t)=N_i(t)/N(t)$ (colored), and typical evolution of the randomly switching $K(t)$ (gray). 
Parameters are: $(s,r_1,r_2,r_3,\nu,K_+, K_-)=(1/20, 1/3,1/3,1/3, 
1/4, 2700, 300)$. %There is a timescale separation:
$N(t)$ quickly settles into  its (quasi) stationary state while $x_i$ vary
much more slowly  until fixation occurs in a time $\sim \langle K \rangle$, see Sec.~4.2.
In all panels:  $N_1(t), x_1(t)$  in red,  $N_2(t), x_2(t)$ in blue, and $N_3(t), x_3(t)$ 
in green, $\epsilon=0$. Initially, all species have the same density $1/3$.
} 
	\label{fig:Fig1}
\end{figure}

Population dynamics is often modeled by assuming a finite population of {\it constant size}
evolving according to a Moran process~\cite{Moran62,Ewens,Blythe07,Antal,Nowak}, see \ref{AppendixA}. 
Here, the population size is {\it not constant but  fluctuates in time}
due to environmental variability modeled by introducing a carrying capacity $K$, see Fig.~\ref{fig:Fig1}. 
Below, we first consider a constant carrying capacity, and 
then focus on the case where $K$ fluctuates in time.
For the fluctuating carrying capacity, we assume that
$K(t)$ continuously switches between two values, $K_+$ and $K_-$.
 This simply models that  available resources continuously and randomly
 change from being scarce ($K=K_-$) to being abundant ($K=K_+>K_-$). 
 The population size thus varies with $K$ and so do the demographic fluctuations, resulting in
 IN being \emph{coupled} to EN.
 For  simplicity, we model the 
 switches  of $K(t)$ with 
a {\it colored  dichotomous Markov noise} (DMN)~\cite{Bena06,HL06}, or ``random telegraph noise'',
$\xi(t)\in \{-1,+1\}$ with  symmetric 
switching rate $\nu$:
\begin{eqnarray}
\label{eq:DMN}
\xi \xrightarrow{\nu} -\xi.
\end{eqnarray}  
Here, the DMN is always at stationarity\footnote{In all our simulations, without loss of generality, $N(0) = 2K_+K_-/(K_++K_-)$.}: Its average vanishes, 
$\langle \xi (t) \rangle = 0$, and its autocorrelation is $\langle \xi (t) \xi (t')\rangle = \exp \left(-2\nu |t-t'|
\right)$~\cite{Bena06,HL06} (here,  $\langle \cdot \rangle$ denotes the ensemble average over the DMN).
The randomly switching carrying capacity therefore reads~\cite{KEM1,KEM2}
\begin{eqnarray}
\label{eq:K}
K(t) =
\frac{1}{2}\left[\left(K_+ + K_-\right) + \xi(t)\left(K_+ - K_-\right) \right],
\end{eqnarray}  
where $\langle K\rangle=(K_+ + K_-)/2$ is its constant average.
The constant-$K$ case is recovered by setting $K_+=K_-$ in (\ref{eq:K}).

In what is arguably its simplest formulation, see \ref{AppendixA1},
the RPS  dynamics subject to $K(t)$ is here defined in terms of the   birth-death process~\cite{Melbinger2010,KEM1} 
\begin{equation}
\label{eq:birtheq}
N_i \xrightarrow{T_i^+} N_i + 1 \quad \text{and} \quad N_i \xrightarrow{T_i^-} N_i -1,
\end{equation}
for the birth ($N_i \to N_i + 1$) and death ($N_i \to N_i - 1$) of an $i$-individual, respectively,
with
the  transition rates 
\begin{equation}\label{eq:transrates}
T^+_i = f_i N_i \quad \text{and} \quad T^-_i = \frac{N}{K(t)}N_i,
\end{equation}
%.
where the randomly switching carrying capacity is given by (\ref{eq:K}), while  
$K(t)=K$ when the  carrying capacity  is constant. It is worth noting that we consider $0\leq s\leq 1/(1+\epsilon)$,
which suffices to  
ensure $T^{\pm}_i\geq 0$.
The master equation (ME) associated with the
continuous-time  birth-death process (\ref{eq:birtheq}),(\ref{eq:transrates})
gives the probability $P(\vec{N}, \xi,t)$ to find the population 
in state  $(\vec{N},\xi)=(N_1,N_2,N_3,\xi)$ at time $t$
~\cite{Gardiner,VanKampen}, and reads:
\begin{eqnarray}\label{eq:ME}
\hspace{-5mm}
\frac{d P(\vec{N},\xi,t)}{dt} &=& \sum_{i=1}^3 \left( \mathbb{E}_i^--1\right)\left[T^+_i P(\vec{N},\xi,t)\right] \\
&+& \sum_{i=1}^3 \left( \mathbb{E}_i^+ -1\right)\left[T^-_i P(\vec{N},\xi,t)\right]\nonumber\\
&+& \nu \left[P(\vec{N},-\xi,t) - P(\vec{N},\xi,t)\right],\nonumber
\end{eqnarray}
where $\mathbb{E}^{\pm}_i$ are shift operators, associated with (\ref{eq:birtheq}), such that
$\mathbb{E}^{\pm}_1 h(N_1,N_2,N_3,t) =h(N_1\pm 1,N_2,N_3,t)$ etc, 
for any $h(\vec{N},\xi,t)$, 
and the last line accounts for the random switching of $K$.
In Eq~(\ref{eq:ME}), $P(\vec{N},\xi,t)=0$
whenever any $N_i<0$. This multidimensional
ME can be simulated exactly to
fully capture the stochastic RPS dynamics~\cite{Gillespie}. 
This is characterized by a first stage in which all species coexist, then
two species compete in a second stage, and, after a time that diverges with the system size, the population finally 
collapses~\footnote{\label{coll}The  population 
eventually collapses into  
 the unique absorbing state of the birth-death process (\ref{eq:birtheq})-(\ref{eq:ME}) which is $N=N_i=0$.
 However, this phenomenon is practically unobservable in a population with a large carrying capacity: it 
 occurs after lingering in the $N$-QSD for a time that diverges with the system size~\cite{Spalding17}, and is here ignored.}. 
 Here,  we focus on the first two stages of the dynamics in which 
 $N(t)$ is characterized by its quasi-stationary distribution ($N$-QSD).
 In the constant-$K$ case, one  drops the last line and 
 sets  $K_+=K_-=K$ in  Eq.~(\ref{eq:ME}), yielding the underpinning three-dimensional ME for  $P(\vec{N},t)$.

\begin{figure}
	\centering
        \includegraphics[width=0.475\linewidth]{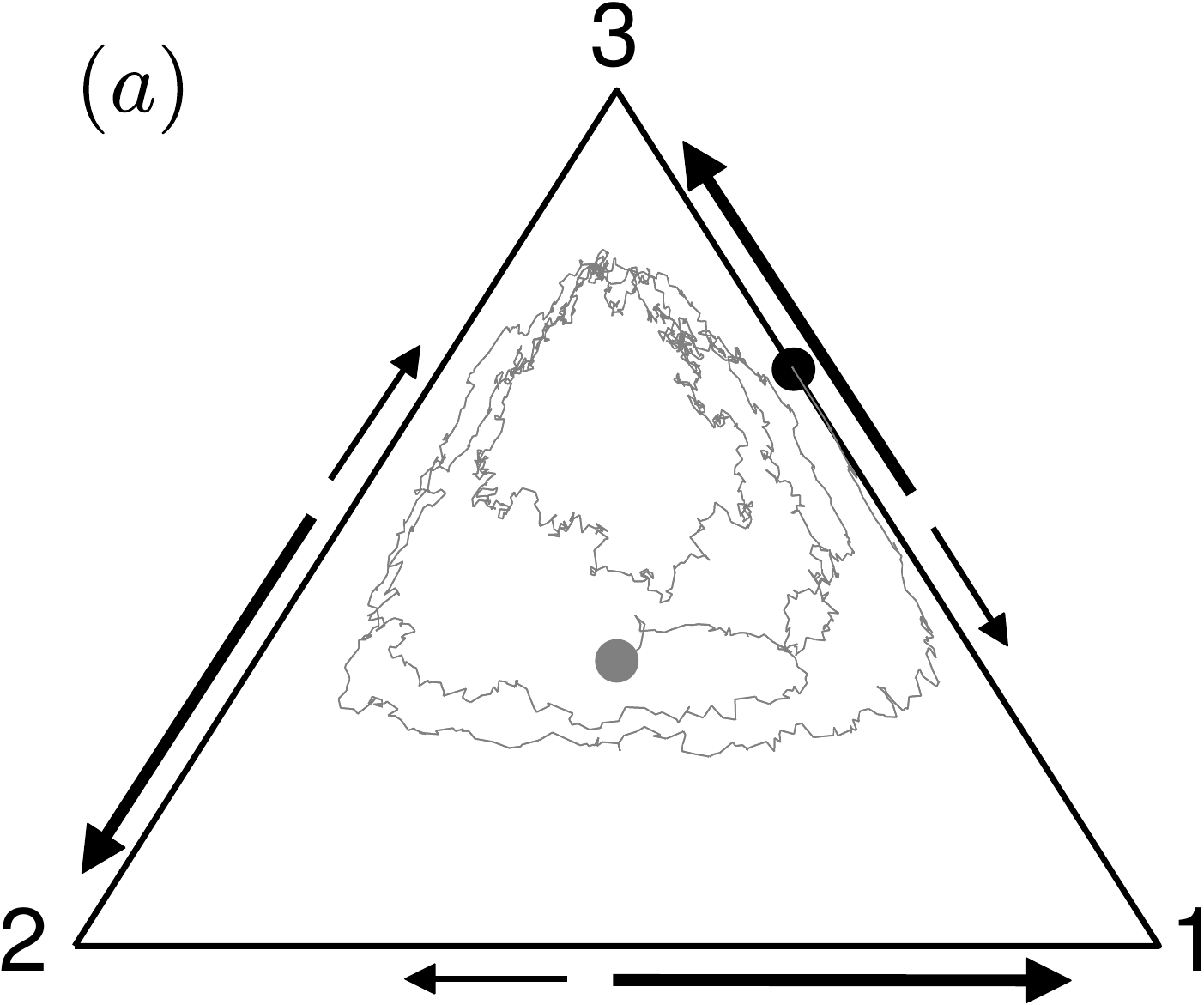} \hspace{2mm}
	\includegraphics[width=0.475\linewidth]{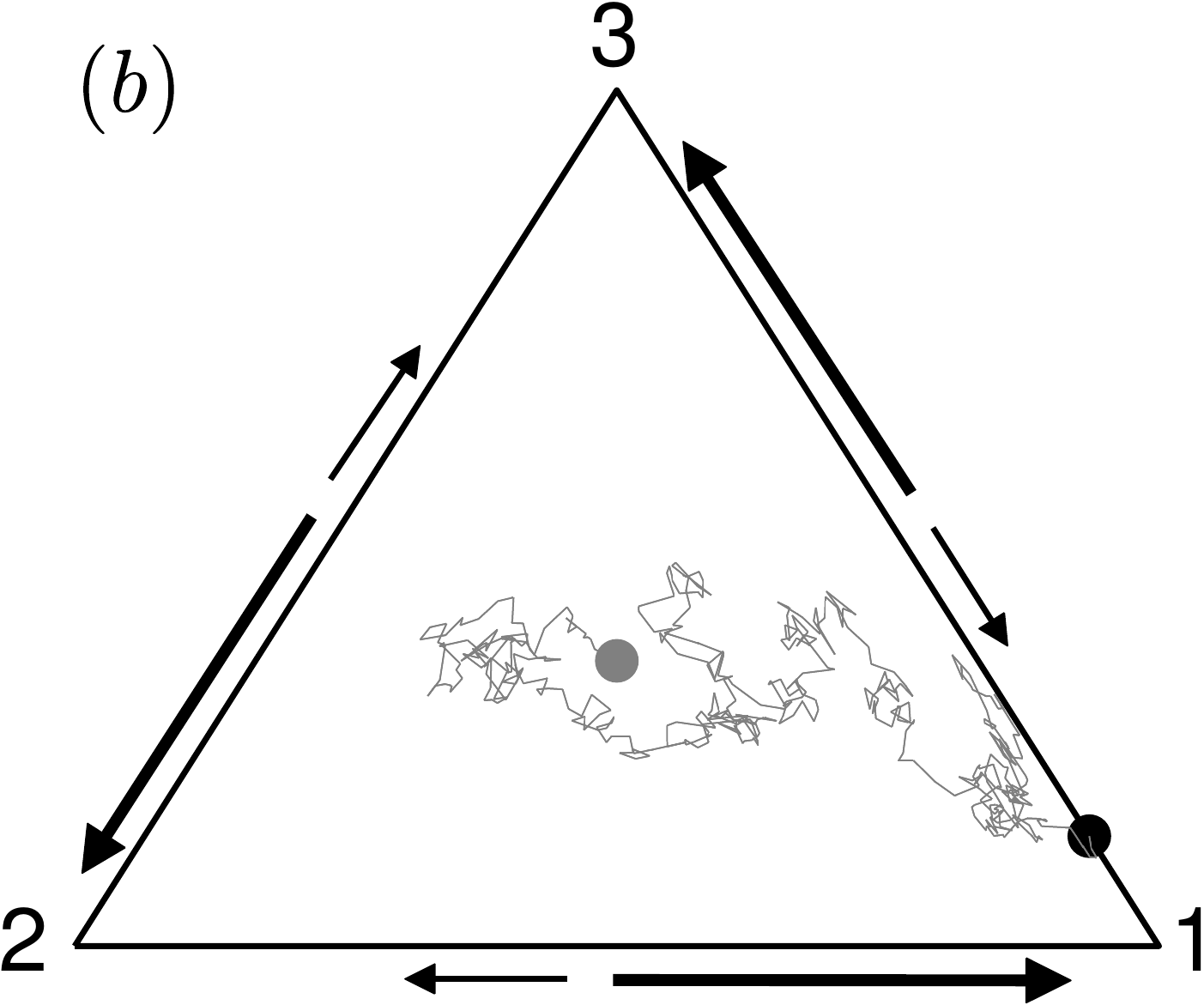}
	\caption{Stochastic orbits in $S_3$ of the constant-$K$ BDCLV ($\epsilon=0$)
	of Fig.~\ref{fig:Fig1}~(a,b), with $(s,r_1,r_2,r_3)=(1/10,3/5,1/5,1/5)$ and illustration of Stages 1 and 2 of 
	dynamics, see text. 
	Initially all species have the same density $1/3$ (gray dot), and (a) $K=10^4$, (b) $K=200$.
	(a) In Stage 1, when $sK\gg 1$, erratic trajectories approach $\partial S_3$ from the 
	outermost orbit (deterministic 
	orbit at a distance $1/K$ from $\partial S_3$, see text). (b)
	When $sK={\cal O}(10)$, in Stage 1,
	stochastic trajectories reach $\partial S_3$ without settling onto the outermost orbit. 
	Stage 2: Once on an edge of $\partial S_3$ (black dot), a  competition (shown as arrows) takes
	place between species $i$ and its weak opponent $i+1$, with the former (long arrows) more likely to win than the latter
	(short arrows), see text.}
	\label{fig:Fig2}
\end{figure}
\section{The birth-and-death cyclic Lotka-Volterra model $(\epsilon=0)$ with constant carrying capacity}\label{constK}
\label{Section3}
In order to understand how environmental variability affects
the RPS dynamics, it is useful to study first the dynamics of the model
defined by (\ref{payoff})-(\ref{eq:transrates}) with  $\epsilon=0$ when the carrying capacity $K$ is constant.
This zero-sum model ($\bar{\Pi}=0, \bar{f}=1$), is referred to as  
the constant-$K$ birth-and-death cyclic Lotka-Volterra model (BDCLV)
and its dynamics is fully described by the underpinning  ME.
Proceeding as  in \ref{AppendixA1}, the 
mean-field description of the constant-$K$ BDCLV is obtained by neglecting all fluctuations, yielding
\begin{eqnarray}
\hspace{-3mm}
\dot{N} & = & \sum_{i=1}^3 (T_i^+ - T_i^-) = N\left(1- \frac{N}{K}\right)  
\label{eq:Ndot},\\
\hspace{-3mm}
\dot{x_i}&=&\frac{T_{i}^{+}-T_{i}^{-}}{N}-x_i\frac{\dot{N}}{N}%=sx_i [r_ix_{i+1}-r_{i-1}x_{i-1}]
%\nonumber\\
=x_i [\alpha_i x_{i+1}-\alpha_{i-1}x_{i-1}],
\label{eq:xidot}
\end{eqnarray}
where $\alpha_i\equiv sr_i$, and the dot stands for the time derivative.
Clearly, the population size  obeys the logistic equation (\ref{eq:Ndot}), and thus 
$N(t)\to K$ after a time $t={\cal O}(1)$.
The rate equations (REs) for $x_i=N_i/N$ describe how the population composition changes 
due to cyclic dominance on a timescale $1/s$. 
 Eqs.~(\ref{eq:Ndot}) and (\ref{eq:xidot}) are decoupled and, when $s\ll 1$, there is a timescale separation: 
 $N$  rapidly approaches $K$ while the $x_i$'s evolve much slower.
When time is rescaled ($t\to st$), the REs (\ref{eq:xidot}) coincide with the celebrated replicator equations of the zero-sum RPS 
game~\cite{Maynard,Hofbauer,Nowak,Szabo07,Broom}. These REs are characterized 
 by a neutrally stable  fixed 
point $\vec{x}^*=\left(r_2, r_3, r_1\right)$ associated with the coexistence of a fraction $r_{i+1}$ of each species $i$, and 
three saddle (unstable) fixed points  $\left\{\vec{e}_1=(1, 0, 0),\vec{e}_2=(0, 1, 0),\vec{e}_3=(0, 0, 1)\right\}$, 
$\vec{e}_i$ corresponding to a state in which only  individuals of species $i$ are present.
In addition to conserving $x_1+x_2+x_3=1$, the REs (\ref{eq:xidot})
also conserve the quantity $\mathcal{R} = \prod_{i=1}^{3}x_i^{r_{i+1}}$. 
The deterministic trajectories  in the 
phase space $S_3$ are therefore  neutrally stable orbits  surrounding $\vec{x}^*$~\cite{Hofbauer}.
The dynamics in a finite population is characterized by 
 noisy oscillations about $\vec{x}^*$, see Fig.~\ref{fig:Fig1}~(a,b), 
with erratic trajectories    performing a random walk between the
deterministic  orbits  until  $\partial S_3$ is hit and  
one species goes extinct. This first stage of the dynamics (Stage 1) where the three species coexist
is followed by Stage 2 where the two surviving species, say $i$ and $i+1$, compete along the edge $(i,i+1)$
of $S_3$ until one them prevails and fixates,  see Fig.~\ref{fig:Fig2}.
The population size $N(t)$ is not constant but, after $t={\cal O}(1)$,  fluctuates 
about $K$, with fluctuation intensity that decreases with $K$, see Fig.~\ref{fig:Fig1}~(a,b).
 It is worth noting that the population size  keeps fluctuating, $N(t)\approx K$, even after Stage 2 
 when it consists of only the species having fixated in Stage 2, see Footnote~\ref{coll}.

The fact that, after a short transient, $N(t)\approx K$ suggests 
a relation between  the
constant-$K$ BDCLV and
the cyclic Lotka-Volterra model evolving according to a  Moran process in a population of 
constant size $N=K$~\cite{Claussen08,Szabo07,Mobilia10,Galla11}, see \ref{AppendixA2}. 
In the Moran cyclic Lotka-Volterra model (MCLV), the birth of 
an  $i$-individual and the death 
of an individual of type $j\neq i$ occurs simultaneously: In  
the MCLV, an $i$ replaces a $j$
 with  rate $T_{j \to i}$ and the population size  remains constant, see, {\it e.g.},~\cite{Claussen08,Mobilia10,Galla11}. 
In \ref{AppendixA2},  the 
constant-$K$ BDCLV is shown to have the same fixation properties 
as the  MCLV with transition rates
$T_{j \to i}=T_i^+ T_j^-/K$ and $N=K$, see Fig.~\ref{fig:FigA1}. 

It is also useful to compare the constant-$K$ BDCLV  with the so-called chemical  cyclic Lotka-Volterra model (cCLV),
see  \ref{AppendixA3}. In the cCLV, the cyclic competition between the three species is of predator-prey type:
An  $i$-individual (predator) kills an $(i+1)$-individual (its prey) and immediately replaces it,
leaving the population size  constant. In \ref{AppendixA3}, we show that the cCLV admits the same mean-field dynamics
as the constant-$K$ BDCLV, see Eq.~(\ref{eq:MF_cCLV}). However, once a species has gone extinct in the cCLV, 
there is a predator-prey competition in Stage 2 won by the predator with a probability $1$. Hence, 
Stage 1 survival and fixation probabilities coincide in the cCLV.
Remarkably, it was found that these quantities
obey two simple laws, the so-called ``law of the weakest'' (LOW) when $N$ is large
and the ``law of stay out'' (LOSO) in smaller populations~\cite{Berr09,Frean01,West18}, see \ref{AppendixA31}
and Fig.~\ref{fig:FigA2}. 

As detailed in \ref{AppendixB}, the stage 1 dynamics of 
the constant-$K$ BDCLV is similar to the stage 1 cCLV dynamics in a population 
of size ${\cal O}(sK)$. The stage 2 dynamics in the constant-$K$ BDCLV and MCLV with $N=K$ are similar, with 
both surviving species having a non-zero probability to fixate, see \ref{AppendixB}. 

In what follows, we exploit the relationships between the BDCLV and  
the  MCLV and cCLV to shed light on its fixation properties when $K$ is constant and randomly switching.
In particular, we study the novel survival  scenarios that can arise when 
$N(t)$ fluctuates.

\subsection{Survival, absorption and fixation probabilities in the constant-$K$ BDCLV}
\label{Sec3.1}
All three species coexist during Stage 1: In the constant-$K$ BDCLV
their fractions erratically oscillate
about $\vec{x}^*$ until $\partial S_3$ is hit, see  Figs.~\ref{fig:Fig1}~(a,b) and \ref{fig:Fig2}.
 Stage 1 ends at this point and is characterized by the probability $\phi_{i,i+1}$ 
to have reached the edge $(i,i+1)$ (survival of species $i$ and $i+1$) or, equivalently,
that species $i-1$ is the first to die out. Once on  $\partial S_3$, Stage 2 starts and two species, say 
$i$ and $i+1$, compete along their edge  until either $i$, with probability $\phi_i$, or 
$i+1$, with probability $1-\phi_i$, get absorbed.
Clearly, the stage 2 dynamics is conditioned by the  outcome of Stage 1 and the overall fixation probability $\widetilde{\phi_i}$ depends
on $\phi_{i,j}$ and $\phi_{i}$, see Eq.~(\ref{eq:PhiTot}).

 Below, we show that $\phi_{i,i+1}, \phi_i$ and $\widetilde{\phi}_i$ are functions of $sK$, see Figs.~\ref{fig:Fig3} and \ref{fig:Fig4},
 and  can respectively be inferred from the well-known properties of the cCLV and MCLV, see 
 \ref{AppendixA}. 
% This will allow us to obtain the overall fixation probabilities.
 In our discussion, we distinguish three regimes:
  (i) {\it quasi-neutrality}, when $sK\ll 1$ and $K\gg 1$;
  (ii) {\it weak selection}, when $sK={\cal O}(10)$, with $s\ll 1$ and $K\gg 1$; and 
  (iii) {\it strong selection}, when $sK\gg 1$, with $s={\cal O}(1)$ and $K\gg 1$.
In the examples below, these three regimes are identified as follows: 
 $s \lesssim 1/K$ in regime (i), $1/K \lesssim  s \lesssim  100/K$ in regime (ii),
and $s \gtrsim 100/K$ in regime (iii), with $K\gg 1$. Furthermore, since the overall fixation 
probability of each species $\widetilde{\phi_i}$ is trivially $1/3$ when $r_1=r_2=r_3=1/3$~\cite{RMF06,Berr09,West18},
we focus on the general case where the $r_i$'s are unequal.
All figures have been obtained with the initial fraction $1/3$ of each species, i.e. $\vec{x}_0\equiv (x_1(0),x_2(0),x_3(0))=\vec{x}_c
\equiv (1,1,1)/3$, and
we consider the following set of parameters:
  $\vec{r}\equiv (r_1,r_2,r_3) =\vec{r}^{(1)}\equiv(1,5,5)/11$ and $\vec{r}=\vec{r}^{(2)}\equiv(3,1,1)/5$. 
These choices suffice to reveal most of the generic properties of the system.
When we study how  $\phi_{i,i+1}$, $\phi_i$ and $\widetilde{\phi_i}$ depend on $sK$, in Figs.~\ref{fig:Fig3} and  \ref{fig:Fig4}
  we consider $K\in \varkappa \equiv\{1000, 450, 250, 90, 50\}$ and 
  $s=1$ for $K=1000$, 
  $s\in \{10^{-k/4}, k=0\dots 3\}$ for $K=450$,
  $s\in \{10^{-(2+k)/4}, k=0\dots 9\}$ for $K=250$,
  $s\in \{10^{-k/4}, k=0\dots 8\}$ for $K=90$, 
  and $s\in \{10^{-(9+k)/4}, k=0\dots 3\}$  for $K=50$. 
  In all figures (except Figs.~\ref{fig:Fig1} and \ref{fig:Fig2}), simulation results have been sampled over 
  $10^4-10^5$ realizations. 

\subsubsection{Stage 1: Survival probabilities in the constant-$K$ BDCLV}
\label{Sec3.1.1}
The stage 1 dynamics of the constant-$K$ BDCLV
and cCLV with $N={\cal O}(sK)$ are similar, see \ref{AppendixB}.
The constant-$K$ BDCLV survival probabilities $\phi_{i,j}$
are therefore similar to the survival/fixation  probabilities in the cCLV.
 These obey the LOW when $N$ is large and the LOSO in smaller populations~\cite{Berr09,Frean01,West18}, see \ref{AppendixA31}.
%
%These quantities obey the so-called law of the weakest (LOW) when $N$ is large
%and the law of stay out (LOSO) in smaller populations~\cite{Berr09,Frean01,West18}, see \ref{AppendixA31}
%and Fig.~\ref{fig:FigA2}. 
The LOW and LOSO are here used to determine $\phi_{i,j}$ in regimes (ii) and (iii).
\begin{figure}
\centering
	\includegraphics[width=0.9\linewidth]{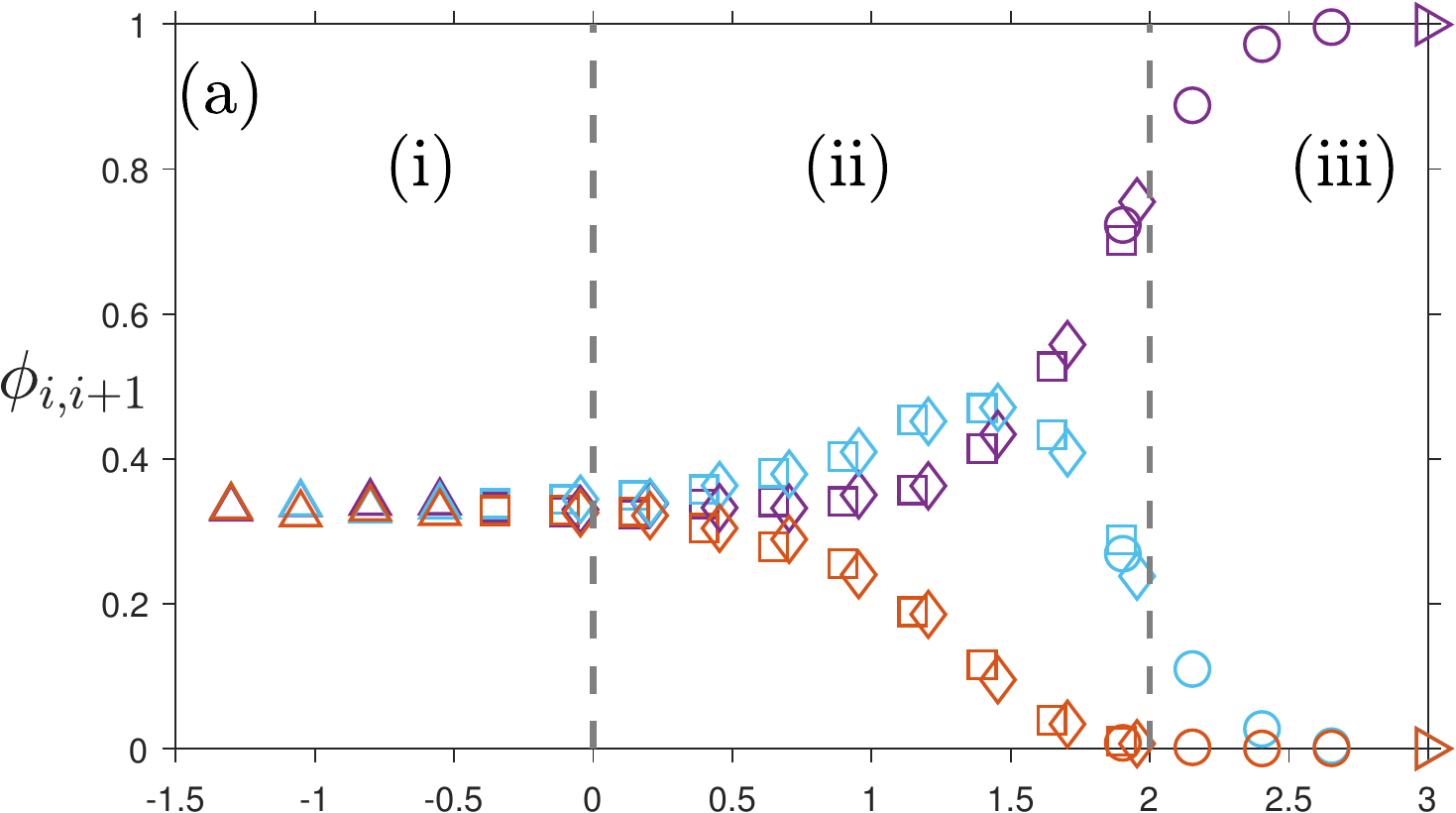}
	\includegraphics[width=0.9\linewidth]{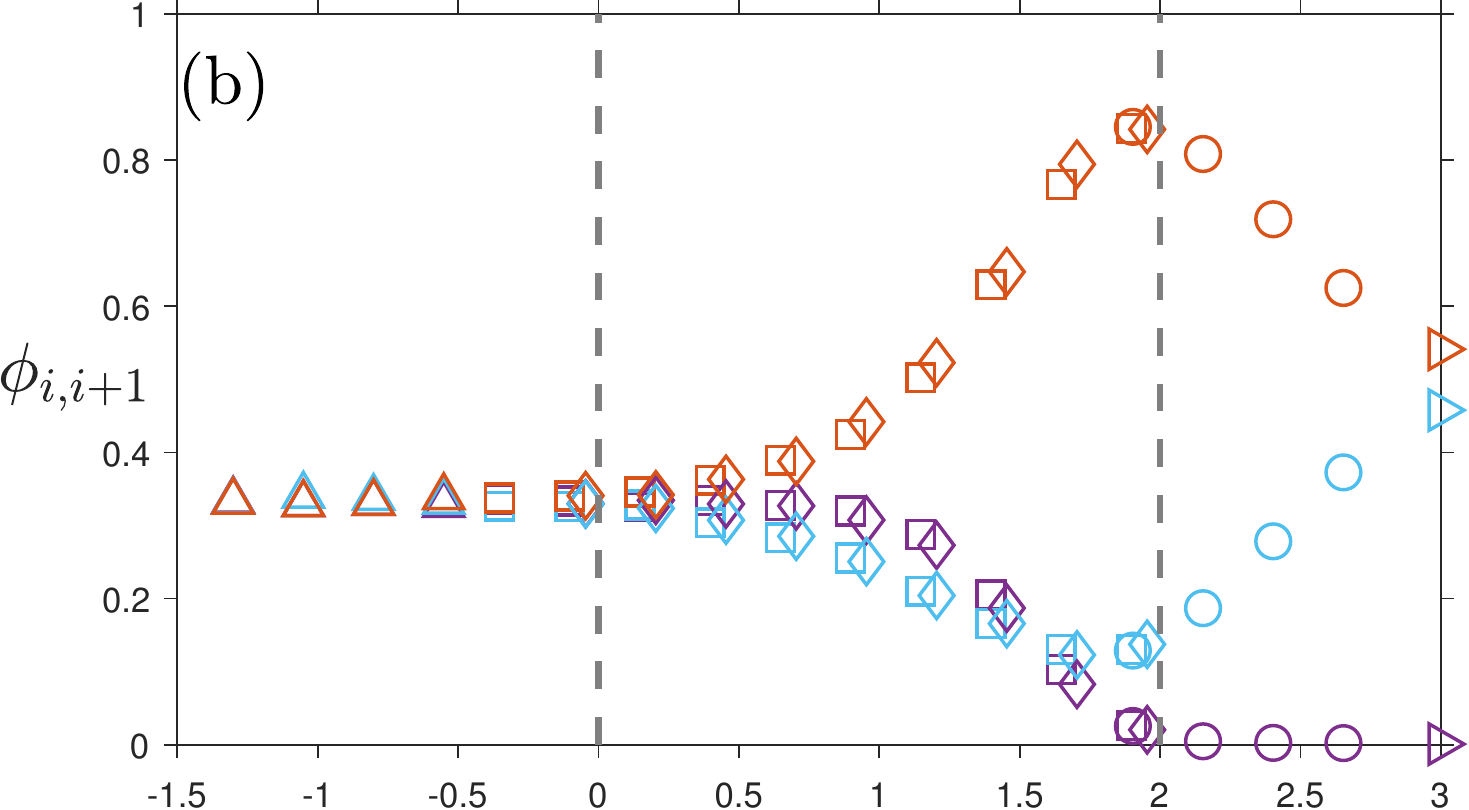}
	\includegraphics[width=0.9\linewidth]{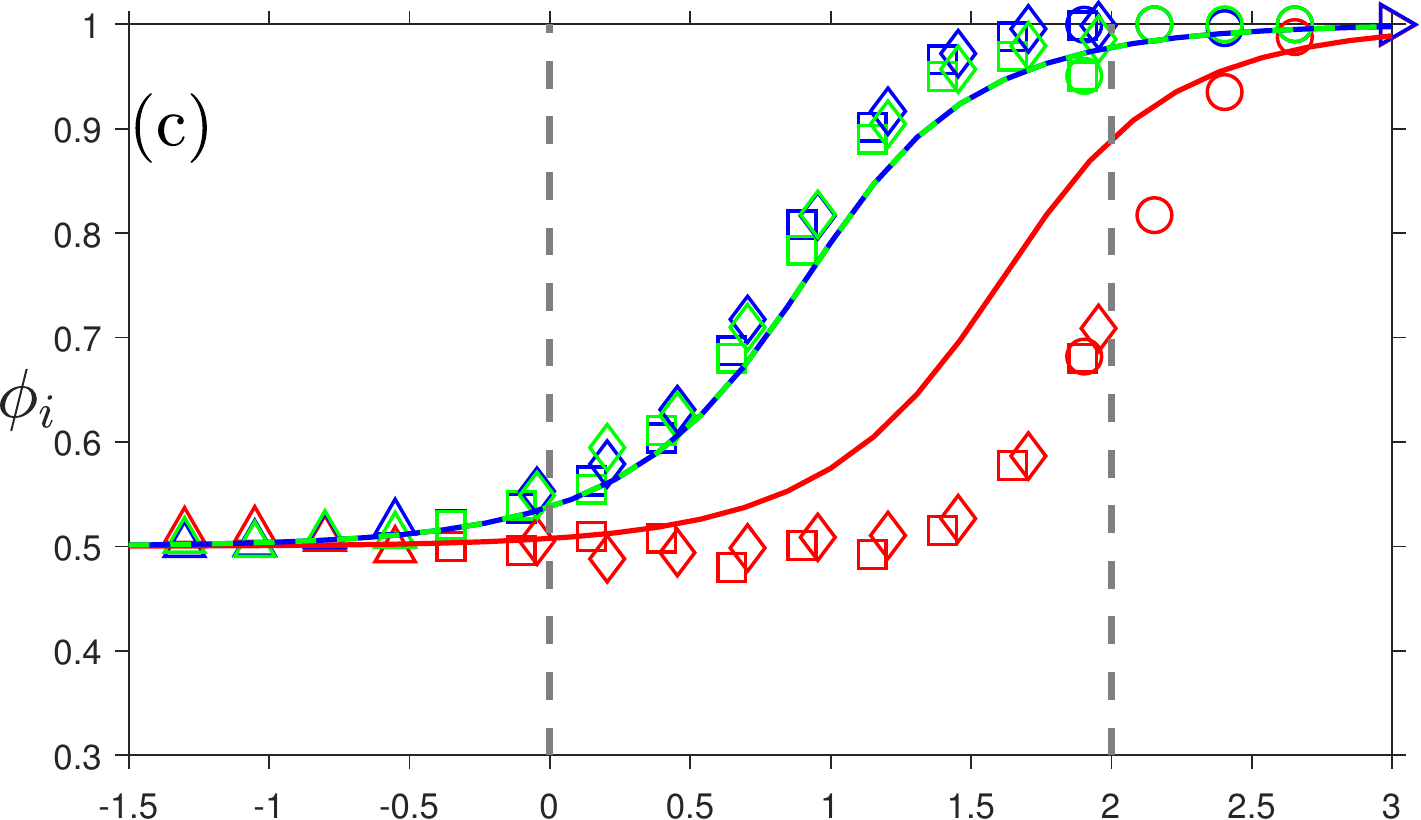}
	\includegraphics[width=0.9\linewidth]{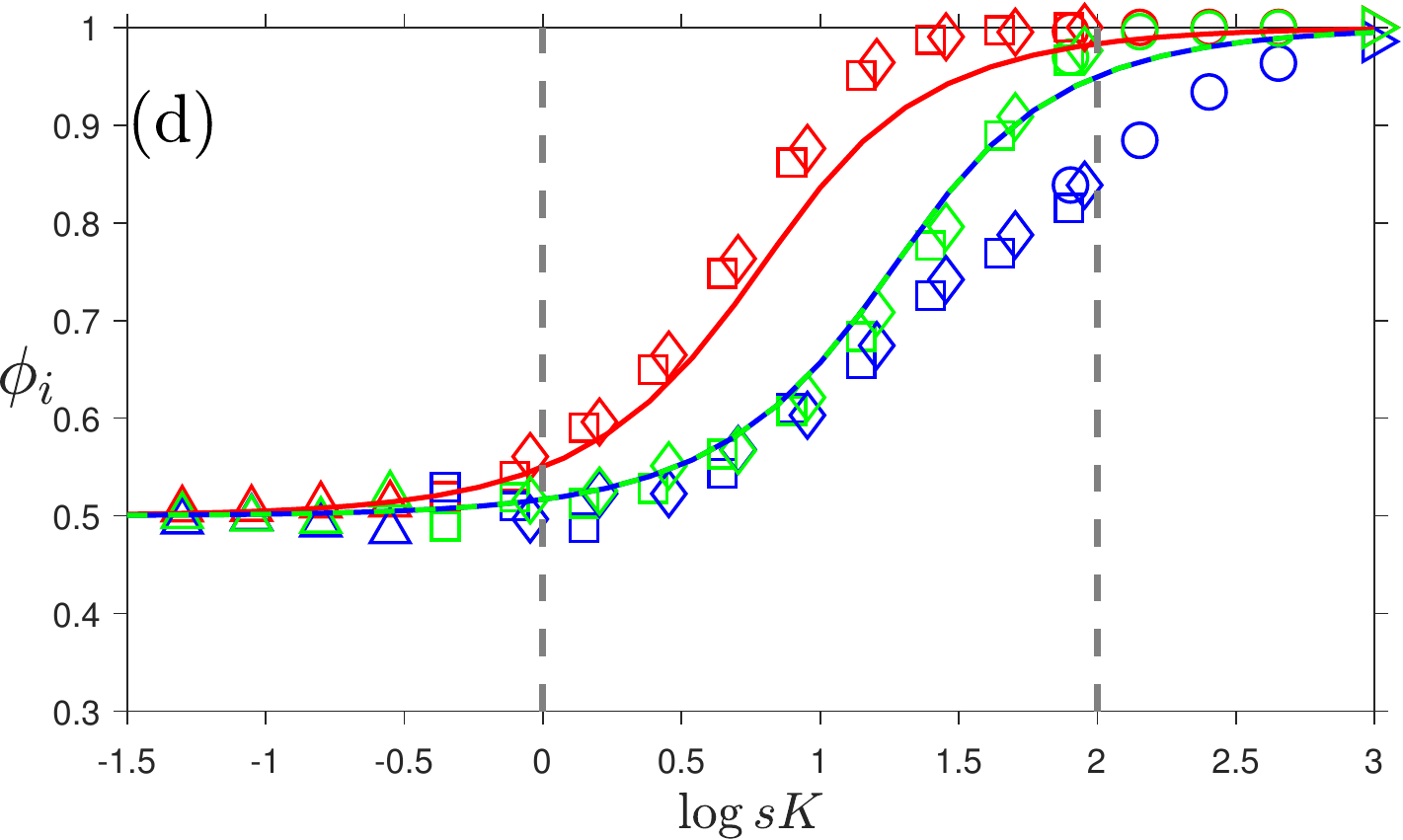}
	\caption{(a,b) Constant-$K$ BDCLV survival probabilities simulation results ($\Diamond$):
	$\phi_{1,2}$ (purple), $\phi_{2,3}$ (light blue) and 
	$\phi_{3,1}$ (orange) vs. $sK$ for   values of  $s\in (10^{-3},1)$  and $K \in \varkappa$ 
	  in regimes (i)-(iii) separated by dashed lines, see text. 
	Non-monotonicity arises across regimes (ii) and (iii) and can be explained in terms of the LOSO
	(regime (ii)) and  
	LOW (regime (iii)), see text.
	(a) $\vec{r}=\vec{r}^{(1)}$;  species $1$ and $3$ are the most likely to die out in regime (ii)
        and (iii), respectively.	
	(b) $\vec{r}=\vec{r}^{(2)}$ ;  species $2$ and $1$ are the most likely to die out in regime (ii)
        and (iii), respectively.	
	(c,d)  Constant-$K$ BDCLV absorption probabilities
	$\phi_i$ vs. $s K$: 
	$\phi_{1}$ (red), $\phi_{2}$ (blue) and 
	$\phi_{3}$ (green) vs. $sK$ for $K = (1000, 450, 250, 50, 20)$, with  
	(c) $\vec{r}=\vec{r}^{(1)}$ and  (d) $\vec{r}=\vec{r}^{(2)}$. The solid line is given by (\ref{eq:condfixprob2})
	and coincide for species $2$ and $3$.
	In all panels $K = 1000~(\rhd) , 450~(\circ), 250,~(\diamond), 90~(\square), 50~(\triangle)$,  $\epsilon=0$, 
	$\vec{x}_0=\vec{x}_c$.
	}  
	\label{fig:Fig3}
	\end{figure}

\vspace{0.1cm}

- Regime (i): When $sK \ll 1$, with $K\gg 1$, the system is at quasi-neutrality. The dynamics is driven by demographic
fluctuations and all species have the same survival probability $\phi_{i,i+1}\approx 1/3$, see  (i) in
Fig.~\ref{fig:Fig3}~(a,b).

\vspace{0.1cm}
- Regime (ii):  When $sK ={\cal O}(10)$ and $K\gg 1$, 
the intensity of selection strength is weak ($s\ll 1$) and comparable to 
that of  demographic fluctuations. From the relation with the cCLV,
we infer that $\phi_{i,i+1}$ is given by the fixation probability $\phi_{i}^{{\rm cCLV}}\vert_{sK}$  of species $i$
in the cCLV in a population of size of order $sK$, i.e.
$\phi_{i,i+1}\approx \phi_{i}^{{\rm cCLV}}\vert_{sK}$. In regime (ii),
 $\phi_{i}^{{\rm cCLV}}\vert_{sK}$ obeys the LOSO,
see \ref{AppendixA31}, and from Eq.~(\ref{eq:A1_LOSO})  we obtain: 
\begin{eqnarray}
\label{eq:BDCLV_LOSO}
&&\phi_{i-1,i}> \phi_{i,i+1},\phi_{i+1,i-1} \quad
\text{if $\quad r_i>r_{i\pm1}$} \\
&& \phi_{i,i+1} \approx \phi_{i+1,i-1}>\phi_{i-1,i} \quad 
\text{if $\quad r_{i+1}=r_{i-1}>r_i$}. \nonumber
 \end{eqnarray}
 Accordingly, when   $r_i> r_{i\pm 1}$  species $i-1$ and $i$
 are the most likely to survive  Stage 1 under weak selection, as confirmed by Fig.~\ref{fig:Fig3}~(b).
 When   
 $r_{i+1}=r_{i-1}>r_i$ and $sK ={\cal O}(10)$, the edges $(i,i+1)$ and $(i+1,i-1)$
 are the most likely to be hit, while species $i-1$ is most likely to die out
 first,  see Fig.~\ref{fig:Fig3}~(a).
While the $\phi_{i,j}$'s obey the LOSO, we notice that $\phi_{i,j}\approx 1/3$ when $s\ll 1$.

\vspace{0.1cm}

- Regime (iii): When $sK \gg 1$, with $s={\cal O}(1)$ and $K\gg 1$,
the stage 1 dynamics is governed by cyclic dominance. An edge of $S_3$
is hit from the system's outermost orbit as in the cCLV, see \ref{AppendixB} and Fig.~\ref{fig:Fig2}~(a).
From the relation between
the constant-$K$ BDCLV and the cCLV, we  have 
$\phi_{i,i+1}\approx \phi_{i}^{{\rm cCLV}}\vert_{sK}$ which obeys the LOW in regime (iii), and therefore
from (\ref{eq:A1_LOW}) we  have: 
\begin{eqnarray}
\label{eq:BDCLV_LOW}
&&\phi_{i,i+1}> \phi_{i+ 1,i-1}, \phi_{i-1,i} \; \text{if $\quad r_i< r_{i\pm 1},$ } \\
&&\phi_{i,i+1}\approx \phi_{i+1,i-1}>\phi_{i-1,i} \;  \text{if $\quad r_i=r_{i+1}<r_{i-1}$.
%and 
%$\phi_{i-1}>\phi_{i},\phi_{i+1}$ if $k_i=k_{i+1}>k_{i-1}$
} \nonumber
 \end{eqnarray}
When $sK\gtrsim 10^3$, the LOW becomes asymptotically a zero-one law: $
\phi_{i,i+1}\to 1, \phi_{i-1,i} \to 0$ and $\phi_{i+1,i-1} \to 0$
if $r_i< r_{i\pm 1}$, and $\phi_{i,i+1}=\phi_{i+1,i-1} \to 1/2, \phi_{i-1,i+1}\to 0$
 if $r_i=r_{i+1}<r_{i-1}$, see Eq.~(\ref{eq:A1_LOW2}).
 Accordingly, when  $sK\gg 1$ and $r_i< r_{i\pm 1}$  
 species $i$ and $i+1$ are most likely to survive and 
 species $i-1$ the most likely to die out 
 in Stage 1, in agreement with Fig.~\ref{fig:Fig3}~(a).
\vspace{0.1cm}

The relations (\ref{eq:BDCLV_LOSO}) and (\ref{eq:BDCLV_LOW}) 
explain that $\phi_{i,i+1}$ is a function of $sK$
that can exhibit a non-monotonic behavior.
For instance, for $\vec{r}=\vec{r}^{(1)}$ as in Fig.~\ref{fig:Fig3}~(a),
the relations (\ref{eq:BDCLV_LOSO}) yield
 $\phi_{1,2}\approx\phi_{2,3}>\phi_{3,1}$ when $sK={\cal O}(10)$,
 and (\ref{eq:BDCLV_LOW})
predict  $\phi_{1,2}>\phi_{2,3},\phi_{3,1}$ when $sK\gg 1$, 
while  $\phi_{1,2}\approx\phi_{2,3}\approx \phi_{3,1}\approx 1/3$ when $sK\ll 1$.
From these results, it is clear that
$\phi_{2,3}$ increases  across the regimes (i)-(ii),
and then decreases with $sK$ across the regimes (ii)-(iii), whereas $\phi_{1,2}$ and 
$\phi_{3,1}$ respectively increases and decreases
with $sK$ across all regimes.

\subsubsection{Stage 2: Absorption probabilities in the constant-$K$ BDCLV}\label{BDCLVP2fp}
At start of Stage 2, species $i$ competes against  $i+1$ (weak opponent),
along the edge $(i,i+1)$ where their fitnesses are $f_{i}=1+\alpha_i(1-x_i)$
and $f_{i+1}=1-\alpha_i x_i$, see (\ref{eq:fi}). Stage 2 ends with the absorption
of either $i$ or $i+1$, respectively with probability 
$\phi_i$ and $1-\phi_i$.

\vspace{0.1cm}

- At quasi neutrality, species $i$'s selective advantage is negligible since  
$f_{i}-f_{i+1}=\alpha_i \ll 1$. In regime (i), 
species $i$ and $i+1$ have therefore almost the same absorption probability $\phi_i\approx 1/2$.

\vspace{0.1cm}

- Under strong selection, species  $i$ has an important selective advantage over
species   $i+1$: $f_{i}-f_{i+1}={\cal O}(1)$.
 In regime (iii), species $i$ is almost certain to be absorbed as in Stage 2 of the cCLV dynamics, and therefore 
$\phi_i\approx 1$ as predicted by the LOW, see Appendices A.3 and B.

\vspace{0.1cm}

- Under weak selection, in regime (ii), 
 $\phi_i$  is nontrivial and can be obtained from the fixation probability $\phi_i\vert_K$ 
of species $i$ in the MCLV with $N=K$, see Appendices A.2 and C. 
When the stage 2 dynamics
starts with a fraction $\hat{x}_i$ of individuals of species $i$,  $\phi_i\vert_K$  
under weak selection
is obtained from the backward Fokker-Planck generator 
\begin{eqnarray}
\hspace{-5mm}
\label{eq:G}
{\cal G}_{(i,i+1)}|_{K}=\frac{x_i(1-x_i)}{K}\left[K\alpha_i\frac{d}{d x_i} +
\frac{d^2}{
d x_i^2}\right],
 \end{eqnarray}
by solving ${\cal G}_{(i,i+1)}|_{K}(\hat{x}_i)\phi_i\vert_K(\hat{x}_i)=0$ with $\phi_i\vert_K(1)=1-\phi_i\vert_K(0)=1$, see
Eq.~(\ref{eq:A3_phi1}), yielding 
\begin{eqnarray}
 \label{eq:phi_S2}
 \phi_i\simeq \phi_i(\hat{x_i})|_{K}=\frac{1-e^{-\alpha_i K \hat{x_i}}}{1-e^{-\alpha_i K}}.
 \end{eqnarray}
A difficulty arises from  $\hat{x_i}$ being a random variable 
depending  on the outcome of Stage 1: $\hat{x_i}$ is
distributed according to the probability density $P_{(i,i+1)}(\hat{x_i})$. 
The absorption  probability 
 is thus  obtained by averaging (\ref{eq:phi_S2}) %$\phi_i|_{K}$
over $ P_{(i,i+1)}$:
\begin{equation}
\label{eq:condfixprob1}
\phi_i\simeq
\phi_i|_{K} = \int_{0}^{1} P_{(i,i+1)}(\hat{x_i})~\phi_i(\hat{x_i})|_{K}~
%\left(\frac{1-e^{-\alpha_i K \hat{x_i}}}{1-e^{-\alpha_i K}}\right)~
d\hat{x_i}.
\end{equation} 
In practice, $P_{(i,i+1)}(\hat{x_i})$ is obtained from stochastic simulations, see \ref{AppendixD}.
Analytical progress can be made by noticing that  in regime (ii) where $s\ll 1$ and $sK\lesssim 10$, 
each  pair $i,i+1$ has approximately the same  survival probability at the end of Stage 1 ($\phi_{i,i+1} \approx 1/3$, 
see Fig.~\ref{fig:Fig3}~(a,b)),
and the  initial distribution along  $(i,i+1)$ can be assumed to be uniform, i.e.
$P_{i,i+1}(\hat{x}_i)\approx 1$, see  Fig.~\ref{fig:FigD3}.  Substituting in Eq.~(\ref{eq:condfixprob1}), 
we obtain  the approximation ($s\ll 1, sK\lesssim 10$):
\begin{equation}
\label{eq:condfixprob2}
\phi_i\simeq \phi_i|_K\approx\frac{e^{-\alpha_i K} +\alpha_i K -1}{\alpha_i K (1-e^{-\alpha_i K})},
\end{equation}
which is an S-shaped function of $\alpha_i K$ that correctly predicts the behaviors
$\phi_i\to 1/2$ when $\alpha_i K \ll 1$ (regime (i)) and $\phi_i\to 1$ when $\alpha_i K \gg 1$ (regime (iii)), 
see Fig.~\ref{fig:Fig3}~(c,d).
Comparison with simulation results of Fig.~\ref{fig:Fig3}~(c,d) confirm  that
$\phi_i$ is sigmoid function of $sK$ and  
Eq.~(\ref{eq:condfixprob2}) provides 
a good approximation of $\phi_i$ when the assumption $P_{(i,i+1)}\approx 1$ holds, see Fig.~\ref{fig:FigD3}.

\begin{figure}
	\centering
	\includegraphics[width=0.98\linewidth]{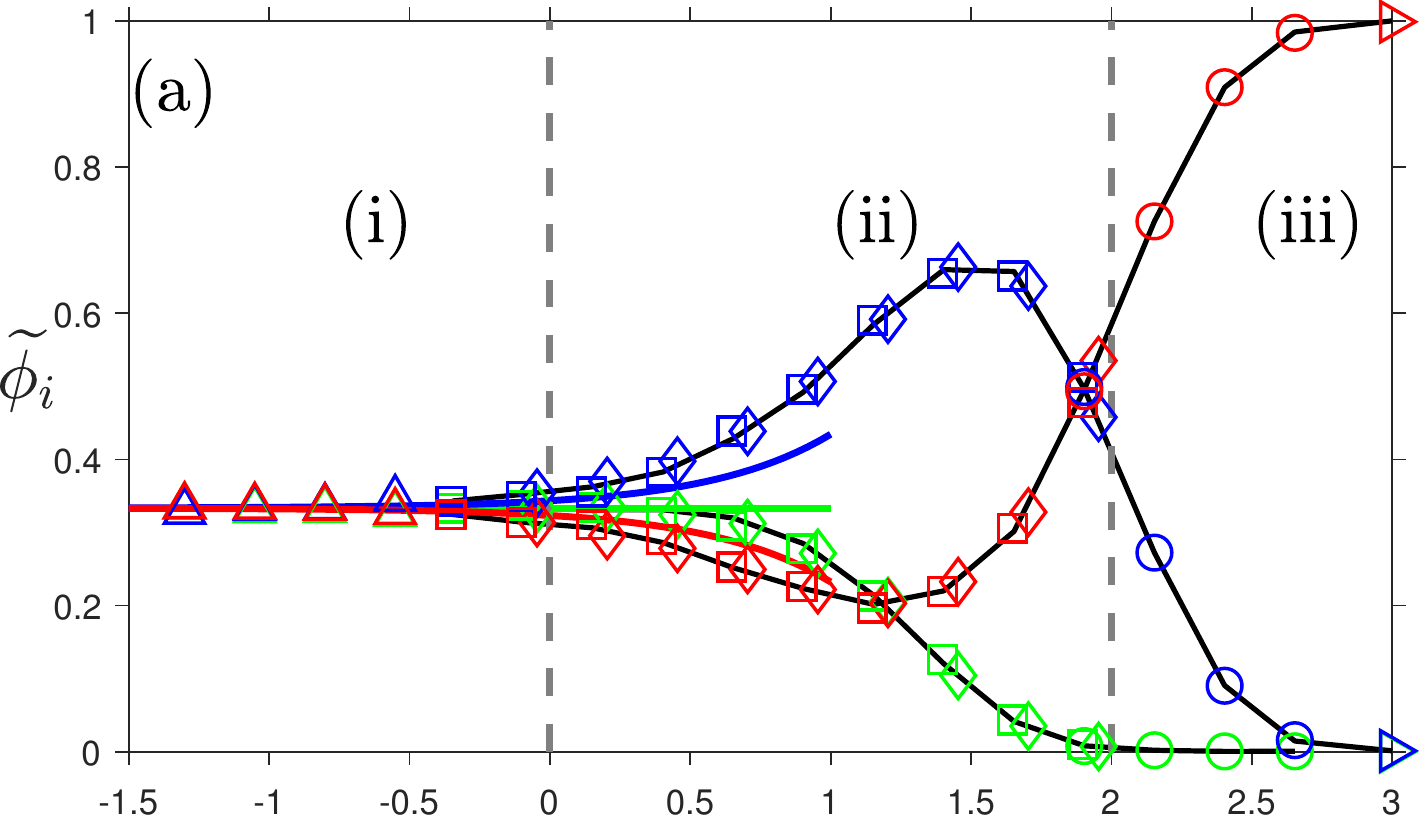}\\
	\includegraphics[width=0.98\linewidth]{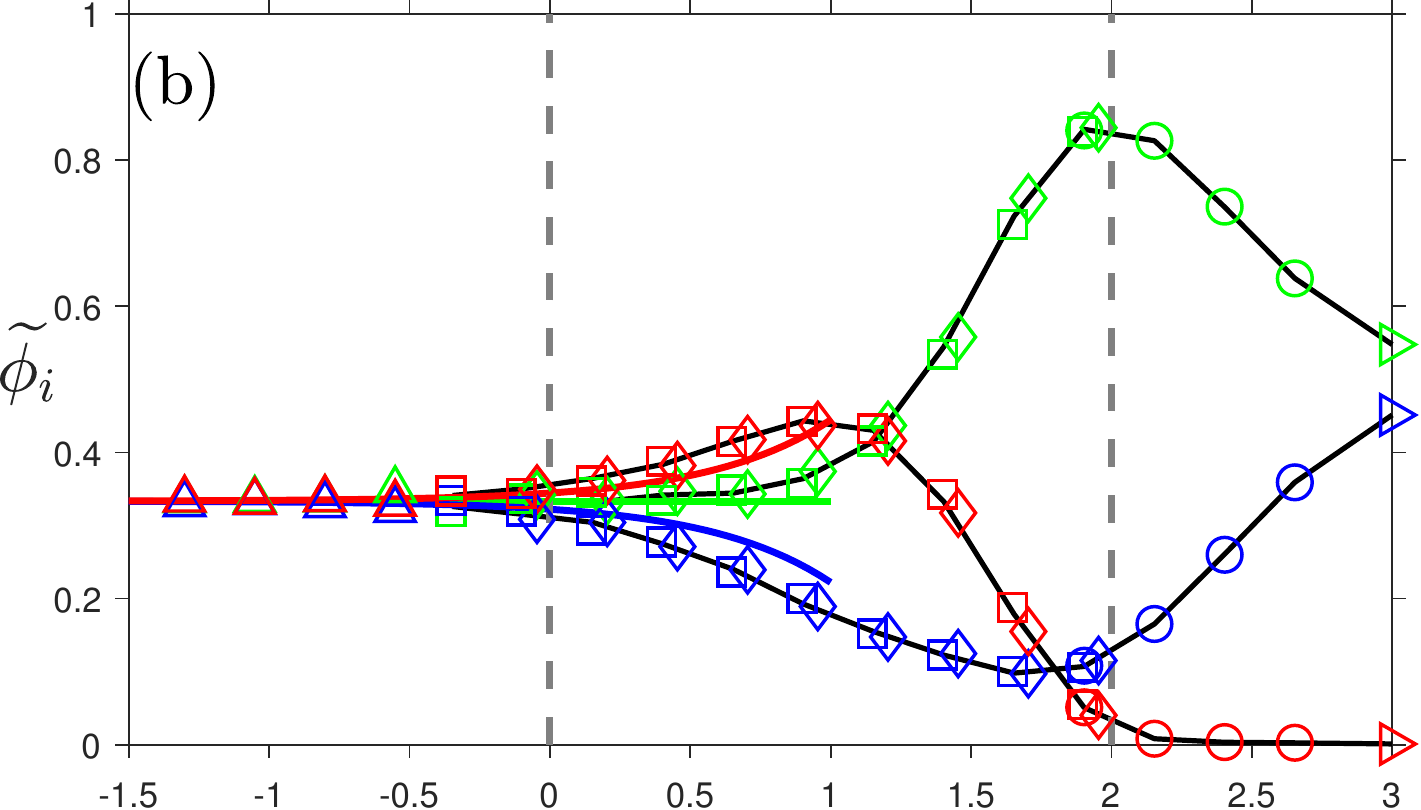}\\
        \includegraphics[width=0.925\linewidth]{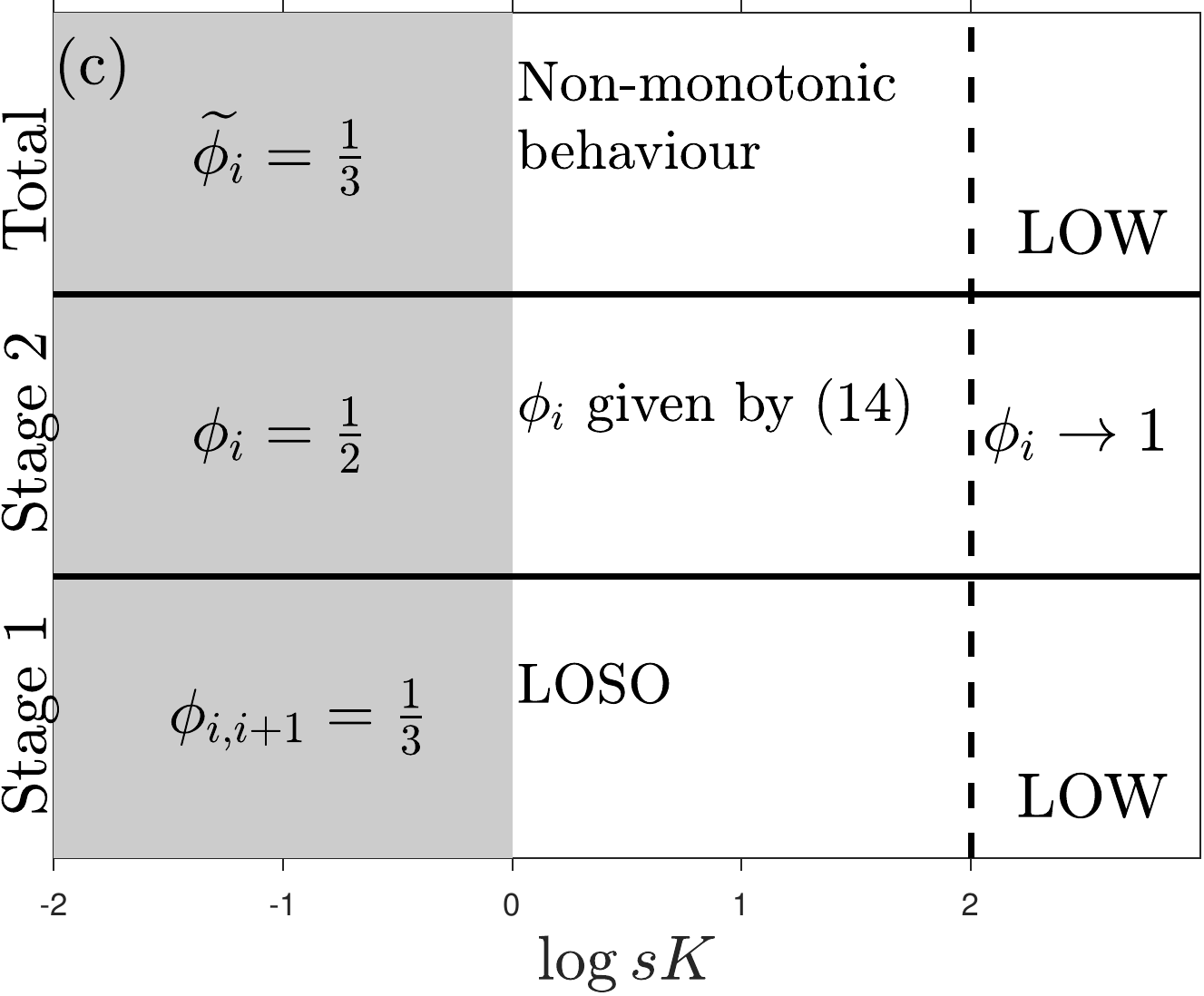}
	\caption{(a,b) Total fixation probabilities $\widetilde{\phi}_1$ (red), 
	$\widetilde{\phi}_2$ (blue), $\widetilde{\phi}_3$ (green) 
	vs. $sK$
	for  values of  $s\in (10^{-3},1)$  and $K \in \varkappa$ 
	with symbols as in Fig.~\ref{fig:Fig3}, see text. Regimes (i)-(iii), from left to right, are indicatively
	separated by dashed gray lines.	
	(a) $\vec{r}=\vec{r}^{(1)}$; (b) $\vec{r}=\vec{r}^{(2)}$. 
	The solid black lines show the predictions of (\ref{eq:PhiTot}) using (\ref{eq:condfixprob1}), with 
	$\phi_{i,i+1}$ and $P_{(i,i+1)}$ inferred from simulations.
	 Predictions from (\ref{eq:TotfixR1}) 
	are shown as solid  colored line. $\widetilde{\phi}_i$ 
	can display a non-monotonic dependence 
	on $sK$ across regimes (ii)-(iii), see text. 
	(c) Chart summarizing the outcome of Stage 1, Stage 2 and the overall 
	fixation probability $\widetilde{\phi}_i$ as function of $sK$ in regimes (i)-(iii), from left to right.
	In all panels: $\vec{x}_0=\vec{x}_c$ and $\epsilon=0$.
	}
	\label{fig:Fig4}
\end{figure}
\subsubsection{Total fixation probabilities in the constant-K BDCLV}\label{BDCLVP2tfp}
Species $i$'s total fixation probability $\widetilde{\phi}_i$  consists of two contributions: $\phi_{i,i+1}\phi_i$
and $\phi_{i-1,i}(1-\phi_{i-1})$.
The first one counts the probability  for $i$ to fixate after 
hitting the edge $(i,i+1)$, with a probability $\phi_{i,i+1}$,
and prevailing against $i+1$ (weak opponent) with a probability $\phi_i$.   
We also need to consider that, after reaching the edge $(i-1,i)$  with a probability
$\phi_{i-1,i}$, species $i$ has a  
 probability $1-\phi_{i-1}$ to win against  $i-1$ (strong opponent),  which yields $\phi_{i-1,i}(1-\phi_{i-1})$. 
With these two contributions, we obtain
\begin{eqnarray}
 \label{eq:PhiTot}
 \widetilde{\phi}_i=\phi_{i,i+1}\phi_i + \phi_{i-1,i}(1-\phi_{i-1}),
\end{eqnarray}
which is also a function of $sK$, see Fig.~\ref{fig:Fig4}~(a,b).
Of particular interest is the situation where the selection intensity is weak, $s\ll 1$, 
in which case (\ref{eq:PhiTot}) can be simplified by noting 
$\phi_{i,i+1}\approx \phi_{i-1,i}\simeq 1/3$ and using the result $\phi_i\simeq \phi_i|_K$,
given by (\ref{eq:condfixprob2}),
 for the absorption probability in the MCLV with $N=K$, see \ref{AppendixA2}, yielding
\begin{eqnarray}
 \label{eq:PhiTot1}
 \widetilde{\phi}_i\simeq\frac{1}{3}\left(1+ \phi_i - \phi_{i-1}\right)
 \approx \frac{1}{3}\left(1+ \phi_i|_K - \phi_{i-1}|_K\right).
\end{eqnarray}
Using the properties of the survival and absorption probabilities  $\phi_{i,j}$  and $\phi_{j}$  discussed above,
we can infer those of $\widetilde{\phi}_i$ in the  regimes (i)-(iii):
\\
\vspace{0.1cm}
- Regime (i): At quasi-neutrality,  all species have the same 
fixation probability to first order: $\widetilde{\phi}_i=1/3 +{\cal O}(sK)$.
An estimate of the subleading correction is obtained by noticing
$ \phi_i|_{K}\simeq
\frac{1}{2}\left(1+ \alpha_i K/6\right) $ when  $\alpha_i K \ll  1$.
This, together with 
Eq.~(\ref{eq:PhiTot1}), gives
\begin{eqnarray}
\label{eq:TotfixR1}
\widetilde{\phi}_i\simeq \frac{1}{3}\left(1+\frac{sK}{12}(r_i-r_{i-1})\right).
\end{eqnarray}
This result allows us to understand which are the species (slightly) favored by selection:
When $r_1<r_2,r_3$, Eq.~(\ref{eq:TotfixR1}) predicts that $\widetilde{\phi}_1$ is less than $1/3$ 
and decreases with $sK$, while
$\widetilde{\phi}_2>1/3$ and increases with $sK$, and $\widetilde{\phi}_3=1/3 +{\cal O}(s^2)$.
These predictions agree with the simulation results of Fig.~\ref{fig:Fig4}~(a) in regime (i).
\\
\vspace{0.1cm}
- Regime (iii): Under strong selection, 
the total fixation probability obeys the LOW, as in the cCLV (see \ref{AppendixB}).
The species overall fixation probabilities are therefore ordered as  follows, see Eqs.~(\ref{eq:A1_LOW}, \ref{eq:A1_LOW2}):
\begin{eqnarray}
\label{eq:TotfixR3}
&&\widetilde{\phi}_i> \widetilde{\phi}_{i+ 1}, \widetilde{\phi}_{i-1} \quad  \text{if \quad $r_i< r_{i\pm 1}, \;\;\;$ and \;\;\;} \nonumber\\
&&\widetilde{\phi}_{i} \approx \widetilde{\phi}_{i+1}>\widetilde{\phi}_{i-1}\quad \text{if} \quad r_i=r_{i+1}<r_{i-1},
%and 
%$\phi_{i-1}>\phi_{i},\phi_{i+1}$ if $k_i=k_{i+1}>k_{i-1}$
 \end{eqnarray}
with $\widetilde{\phi}_i\approx \phi_{i,i+1} \overset{sK \gg 1}{\longrightarrow}1$, $1/2$ or $0$.
These predictions agree with the simulations results of Fig.~\ref{fig:Fig4}~(a,b).
\\
\vspace{0.1cm}
- Regime (ii): Under weak selection, $\widetilde{\phi}_i$
can vary non-monotonically with $sK$, see  Fig.~\ref{fig:Fig4}~(a,b).
This behavior can be understood by noticing that near the boundary of regimes (i)-(ii),
we have $\phi_i\approx 1/3$ that increases with $sK$
if $r_i>r_{i-1}$ and decreases when $r_i<r_{i-1}$, see Eq.~(\ref{eq:TotfixR1}) and Fig.~\ref{fig:Fig4}~(a,b).
As $sK$ approaches the boundary of regimes (ii)-(iii), the dynamics is increasingly governed by the 
LOW with $\widetilde{\phi}_i\approx \phi_{i,i+1} \overset{sK \gg 1}{\longrightarrow}1$, $1/2$ or $0$. This can lead to a non-monotonic dependence on 
$sK$: For instance, if $r_1<r_2,r_3$, $\widetilde{\phi}_1$ decreases and $\widetilde{\phi}_2$ increases
about the value $1/3$ near the (i)-(ii) boundary,
and then respectively increases and decreases as $sK$ approaches the boundary (ii)-(iii),
and through regime (iii) where
$\widetilde{\phi}_1 \to 1$
while $\widetilde{\phi}_2 \to 0$, see  
Fig.~\ref{fig:Fig4}~(a).

The main features of the survival, absorption and overall fixation probabilities in
the constant-$K$ BDCLV are summarized in the chart of
Fig.~\ref{fig:Fig4}~(c).

\subsection{Mean fixation time in the constant-$K$ BDCLV}
\label{Section3.2}
The overall mean fixation time $T_F$ is the average
time after which one of the species takes over the entire population.
This quantity consists of one contribution arising from Stage 1, referred to as the 
 {\it mean extinction time} $T_1$, and the  {\it mean absorption time}  $T_2$ arising from Stage 2.
 In \ref{AppendixE1}, we study $T_1$ and $T_2$ in the regimes (i)-(iii) and show 
  that, when $\vec{x}_0=\vec{x}_c$, the overall mean fixation time 
 $ T_F=T_1+T_2= {\cal O}(K)$, see Fig.~\ref{fig:FigA4}(c).
 Since $N(t)\simeq K$ after a short transient, this
means that species coexistence is lost after  a mean time scaling linearly with the population size. 
We  also show that $T_1$ and $T_2$ are both of order ${\cal O}(K)$ in regimes (i)-(ii) and
$T_1 \gg T_2$ in the regime (iii), see Figs.~\ref{fig:FigA4}~(a,b) and \ref{fig:Fig1}.

\section{CLV with randomly switching carrying capacity}
\label{swK}
In many biological applications, 
the population is subject to sudden and extreme 
environmental changes dramatically affecting its size~\cite{Brockhurst07a,Wahl02,Gore13a,Gore13b,Harrington14}. 
The variation of $N(t)$ leads to a coupling between demographic fluctuations 
which greatly influence the population's evolution
~\cite{KEM1,KEM2,Gore13a,Gore13b,Harrington14}.

 Here, we study the \emph{coupled} effect of demographic and environmental fluctuations on the BDCLV fixation properties
 by considering the  randomly-switching carrying capacity (\ref{eq:K}), modeled
 in terms of the stationary DMN (\ref{eq:DMN}), 
 that can also be written as
 \begin{eqnarray}
\label{eq:Kbis}
K(t)% 
=\langle K \rangle (1+\gamma \xi(t)), \quad \text{with  $\gamma \equiv \frac{K_+ - K_-}{2\langle K \rangle} $} \nonumber 
\end{eqnarray}  
where $0<\gamma<1$ is a parameter 
measuring the intensity of the environmental variability.
In fact, the variance of $K(t)$
is ${\rm var}(K(t))=(\gamma \langle K \rangle)^2$, and we can 
write $K_{\pm}=(1\pm \gamma)\langle K \rangle$.
In order to study the influence of  environmental variability on the population dynamics, we 
 consider $\gamma={\cal O}(1)$ and $\langle K \rangle\gg 1$. This ensures that the population is 
subject to significant  environmental variability (${\rm var}(K)\gg 1$), and  its typical size
is large enough to avoid that 
demographic fluctuations (internal noise, IN) alone are the main source of randomness.
In all our simulations,  the initial value of $K(t)$ is either 
$K_+$ or $K_-$ with  probability $1/2$.

From the ME (\ref{eq:ME}), proceeding as  in \ref{AppendixA1}, 
the population composition is found to still evolve according to the REs  (\ref{eq:xidot}) when all 
 demographic fluctuations are neglected. However, now the  random switching of $K(t)$
 drives the stochastic evolution of the 
population size which, when IN is ignored, obeys $\dot{N}=N(1-N/K_{\pm})$ if $\xi=\pm 1$, see Eq.~(\ref{eq:N_MF2}).
This can be rewritten as 
\begin{eqnarray}
\dot{N}&=& N\left(1-\frac{N}{{\cal K}}\{1-\gamma \xi(t)\}
\right)\,,
\label{eq:PDMP}
\end{eqnarray}
where 
\begin{eqnarray*}
%\label{eq:K}
{\cal K}\equiv
(1-\gamma^2)\langle K \rangle=\frac{2K_+K_-}{K_+ + K_-} 
\end{eqnarray*}
is the harmonic mean of $K_{\pm}$ and $\xi$ is the multiplicative dichotomous noise (\ref{eq:DMN}).
The DMN intensity being $N^2\gamma/{\cal K}$,
the environmental fluctuations increase with $\gamma$ together with ${\rm var}(K)=(\gamma \langle K \rangle)^2$. Eq.~(\ref{eq:PDMP}) 
defines a  piecewise-deterministic Markov process
(PDMP)~\cite{Davis84}. When $\nu \to \infty$,  the DMN self averages, with
 $\xi \to \langle \xi \rangle=0$ in (\ref{eq:PDMP})
which reduces to the  logistic equation (\ref{eq:Ndot})
with a renormalized  carrying capacity $K \to {\cal K}$~\cite{KEM1,KEM2}. 
Again, a timescale separation arises when $s\ll 1$, with $N$ evolving faster than $x_i$'s:
$N$ settles in its $N$-QSD in a time $t={\cal O}(1)$, while
the $x_i$'s change on a timescale $t={\cal O}(1/s)$, see Fig.~\ref{fig:Fig1}~(c). 

\begin{figure}
	\centering
	\includegraphics[width=0.48\linewidth]{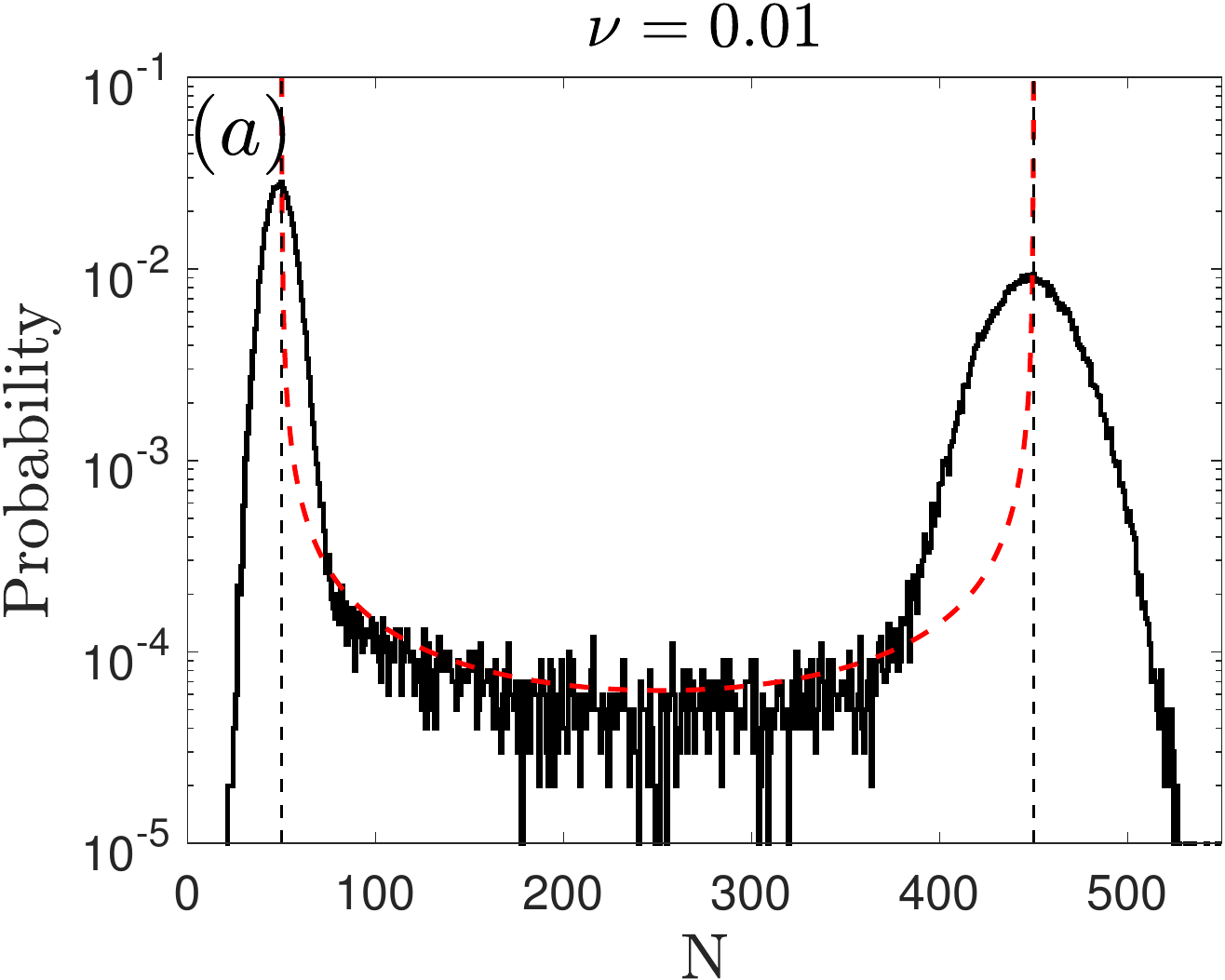}
	\includegraphics[width=0.48\linewidth]{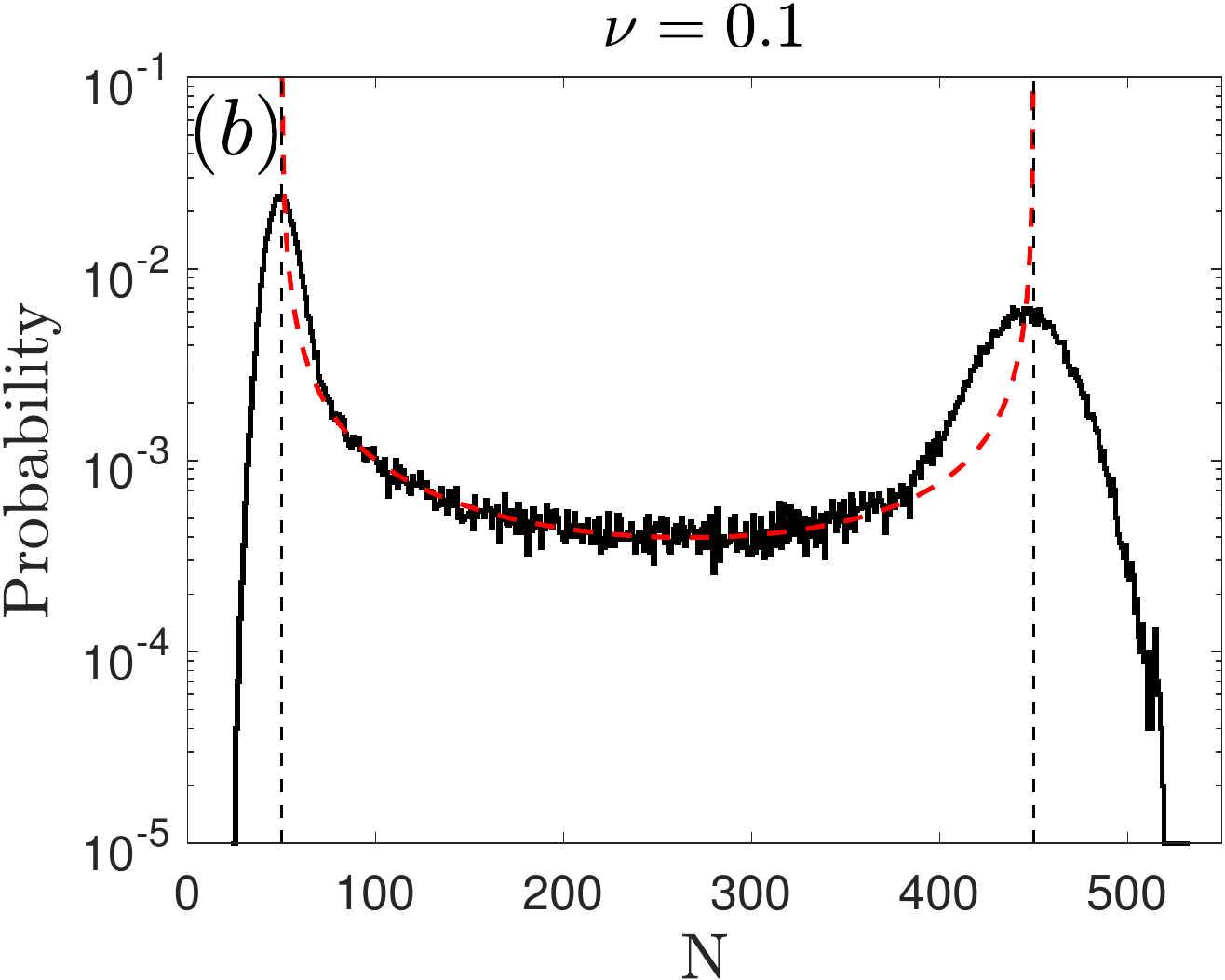}
	\includegraphics[width=0.48\linewidth]{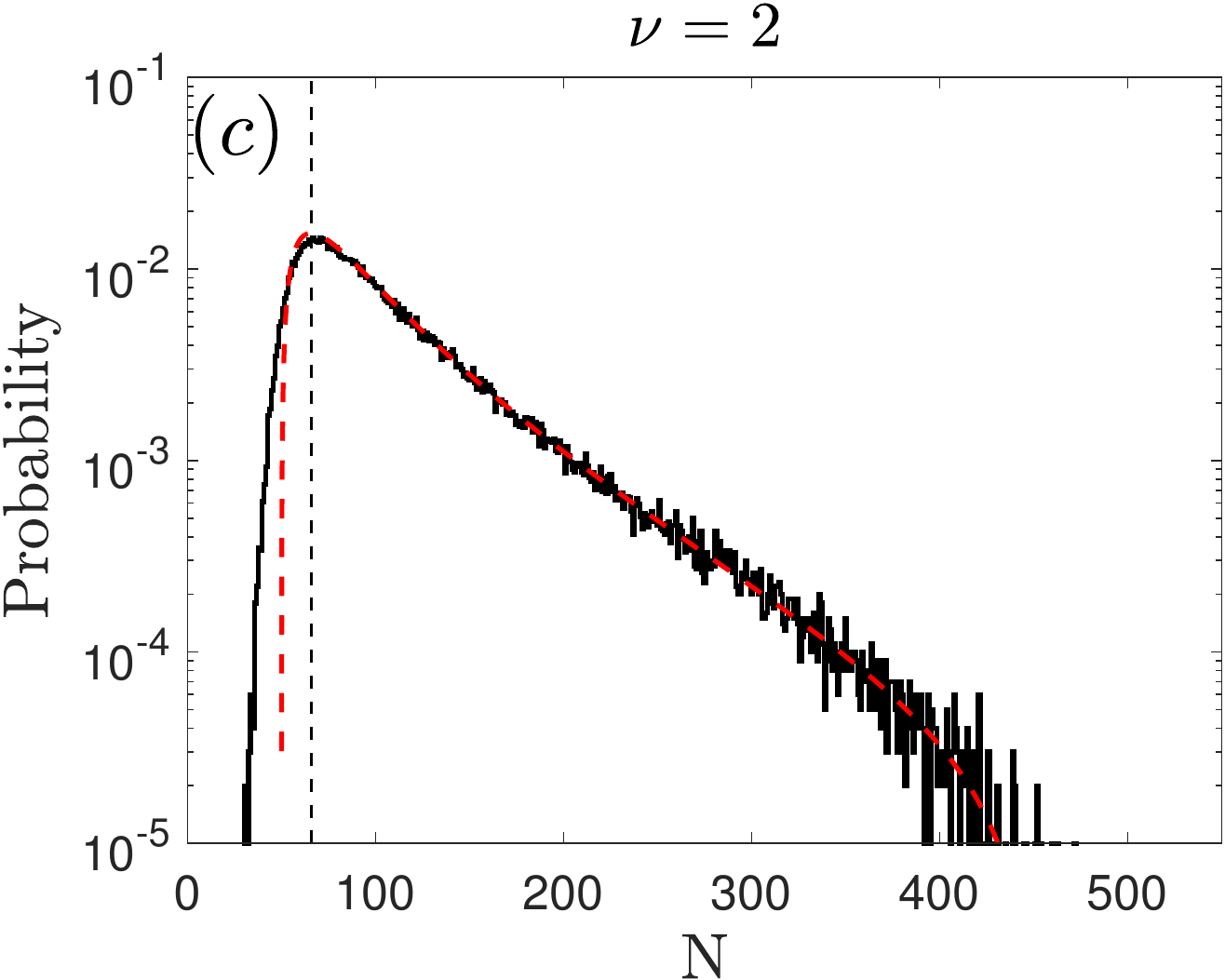}
	\includegraphics[width=0.48\linewidth]{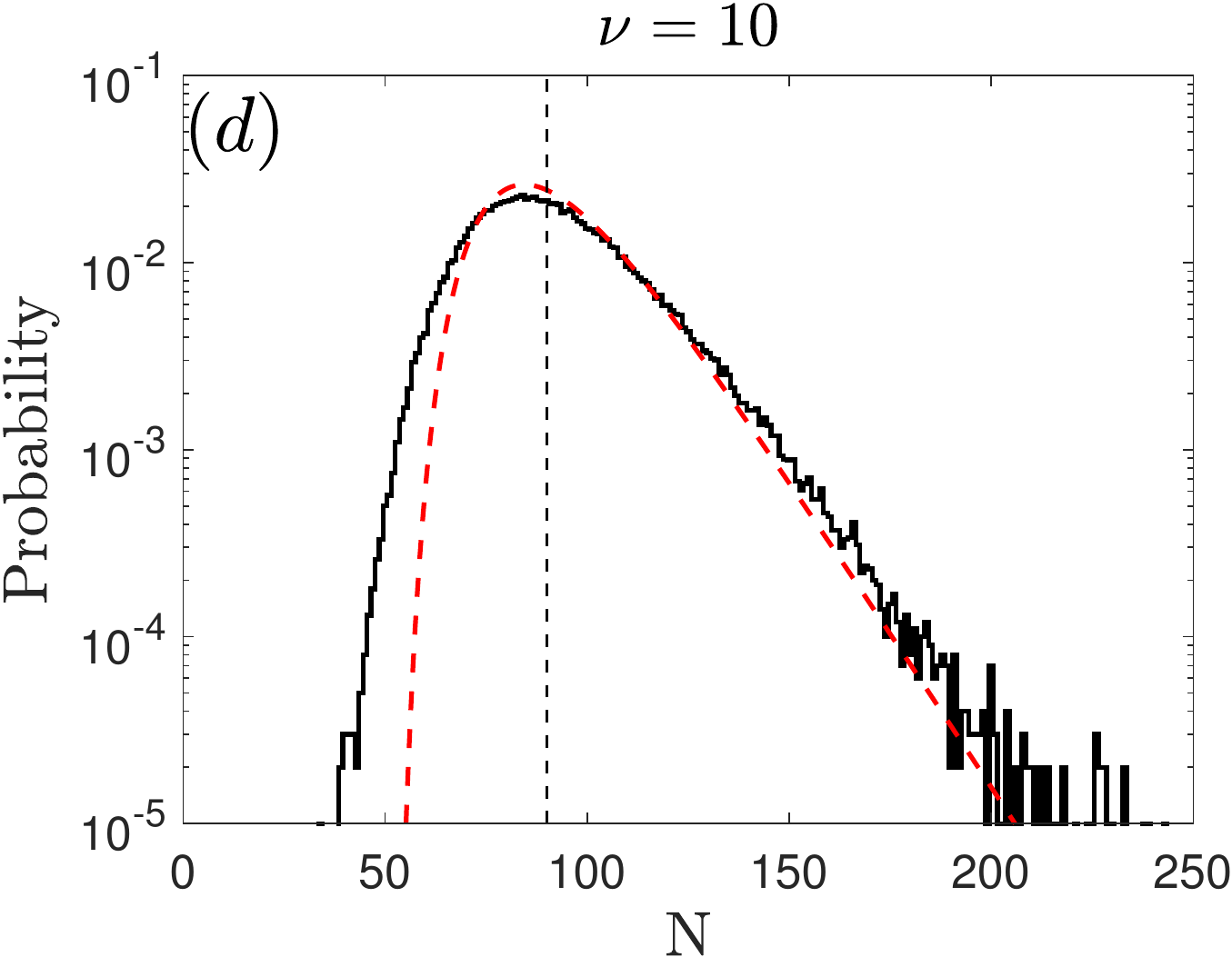}
	\caption{$N$-QSD and $p^*_{\nu}(N)$ for  
	(a) $\nu=0.01$, (b) $\nu=0.1$, (c) $\nu=2$, (d) $\nu=10$.
	Parameters are $(s,K_+, K_-)=(0.02, 450, 50)$.	Solid lines are histograms from
	stochastic simulations and colored dashed lines are PDMP predictions from  (\ref{eq:pnustar}), see text.
	Black dashed lines indicate $N=K_{\pm}$ in (a) and (b), $N=N^*$ in (c),
	and $N={\cal K}$ in 
	(d), see text.
	}
	\label{fig:Fig5}
\end{figure}

The PDMP defined by Eq.~(\ref{eq:PDMP})~\cite{HL06,Bena06,Hufton16,KEM1,KEM2} is
characterized by the following stationary marginal probability density function (pdf)~\cite{KEM1}:
\begin{eqnarray}
\hspace{-5mm}
\label{eq:pnustar}
p_{\nu}^{*}(N) = \frac{\mathcal{Z}}{N^2} \left[\frac{(K_+ - N )
(N -K_-) }{N^2}\right]^{\nu-1},
\end{eqnarray} 
where $\mathcal{Z}$ is the normalization constant. The PDMP pdf $p_{\nu}^{*}$
gives the long-time probability density of $N$ on the support $N\in [K_-, K_+]$ regardless of the 
environmental state $\xi$~\cite{HL06,Bena06}. When $\gamma={\cal O}(1)$ and $\langle K \rangle\gg 1$, $p_{\nu}^{*}$
is a good approximation of the $N$-QSD even if it ignores the effect of the IN,
see Fig.~\ref{fig:Fig5}. 
In fact, the comparison of $p_{\nu}^{*}$  and   $N$-QSD shown in Fig.~\ref{fig:Fig5} reveals that  $p_{\nu}^{*}$
correctly captures the main features of the $N$-QSD, such as  the location of the peak(s) and its right-tailed skewness, whereas it fails 
to capture the width about the peak(s)~\footnote{This  stems from the  
demographic fluctuations being ignored by the PDMP approximation: These cause a ``leakage'' of the distribution of $N$ 
outside  $[K_-, K_+]$. 
This is particularly visible  when $\nu<1$, see Fig.~\ref{fig:Fig5}~(a). As shown in Ref.~\cite{KEM2}, the actual width  of the $N$-QSD
can be accurately 
computed with a linear-noise approximation about the PDMP process  (\ref{eq:PDMP}).}. However,  
for our purposes here the PDMP approximation is sufficient to characterize the system's fixation properties~\cite{KEM1,KEM2}.
It is noteworthy that  $p_{\nu}^{*}$ and the $N$-QSD
 are bimodal if $\nu<1$, with peaks at $N\simeq K_{\pm}$, see
  Fig.~\ref{fig:Fig5}~(a,b).  When $\nu>1$,  
  $p^*_{\nu}$ and $N$-QSD are unimodal and $N$ fluctuates
  about the maximum of $p^*_{\nu}$ given by $N^{*}= \langle K \rangle(1+\nu)\left(1-\sqrt{1-
  4\nu(1-\gamma^2)/(1+\nu)^2}\right)/2$. The value of $N^{*}$
increases with $\nu$ at $\gamma$ fixed,
 see Fig.~\ref{fig:Fig5}~(c,d), and decreases 
with $\gamma$ (environmental variability) at  $\nu$ fixed.
When $\nu \to \infty$, we have $N^{*} \to {\cal K}$
and $p^*_{\nu}$ is  sharply peaked about ${\cal K}$, as expected from the self-averaging of $\xi(t)$
when $\nu \gg 1$, see Fig.~\ref{fig:Fig5}~(d). In this case,
we recover the constant-$K$ BDCLV dynamics with  $K \to {\cal K}$.

\begin{figure}
	\centering
	\includegraphics[width=0.9\linewidth]{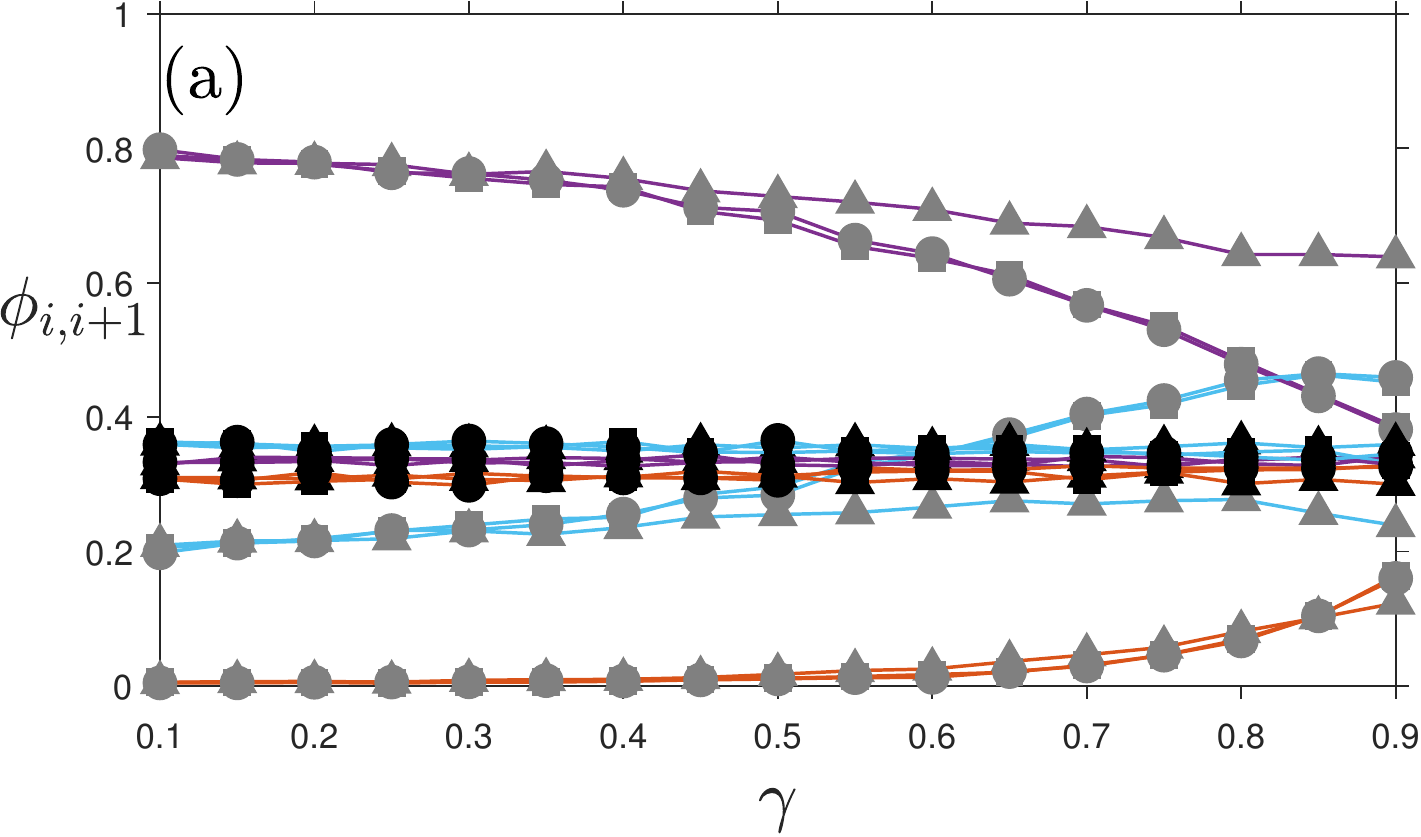}
	\includegraphics[width=0.9\linewidth]{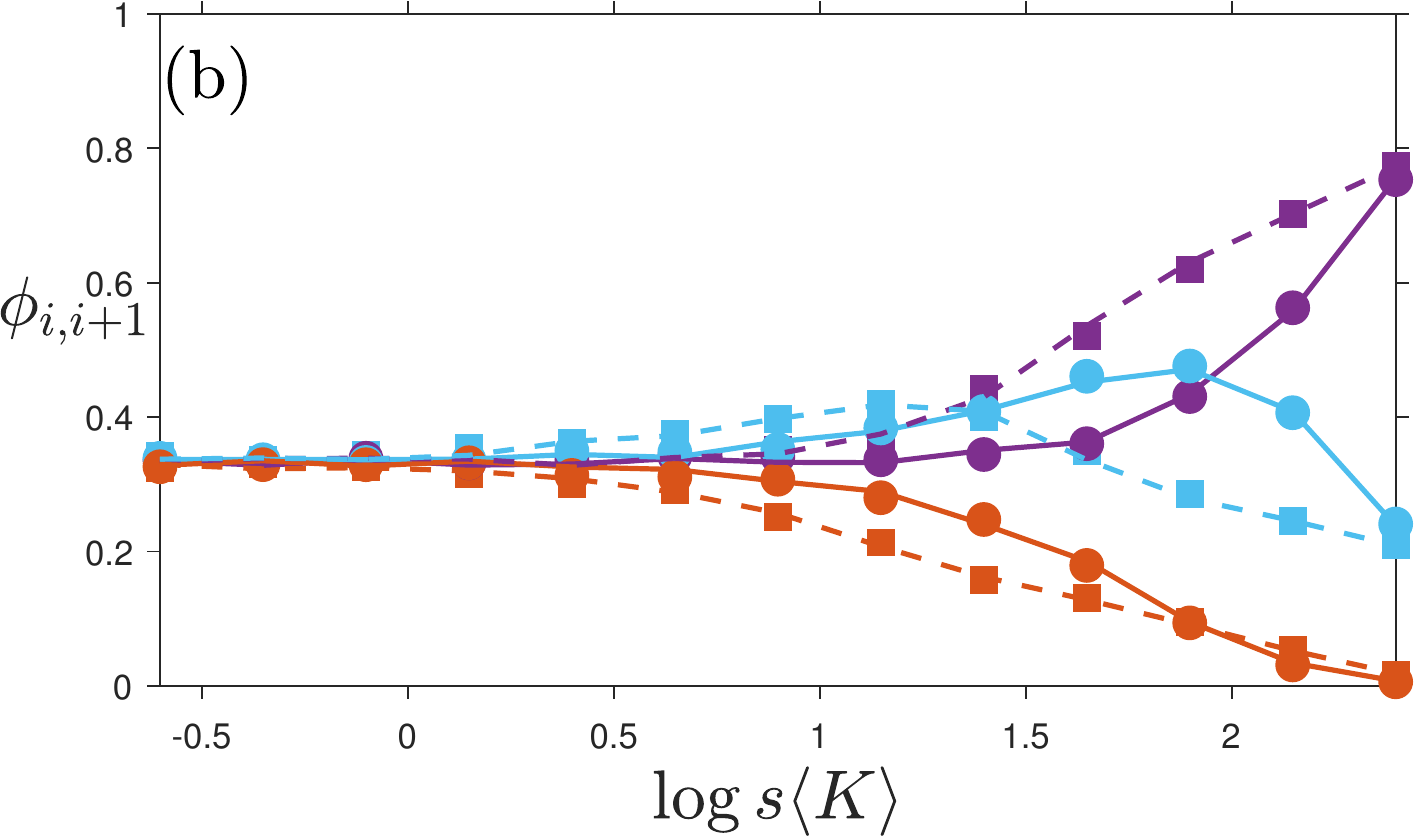}
	\includegraphics[width=0.9\linewidth]{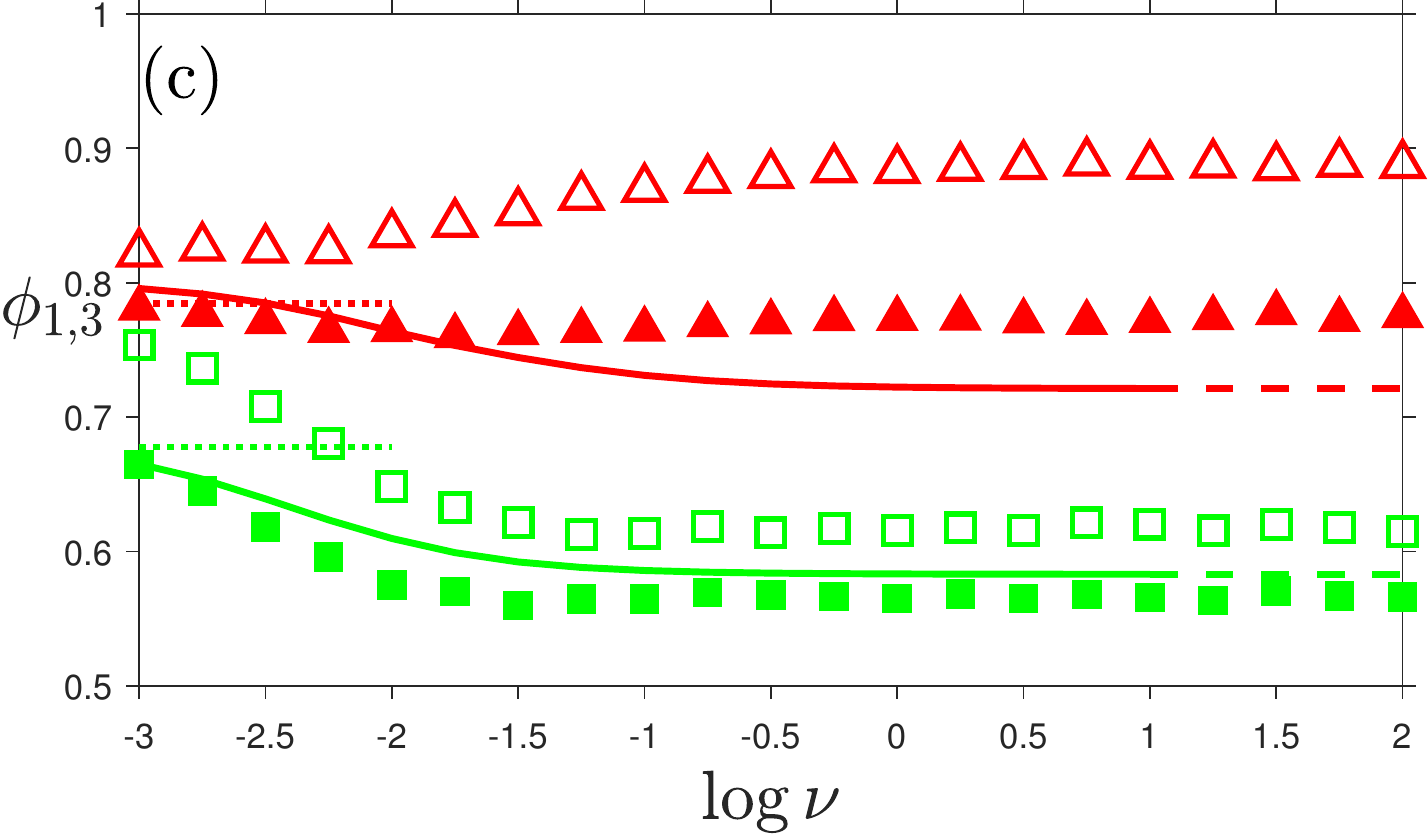}
	\includegraphics[width=0.9\linewidth]{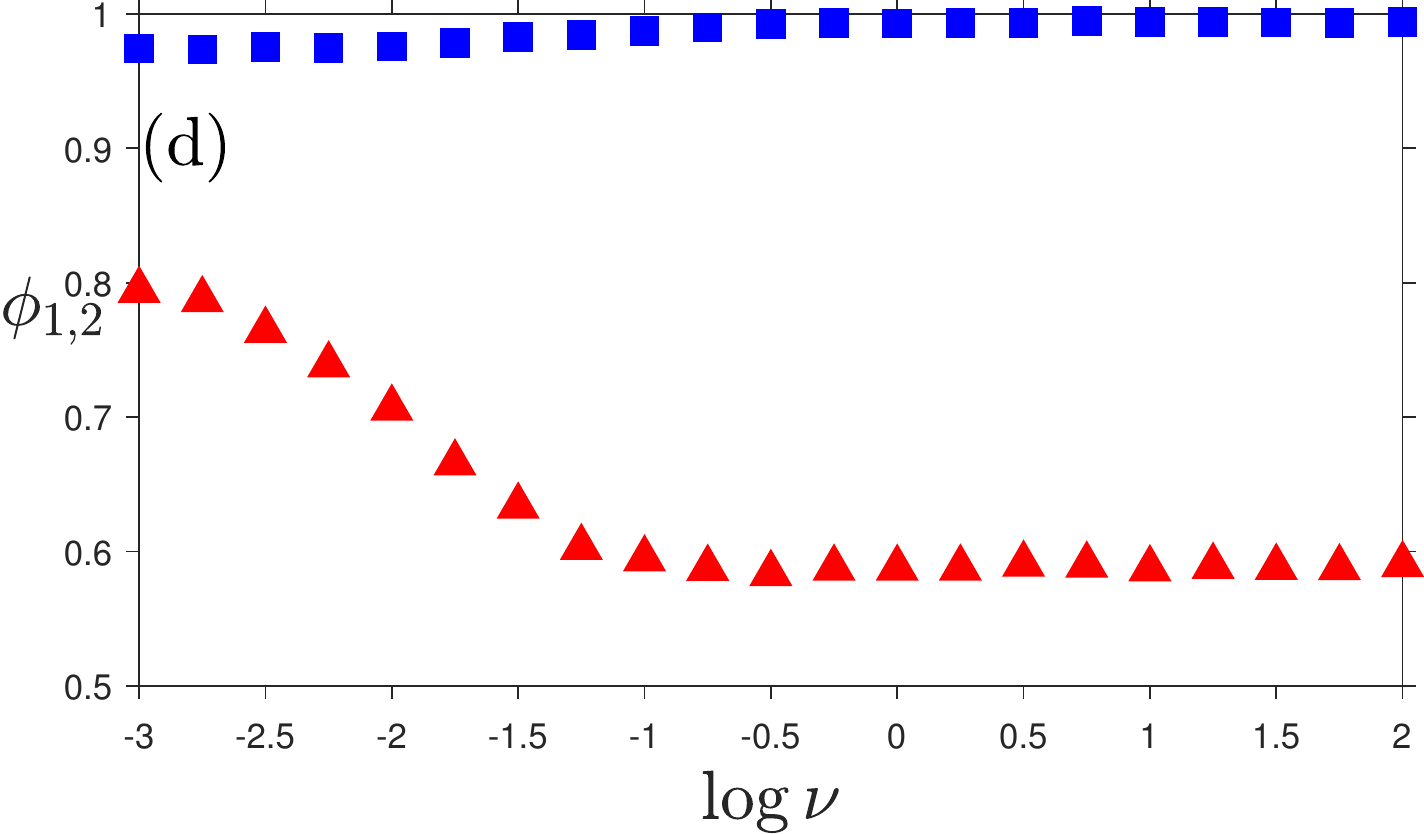}
	\caption{(a) Stage 1 survival probability $\phi_{i,i+1}$  vs. $\gamma$
	for $\langle K\rangle=250$ kept fixed ($K_+\in [275,475]$ and $K_-\in [25,225]$).
	and $s=0.01$ (black), $s=0.4$ (gray).
	Simulation results for $\nu=10$ (circles),  $\nu=1.2$ (squares) and $\nu=0.001$ (triangles).
%	Significant deviations from $\phi_{i,j}\vert_{\langle K\rangle}$ reported in
%	 Fig.~\ref{fig:Fig3}(a) are found for $\gamma\gtrsim 0.4$ 
	 (b) $\phi_{i,i+1}$ vs. $s\langle K \rangle$  
	for $\langle K\rangle=250, \gamma=0.8$ and  $s\in \{10^{-k/4}, k=0,\dots,12\}$ kept fixed, with 
	 $\nu=2$ (circles) and $\nu=0.001$ (squares); lines are 
	  $\phi_{i,j}\vert_{(1-\gamma^2)\langle K\rangle}$ (solid) and $\frac{1}{2}
	 (\phi_{i,j}\vert_{(1+\gamma)\langle K\rangle}+\phi_{i,j}\vert_{(1-\gamma)\langle K\rangle})$ (dashed) 
	 are from the constant-$\langle K\rangle$ BDCLV. In panels (a,b) $\vec{r}=\vec{r}^{(1)}$, 
	 $\phi_{1,2}$ in purple, $\phi_{2,3}$ in light blue, $\phi_{3,1}$ in orange.
	 (c) Stage 2 absorption probabilities $\phi_{1}$ (red triangles) and $\phi_{3}$ (green squares) vs. $\nu$
	for $\langle K\rangle=250$ and $\gamma=0.8$ kept fixed and $\vec{r}=\vec{r}^{(2)}$.
	Symbols are from simulations with  $s=0.1$~(open) and $s=10^{-5/4}\approx 0.056$~(filled).
	 Lines are from (\ref{eq:phi3}) (solid), (\ref{eq:phi2}) (dashed), (\ref{eq:phi1}) (dotted), and
	 assume $P_{i,i+1}\approx 1$; they 
	 capture reasonably well the $\nu$-dependence of $\phi_{1}$ and $\phi_{3}$ 
	 when $s\langle K\rangle\lesssim 10$, see text.
	(d) Same as in panel (c) for $\phi_{1}$ (red  triangles) and $\phi_{2}$ (blue squares) vs. $\nu$
	with  $s=10^{-1/4}$ and $\vec{r}=\vec{r}^{(1)}$.
	In all panels $\vec{x}_0=\vec{x}_c$, $\epsilon=0$.}
	\label{fig:Fig6}
\end{figure}%

\subsection{Survival, absorption and fixation probabilities in the switching-$K$ BDCLV}
As in the  constant-$K$ BDCLV, the total fixation probability $\widetilde{\phi}_i$
 depends on the stage 1 survival  and stage 2 absorption probabilities. Here, we analyze 
 the effect of the environmental randomness on these quantities,
 by distinguishing again the  regimes of 
  (i)  quasi-neutrality, where $s\ll 1$ and $s\langle K\rangle \ll 1$;
  (ii) weak selection, where $s\ll 1$ and $s\langle K\rangle={\cal O}(10)$; and 
  (iii)  strong selection, where $s={\cal O}(1)$ and $sK\gg 1$.
\subsubsection{Stage 1: Survival probabilities in the switching-$K$ BDCLV}
To analyze the survival probability $\phi_{i,i+1}$ in the switching-$K$ BDCLV, it is convenient
to consider this quantity in the limits $\nu\to \infty$ and $\nu\to 0$, where 
$\phi_{i,i+1}$ can be expressed  in terms of $\phi_{i,i+1}|_{K}$, the survival probability in 
the constant-$K$ BDCLV studied in Sec.~\ref{Sec3.1.1}.

When $\nu\to \infty$,  many switches occur in Stage 1 and the DMN self averages,  $\xi\to \langle 
\xi\rangle=0$~\cite{KEM1,KEM2}. The  population thus rapidly 
settles in its $N$-QSD
that is delta-distributed at $N=(1-\gamma^2)\langle K\rangle$ when $\langle K\rangle\gg 1$.
Hence, the stage 1 dynamics under fast switching is similar to the 
cCLV dynamics in a population of size $(1-\gamma^2)s\langle K\rangle$, see \ref{AppendixB}.
This yields 
$\phi_{i,i+1} \overset{\nu \to \infty}{=}\phi_{i,i+1}|_{(1-\gamma^2)\langle K\rangle}$.

When $\nu\to 0$, there are no switches in  Stage 1, and 
the extinction of the first species is equally likely to occur in each environmental state $\xi=\pm 1$ (with $K=
(1\pm\gamma)\langle K \rangle$).
This gives 
$\phi_{i,i+1} \overset{\nu \to 0}{=} 
\left(\phi_{i,i+1}|_{(1+\gamma)\langle K \rangle} + \phi_{i,i+1}|_{(1-\gamma)\langle K \rangle} \right)/2$.

The case of intermediate $\nu$ can be inferred from the above
by noting that the average number of switches 
occurring in Stage 1 is ${\cal O}(\nu \langle K\rangle)$, see Fig.~\ref{fig:FigA6}~(a). As the population 
experiences a large number of switches
in Stage 1  when $\nu={\cal O}(1)$ and $\langle K\rangle\gg 1$, 
the DMN  effectively
self-averages, $\xi(t) \simeq \langle \xi \rangle=0$,
and therefore
\begin{eqnarray}
 \label{eq:phijK2}
\phi_{i,i+1} \overset{\nu ={\cal O}(1)}{\approx} \phi_{i,i+1}|_{(1-\gamma^2)\langle K\rangle}.
\end{eqnarray}
When $\nu\ll 1/\langle K\rangle$,
there are very few or no switches after a time of order 
${\cal O}(\langle K \rangle)$ prior to extinction the first species, and therefore
 \begin{eqnarray}
 \label{eq:phijK1}
\phi_{i,i+1} \overset{\nu \ll 1/\langle K\rangle}{\approx}
\frac{1}{2}\left(\phi_{i,i+1}|_{(1+\gamma)\langle K \rangle} + \phi_{i,i+1}|_{(1-\gamma)\langle K \rangle} \right).
\end{eqnarray}
Eq.~(\ref{eq:phijK2}) implies that for any  $\nu ={\cal O}(1)$,
the survival probability of species $i, i+1$, i.e the probability that species $i-1$
dies out first, is given by the survival probability in the constant-$K$ BDCLV 
with $K=\langle K\rangle$ (same average carrying capacity) and a rescaled selection intensity $(1-\gamma^2)s$.
The effect of random switching is therefore to effectively reduce 
the selection intensity by a factor $1-\gamma^2=1-({\rm var}(K(t))/\langle K\rangle^2)$ proportional to the variance of the carrying
capacity. The
 $s\langle K\rangle$-dependence of $\phi_{i,i+1}$ can thus readily be obtained from Fig.~\ref{fig:Fig3}~(a,b)
by rescaling $s\to (1-\gamma^2)s$ as shown in  Fig.~\ref{fig:Fig6}~(a,b).
Hence, when  there is enough environmental variability
($\gamma$ large enough) the 
survival scenarios differ from those of 
the constant-$K$ BDCLV  and 
depend on 
the switching rate:
\vspace{2mm}
\\
- When $\nu\gg 1/\langle K \rangle$,   switching reduces the selection by a factor $1-\gamma^2$,
see Fig.~\ref{fig:Fig6}~(b). Hence, there is a critical $\gamma^*$,  estimated as
$\gamma^*\approx (1-50/s\langle K\rangle)^{1/2}$,
such that 
$\phi_{i,i+1}$ obeys the LOSO  when 
$\gamma>\gamma^*$
and $s\langle K \rangle\gg 1$, while the LOW still applies when $\gamma<\gamma^*$. Therefore, when $\gamma >\gamma^*$,  
all species have a finite chance to survive Stage 1, with probabilities ordered according to the LOSO, 
($\phi_{1,2}\approx \phi_{2,3}>\phi_{3,1}$ with $\gamma^*\approx 0.7$, 
in Fig.~\ref{fig:Fig6}~(a)). Fig.~\ref{fig:Fig6}~(a), also shows that
the exact value $\nu$ has little influence on $\phi_{i,i+1}$ provided that $\nu \langle K \rangle \gg 1$
(circles and squares almost coincide).
\vspace{1mm}
\\
- When $\nu\ll 1/\langle K \rangle$, we have 
$\phi_{i,i+1}\approx ( \phi_{i,i+1}|_{K_+}+ \phi_{i,i+1}|_{K_-})/2$. Hence, if 
$s\langle K\rangle \gg 1$ and 
$\gamma > \hat{\gamma}$, where  
$\hat{\gamma}\approx 1-50/s\langle K\rangle$,
$\phi_{i,i+1}|_{K_+}$ follows the LOW whereas $\phi_{i,i+1}|_{K_-}$ obeys the LOSO, and  
the $\phi_{i,i+1}$'s 
therefore interpolate between LOW and LOSO values: For $\gamma > \hat{\gamma}$, the 
survival probabilities under strong selection and slow switching
deviate markedly from the purely LOW values of  
$\phi_{i,i+1}|_{\langle K \rangle}$ 
which asymptotically approach $0$, $1$ or $1/2$ (see triangles in Fig.~\ref{fig:Fig6}~(a) where  $\hat{\gamma}\approx 0.5$).

\vspace{1mm}

 When $s\ll 1$ and $s\langle K \rangle={\cal O}(10)$ in regime (ii),
changing $\gamma$ has little effect on the survival probabilities: the survival probabilities 
$\phi_{i,i+1}\approx 1/3$,  and remain ordered according to the LOSO (see black symbols in Fig.~\ref{fig:Fig6}~(a)).
\vspace{-1mm}

These results show that environmental variability leads to new survival scenarios in the BDCLV under strong selection: When there is enough variability, 
all species have a finite probability to survive even when  $s\langle K\rangle \gg 1$. The departure from
the pure LOW survival scenario is most marked in the generic case of a finite  switching rate ($\nu\gg 1/\langle K \rangle$). 
With  respect to the constant-$K$ BDCLV, the general effect of random switching in Stage 1 is therefore
to ``level the field'' by hindering the onset of the zero-one LOW. Since BDCLV survival probability $\phi_{i,i+1}$ coincides
with the fixation probability of species $i$ in the cCLV, see \ref{AppendixB}, it is noteworthy that these results also 
show that random switching can
lead to new survival/fixation scenarios in the cCLV when the variance of the carrying capacity is sufficiently high.

\subsubsection{Stage 2: Absorption probabilities in the switching-$K$ BDCLV}
\label{sec:4.1.2}
%
%At the end of Stage 1,  two species survive, say $i$ and $i+1$ with $r_i>r_{i+1}$. 
Stage 2  consists of the  competition between types $i$ and $i+1$ 
along the edge $(i,i+1)$ of $S_3$. This starts with 
an initial fraction $\hat{x}_i$ of $i$ individuals and ends up with the absorption  of one of the species  
 with probabilities $\phi_i$ (for species $i$) and $1-\phi_i$ (for $i+1$). Again  
$\hat{x}_i$ is randomly distributed according to a probability density
$P_{(i,i+1)}$ resulting from  Stage 1, see \ref{AppendixD}\footnote{The probability density function
of $\hat{x}_i$ is generally different in the constant-$K$ and switching-$K$ BDCLV, see Fig.~\ref{fig:FigD3}. 
Yet, for the sake of simplicity, with a slight abuse of notation, we denote these two quantities by $P_{i,i+1}(\hat{x}_i)$.}.
Since  $\phi_i\approx 1/2$ at quasi-neutrality and $\phi_i\approx 1$ under strong selection, see 
Fig.~\ref{fig:Fig6}~(c,d),  Stage 2 dynamics is nontrivial in regime (ii). 
To analyze the stage 2 dynamics under weak selection $s\ll 1$ and $\langle K \rangle \gg 1$,
it is again useful to consider the limits $\nu\to 0$ and $\nu\to \infty$:
\vspace{1mm}

- When $\nu \to 0$, there are no switches in 
Stage 2 and  absorption is equally likely to occur
in the static environment $K=K_-$ or $K=K_+$. Hence, if the fraction $\hat{x}_i$ is known, 
we have $\phi_{i}(\hat{x}_i) \overset{\mathrm{\nu \to 0}}{=} \phi_{i}^{(0)}(\hat{x}_i) =\frac{1}{2}\left(\phi_i(\hat{x}_i)\vert_{K_-}
+ \phi_i(\hat{x}_i)\vert_{K_+}\right)$, where 
$\phi_i(\hat{x_i})|_{K}=(1-e^{-\alpha_i K \hat{x_i}})/(1-e^{-\alpha_i K})$, see (\ref{eq:phi_S2}).
Since $\hat{x}_i$ is randomly distributed, one needs to integrate over $P_{(i,i+1)}$:
$ \phi_{i} \overset{\mathrm{\nu \to 0}}{\equiv} \phi_{i}^{(0)} =\int_0^1 \phi_{i}^{(0)}(\hat{x}_i)P_{(i,i+1)}(\hat{x}_i) ~d\hat{x}_i$. 
In general, $P_{(i,i+1)}$ is obtained from stochastic simulations and has been found to be mostly independent of $\nu$, 
see Fig.~\ref{fig:FigD3}~(c,d). When $s\ll 1$ with $s\langle K \rangle \lesssim 10$, we can
again assume $P_{(i,i+1)}\approx 1$ (uniform distribution), which allows us to obtain
\begin{eqnarray}
%\hspace{-5mm}
 \label{eq:phi1}
 \phi_{i} \overset{\mathrm{\nu \to 0}}{=} \phi_{i}^{(0)} \simeq \frac{1}{2}~\left(\phi_i\vert_{K_-}
+ \phi_i\vert_{K_+}\right), \quad \text{where}
\end{eqnarray}
$\phi_i|_K\equiv (e^{-\alpha_i K} +\alpha_i K -1)/(\alpha_i K (1-e^{-\alpha_i K}))$, see (\ref{eq:condfixprob2}).
\vspace{1mm}

- When $\nu \to \infty$,  the DMN self averages ($\xi \to \langle \xi
\rangle=0$)~\cite{KEM1,KEM2}, and  the absorption occurs subject to 
the  effective  $K(t)={\cal K}$, see Eq.~(\ref{eq:PDMP}).
Hence, when $\hat{x}_i$ is known, 
$\phi_{i}(\hat{x}_i) \overset{\mathrm{\nu \to \infty}}{=} \phi_{i}^{(\infty)}(\hat{x}_i)
=\phi_i(\hat{x}_i)\vert_{{\cal K}}$, whose integration over 
$P_{(i,i+1)}$ gives the absorption probability:
$\phi_{i} \overset{\mathrm{\nu \to \infty}}{\equiv}\phi_{i}^{(\infty)} =\int_0^1~\phi_{i}^{(\infty)}(\hat{x}_i)~P_{(i,i+1)}(\hat{x}_i)~d\hat{x}_i$.
When $s\ll 1$ with $s\langle K \rangle \lesssim 10$, and $P_{(i,i+1)}\approx 1$, 
we have 
\begin{eqnarray}
%\hspace{-5mm}
 \label{eq:phi2}
 \phi_{i} \overset{\mathrm{\nu \to \infty}}{=} \phi_{i}^{(\infty)} \simeq \phi_i\vert_{{\cal K}}=
 \frac{e^{-\alpha_i {\cal K}} +\alpha_i {\cal K} -1}{\alpha_i {\cal K} (1-e^{-\alpha_i {\cal K}})}.
\end{eqnarray}
\vspace{1mm}

- When the switching rate $\nu$ is finite and $s\ll 1$, with $s\langle K \rangle ={\cal O}(10)$,
the probability $\phi_i$
can be computed 
as in Ref.~\cite{KEM1}
by exploiting 
the time scale separation between $N$ 
and $x_i$, and by approximating the $N$-QSD by the 
PDMP marginal stationary probability  density (\ref{eq:pnustar}).
In this framework,  $\phi_i$
  can be computed
by averaging  $\phi_{i}(\hat{x}_i)|_{N}
=(1-e^{-\alpha_i N \hat{x_i}})/(1-e^{-\alpha_i N})$
over the rescaled PDMP probability (\ref{eq:pnustar})~\cite{KEM1,KEM2}:
\begin{eqnarray*}
% \label{eq:phi1}
 \phi_i(\hat{x}_i)\simeq \phi_i^{(\nu)}(\hat{x}_i)=\int_{K_-}^{K_+} \phi_i(\hat{x}_i)|_{N}~p^{*}_{\nu/\alpha_i}(N) ~dN,
\end{eqnarray*}
where $p^{*}_{\nu/\alpha_i}$ is given by (\ref{eq:pnustar}) with  a rescaled
switching rate $\nu \to \nu/\alpha_i$ due to an average number ${\cal O}(\nu/\alpha_i)$
of switches occurring in Stage 2, see \cite{KEM2} and Sec.~\ref{AppendixE5}. As above, the  absorption probability
is obtained by formally integrating over $P_{(i,i+1)}$, i.e. 
$\phi_i\simeq \phi_i^{(\nu)}\equiv\int_0^1\phi_i^{(\nu)}(\hat{x}_i)~P_{(i,i+1)}({x}_i)~d\hat{x}_i$.
Under weak selection, we can approximate $P_{(i,i+1)}\approx 1$, see Sec.~S4, and, using 
(\ref{eq:condfixprob1}) and (\ref{eq:condfixprob2}), we obtain
\begin{eqnarray}
\label{eq:phi3}
\hspace{-4mm}
\phi_i\simeq \phi_i^{(\nu)}
\approx \int_{K_-}^{K_+} \left\{\frac{e^{-N\alpha_i}+\alpha_i N -1}{\alpha_i N \left(1-e^{-\alpha_i N}\right)}\right\}
 ~p^{*}_{\nu/\alpha_i}(N)~dN.
\end{eqnarray}
The uniform approximation of $P_{(i,i+1)}\approx 1$  is legitimate when $s\langle K \rangle={\cal O}(10)$, 
and has broader range applicability than in the constant-$K$ case,
see Sec.~S4 and Fig.~S3. Hence, Eq.~(\ref{eq:phi3}), along with (\ref{eq:phi1}) and (\ref{eq:phi2}), 
captures the $\nu$-dependence of $\phi_{i}$ over a broad range of values $\nu$ when $s\ll 1$.
In fact, simulation results of Fig.~\ref{fig:Fig6}~(c,d) 
 show that the $\phi_i$'s generally have a  non-trivial $\nu$-dependence. When 
 $s\ll 1$ and $s\langle K \rangle={\cal O}(10)$, 
 this is satisfactorily captured by (\ref{eq:phi1})-(\ref{eq:phi3}),
with $\phi_i^{(\nu)}\approx \phi_i^{(0)}$
when $\nu\ll 1$, and $\phi_i^{(\nu)}\approx \phi_i^{(\infty)}$ when $\nu\gg 1$, see Fig.~\ref{fig:Fig6}~(c, filled symbols).
Clearly, the assumption $P_{(i,j)}\approx 1$ and the timescale separation 
break down when $s={\cal O}(1)$~\cite{KEM2},  and 
the approximations (\ref{eq:phi1})-(\ref{eq:phi3}) are then no longer valid.

\subsubsection{Overall fixation probabilities in the switching-$K$ BDCLV}
\label{sec:Sec4.1.3}
The overall fixation probability $\widetilde{\phi}_{i}$ is obtained from the survival and absorption probabilities 
according to $\widetilde{\phi}_i=\phi_{i,i+1}\phi_i + \phi_{i-1,i}(1-\phi_{i-1})$, 
see Eq.~(\ref{eq:PhiTot}). 

\begin{figure}[t]
	\centering
	\includegraphics[width=0.98\linewidth]{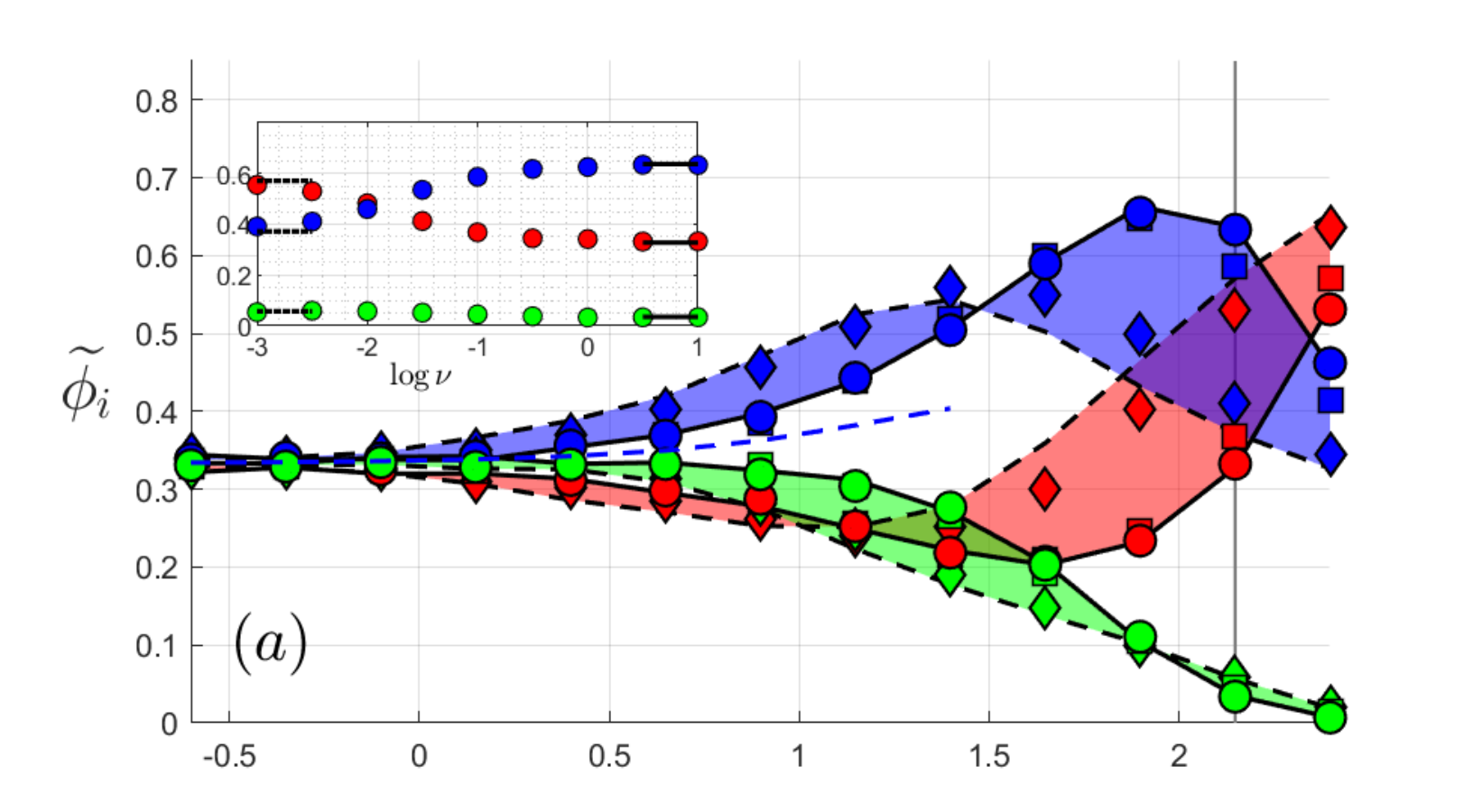}\\
	\includegraphics[width=0.98\linewidth]{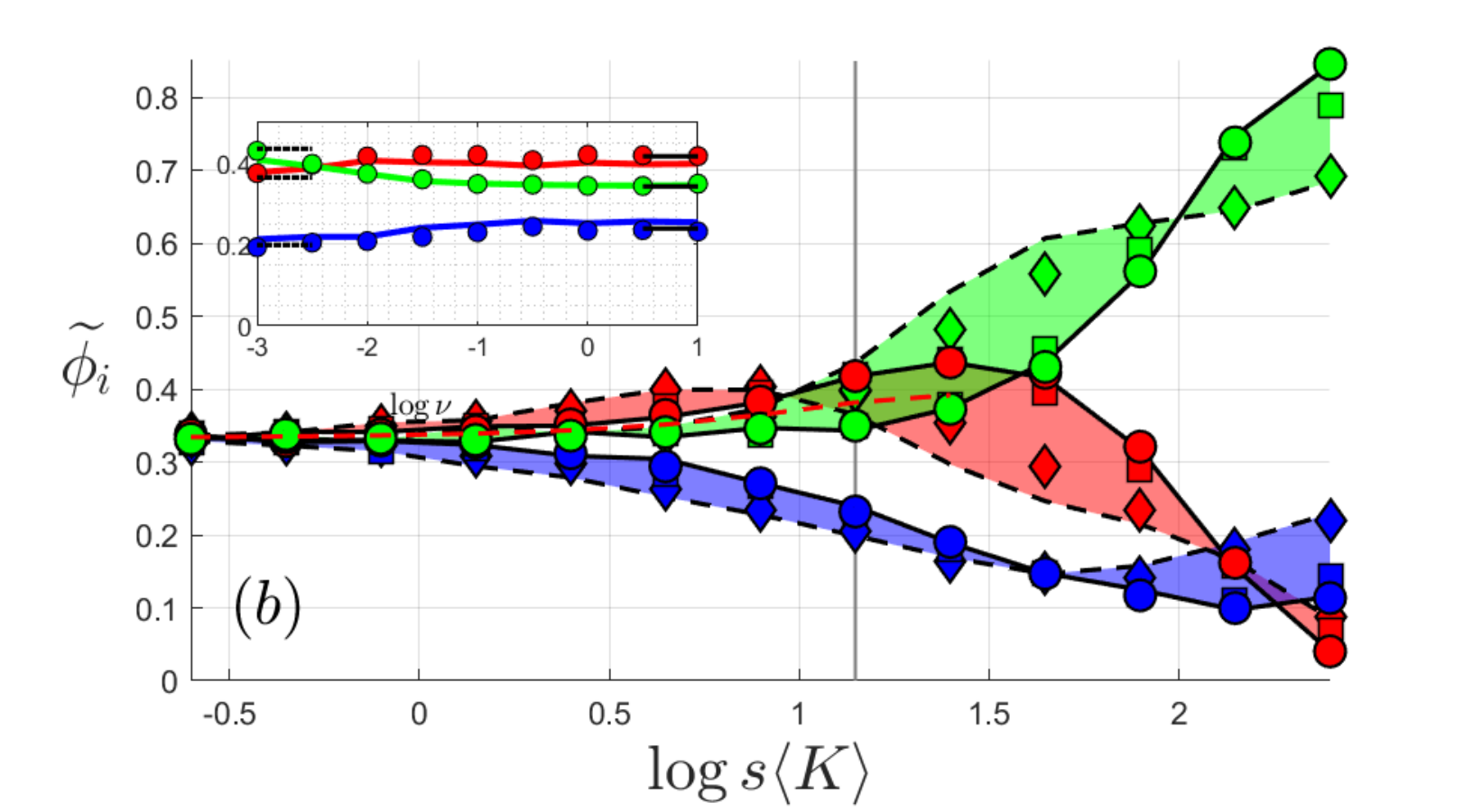}
	\caption{
		Total fixation probabilities 
	 $\widetilde{\phi}_{i}$
	 vs. $s\langle K \rangle$  
	 for  values of  $s\in (10^{-3},1)$  and 
	with $\langle K \rangle=250$ and $\gamma=0.8$   kept fixed, see text.
	(a) $\vec{r}=\vec{r}^{(1)}$; 
	(b) $\vec{r}=\vec{r}^{(2)}$. 
	Shaded areas and symbols are from stochastic simulations
	with $\nu=10$ ($\circ$), $\nu=0.1$ ($\square$), $\nu=10^{-5/2}$ ($\diamond$).
	Solid and dashed black lines show respectively $\widetilde{\phi}_{i}\vert_{{\cal K}}$
	and $(\widetilde{\phi}_{i}\vert_{K_-}+ \widetilde{\phi}_{i}\vert_{K_+})/2$ in both panels and insets, see text.
	Vertical light gray lines indicate $\widetilde{\phi}_{i}$ for
	$s=10^{-1/4}$ (a) and $s=10^{-5/4}$ (b). $\widetilde{\phi}_{i}$ increases with $\nu$
	when the solid black line is above the dashed black line, otherwise  $\widetilde{\phi}_{i}$ decreases with $\nu$, see text.
	Dashed colored lines show $\widetilde{\phi}_{2}$ in (a) and $\widetilde{\phi}_{1}$ in (b)
	obtained from $\widetilde{\phi}_i\approx (1+ \phi_i -\phi_{i-1})/3$, with (\ref{eq:phi3}) and $\nu=10$.
	Insets: $\widetilde{\phi}_{i}$ vs. $\nu$ for $s=10^{-1/4}$ (a) and  $s=10^{-5/4}$ (b);
	symbols are from  stochastic simulations and solid lines in  inset (b)  are predictions of
	(\ref{eq:PhiTot}) obtained using (\ref{eq:phi3}), with $\phi_{i,i+1},\phi_{i-1,i}$ inferred from simulations.  
	Fixation scenario  changes at $\nu=\nu^*(s)$ with $\nu^*\approx 10^{-2}$ in (a) and $\nu^*\approx 10^{-5/2}$ in (b), 
	see text.
	In all panels and insets:  species 1 in red, species 2 in blue, species 3 in green;
	$\vec{x}_0=\vec{x}_c$, $\epsilon=0$. 
	}
	\label{fig:Fig7}
\end{figure}

In order to study the influence of the environmental variability on  $\widetilde{\phi}_{i}$,
it is again useful to consider the limiting cases of fast/slow switching.
In fact, as shown in Fig.~\ref{fig:Fig7}, when $\nu\to \infty, 0$, the overall fixation probability
is given by $\widetilde{\phi}_i\to\widetilde{\phi}_i^{(\infty)}$
when $\nu\to \infty$ and $\widetilde{\phi}_i\to\widetilde{\phi}_i^{(0)}$
when $\nu\to 0$,
with 
\begin{eqnarray}
%\hspace{-18mm}
 \label{eq:phi_lim_inf}
 \widetilde{\phi}_i^{(\infty)}&\equiv&\widetilde{\phi}_i\vert_{{\cal K}}=\widetilde{\phi}_i
 \vert_{(1-\gamma^2)\langle K\rangle}\\
 \label{eq:phi_lim_zero}
  \widetilde{\phi}_i^{(0)}&\equiv&\frac{1}{2}\left(\widetilde{\phi}_i
 \vert_{(1+\gamma)\langle K\rangle} + \widetilde{\phi}_i
 \vert_{(1-\gamma)\langle K\rangle}\right),
 \end{eqnarray}
 where $\widetilde{\phi}_i
 \vert_{K}$ is the overall fixation probability in the BDCLV with constant carrying capacity $K$, see Fig.~\ref{fig:Fig4}~(a,b). 
 These results stem from the outcomes of Stage 2 when  $\alpha_i \langle K\rangle \ll 1$ and from
 Stage 1 when  $\alpha_i \langle K\rangle \gg 1$:
 
 \vspace{1mm}
 
 - When $s \langle K\rangle \ll 1$, in regime (i) and about the boundary of regimes (i)-(ii): 
$\phi_{i,i+1}\approx 1/3$ for all species and $P_{(i,i+1)}\approx 1$, see \ref{AppendixD}. The overall fixation probabilities 
are thus given by  
  $\widetilde{\phi}_i\approx (1+\phi_i -\phi_{i-1})/3$, where $\phi_i\approx \phi_i^{(\infty)}$
  if $\nu/s\gg 1$ and  $\phi_i\approx \phi_i^{(0)}$ if $\nu/s\ll 1$, yielding (to leading order in $s\langle K\rangle$)
  \begin{eqnarray}
 \label{eq:phitotsw1}
 \widetilde{\phi}_{i}&\approx& \widetilde{\phi}_{i}\vert_{\kappa}=\frac{1}{3}\left[1+\frac{s{\kappa}}{12}(r_i-r_{i-1})\right],
 \end{eqnarray}
where  $\kappa=(1-\gamma^2)\langle K \rangle$ if $\nu/s \gg 1$ and $\kappa=\langle K \rangle$ if $\nu/s \ll 1$. 
In agreement with Fig.~\ref{fig:Fig7},
Eq.~(\ref{eq:phitotsw1}) predicts that 
$\widetilde{\phi}_{i}$ is greater than $1/3$ and increases with $s\langle K\rangle$ (at $\nu$ fixed)
if $r_i> r_{i-1}$, whereas  $\widetilde{\phi}_{i}$ is less than $1/3$ and is a decreasing function of  $s\langle K\rangle$ (at $\nu$ constant)
when $r_i< r_{i-1}$.

 \vspace{1mm}
 
 - When $\alpha_i \langle K\rangle \gg 1$, about the boundary of regimes (ii)-(iii) and in regime (iii):
  Selection strongly favors species $i$ on edge $(i,i+1)$ in Stage 2, 
   and the fixation probability is  determined by the outcome of Stage 1:
   $\widetilde{\phi}_i\approx 
   \widetilde{\phi}_i^{(\infty)}$
   if $\nu\gg 1/\langle K \rangle$ and 
   $\widetilde{\phi}_i\approx \widetilde{\phi}_i^{(0)}$  when $\nu\ll 1/\langle K \rangle$.
    
    \vspace{2mm}

    Hence, in regime (i) and about the boundary of regimes (i)-(ii) and (ii)-(iii), as well as 
in regime (iii) 
we have  $\widetilde{\phi}_{i}\to \widetilde{\phi}_{i}^{(\infty)}
$ when $\nu \to \infty$
and $\widetilde{\phi}_{i}\to \widetilde{\phi}_{i}^{(0)}
$ when $\nu \to 0$. We have found 
that the fixation probabilities of the species surviving
Stage 1 vary monotonically with $\nu$, whereas the fixation probability of the species
 most likely to die out first varies little with $\nu$, see the insets of Fig.~\ref{fig:Fig7}.
Therefore, as corroborated by Fig.~\ref{fig:Fig7}, for finite switching rates, we have 
\begin{eqnarray}
%\hspace{-18mm}
 \label{eq:phi_fin}
 {\rm min}\left(\widetilde{\phi}_i^{(0)},\widetilde{\phi}_i^{(\infty)}\right)<
 \widetilde{\phi}_i<{\rm max}\left(\widetilde{\phi}_i^{(0)},\widetilde{\phi}_i^{(\infty)}\right).
 \end{eqnarray}
 Taking into account the average number of switches arising in Stages 1 and 2, see \ref{AppendixE5},
we have $\widetilde{\phi}_i\approx \widetilde{\phi}_i^{(\infty)}$ when $\nu\gg {\rm max}(s,1/\langle K\rangle)$
 and  $\widetilde{\phi}_i\approx \widetilde{\phi}_i^{(0)}$ if $\nu\ll  {\rm min}(s,1/\langle K\rangle)$, see Fig.~\ref{fig:Fig7}. 

According to Eqs.~(\ref{eq:phi_lim_inf})-(\ref{eq:phi_fin}),  
 the fixation probabilities under random switching can be inferred from $\widetilde{\phi}_i\vert_K$ obtained in 
 the constant-$K$ BDCLV with a suitable value of $K$:\\
 - Under fast switching, $\widetilde{\phi}_i$ coincides with
  $\widetilde{\phi}_i\vert_{(1-\gamma^2)\langle K\rangle}$. Since $\widetilde{\phi}_i\vert_K$ is a function of $sK$,
  when the average carrying capacity $\langle K\rangle$ is kept fixed, 
   $\widetilde{\phi}$ is thus given by $\widetilde{\phi}_i\vert_{\langle K\rangle}$ subject to a 
 rescaled  
  selection intensity $(1-\gamma^2)s$. Hence, when $\nu\gg {\rm max}(s,1/\langle K\rangle)$
  and $\langle K\rangle$ is kept fixed, the effect of random switching is to reduce the selection intensity by
  a factor $1-{\rm var}(K(t))/\langle K\rangle^2$.
  \\
 - Under slow switching,  $\widetilde{\phi}_i$ is given by the arithmetic average of $\widetilde{\phi}_i\vert_{K_+}$
and $\widetilde{\phi}_i\vert_{K_-}$.
 When the average carrying capacity $\langle K\rangle$ is kept fixed, 
 $\widetilde{\phi}$ is thus given by the  average of $\widetilde{\phi}\vert_{\langle K\rangle}$
 subject to a selection intensity $(1+\gamma)s$ and $(1-\gamma)s$. 
These predictions, agree with the results of Fig.~\ref{fig:Fig7}, and imply that 
 the $s\langle K \rangle$-dependence of $\widetilde{\phi}_i$ can be readily  obtained from
 Fig.~\ref{fig:Fig4}~(a,b). 

\begin{figure}[t]
	\centering
	\includegraphics[width=0.98\linewidth]{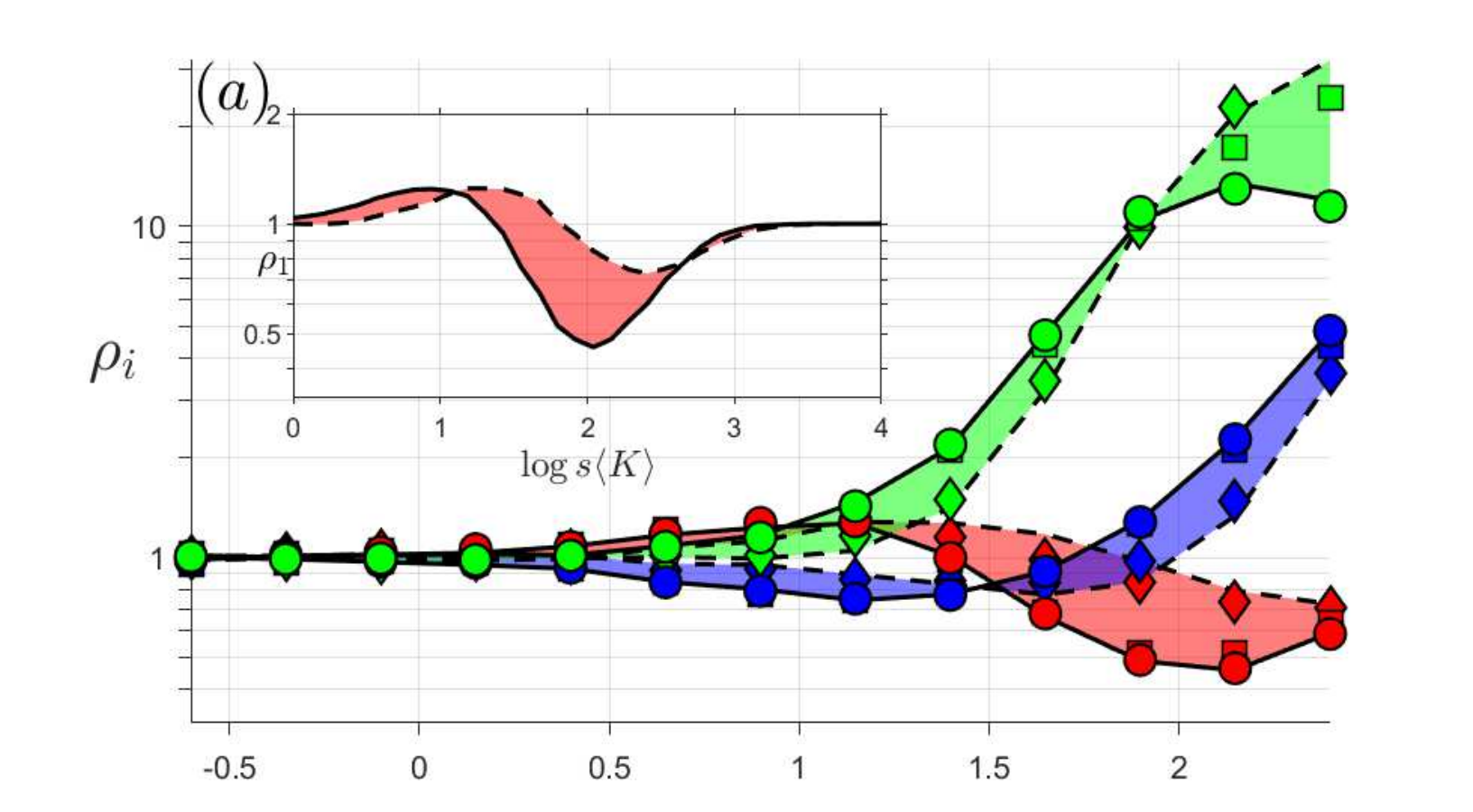}\\
	\includegraphics[width=0.98\linewidth]{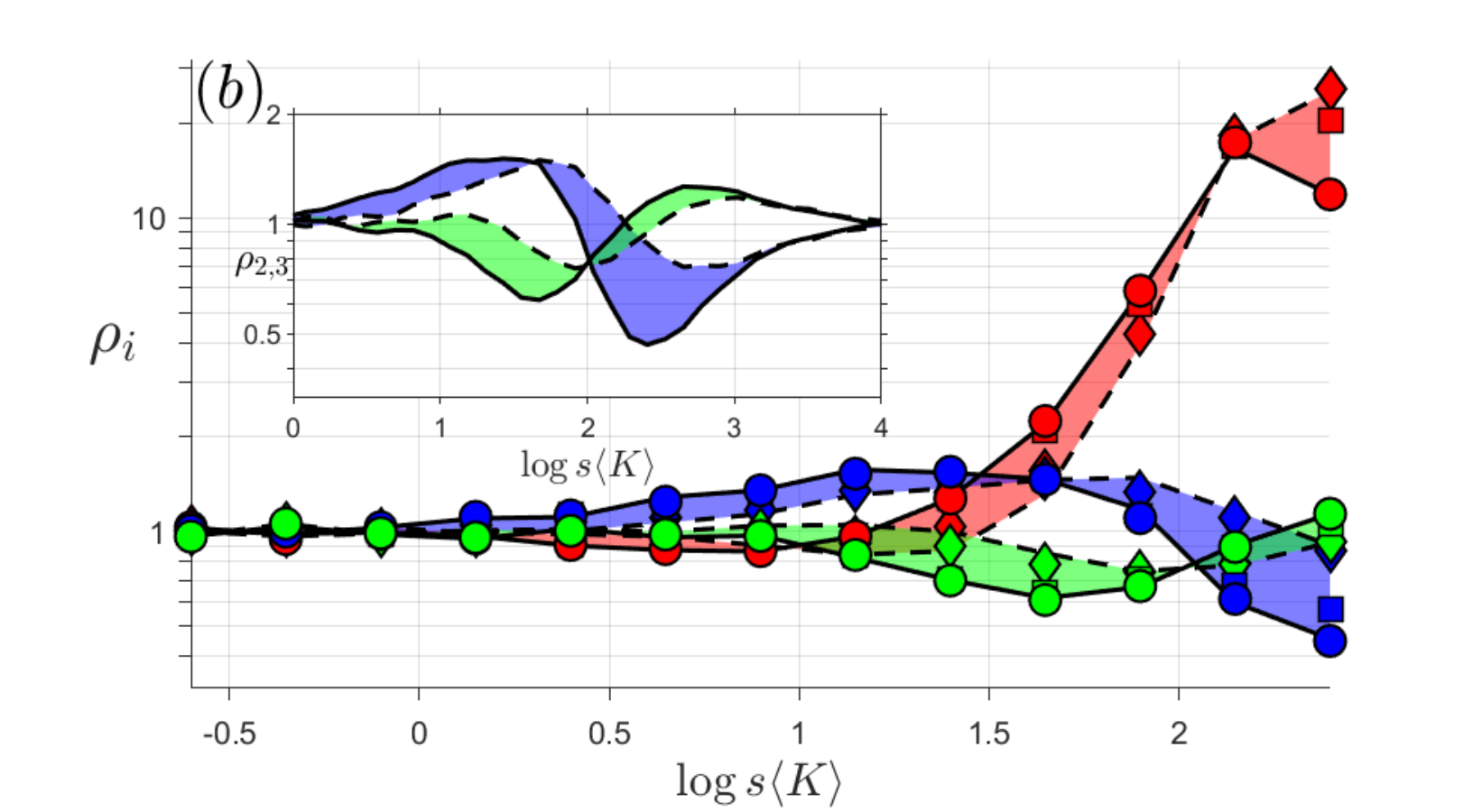}\\
	\caption{$\rho_i$ vs. $s\langle K\rangle$ 
	for  values of  $s\in (10^{-3},1)$  and 
	with $\langle K \rangle=250$ and $\gamma=0.8$   kept fixed, see text.
	(a) $\vec{r}=\vec{r}^{(1)}$; (b) $\vec{r}=\vec{r}^{(2)}$.
Shaded areas and symbols are from stochastic simulations
	with $\nu=10$ ($\circ$), $\nu=0.1$ ($\square$), $\nu=10^{-5/2}$ ($\diamond$); 
	 lines show $\rho_i^{(\infty)}$ 
	 (fast switching, solid) 
	 and  $\rho_i^{(0)}$ (slow switching, dashed), see text.
	 %In panel (b), we notice that, in agreement with (\ref{eq:rhoi}), 
	 %$\rho_2<1$ for $s\langle K\rangle \lesssim 10$ and $\rho_2>1$ for $s\langle K\rangle \gtrsim 30$.
	 Insets: (a) $\rho_1^{(\infty)}$ (solid) and $\rho_1^{(0)}$ (dashed) vs. $s\langle K\rangle$; 
	 (b) $\rho_2^{(\infty)}$ and $\rho_3^{(\infty)}$ (solid), $\rho_2^{(0)}$ and $\rho_3^{(0)}$ (dashed) 
	 vs. $s\langle K\rangle$ with $\gamma=0.8$ and $\langle K \rangle =10000$ fixed and $s$ varies  between $1/\langle K \rangle$ and $1$. 
	When  $s\langle K\rangle = 10^3 - 10^4$, $\rho_i \to 1$.
        In both panels and insets:  species 1 in red, species 2 in blue, 
        and species 3 in green;
	$\vec{x}_0=\vec{x}_c$; $\epsilon=0$. 
	}
	\label{fig:Fig8}
\end{figure}

At this point, we can discuss the effect of random switching on $\widetilde{\phi}_i$
by comparison with $\widetilde{\phi}_i\vert_{\langle K\rangle}$ in the constant-$K$ BDCLV, when $\langle K\rangle$
is kept fixed:

\begin{itemize}
\item {\it Random switching ``levels the field'' of competition and balances the effect of selection}:
 The species that is the least likely to fixate has a higher 
fixation probability under random switching than under a constant $K=\langle K\rangle$, compare
Figs.~\ref{fig:Fig4}~(a,b) and \ref{fig:Fig7} (see also Fig.~\ref{fig:Fig8}).
The DMN therefore balances the selection pressure that favors the fixation of the other species, 
and hence levels the competition. 
\item {\it Random switching effectively reduces the selection intensity under fast switching}:
When $\nu \gg {\rm max}(s,1/\langle K\rangle)$, 
we have seen $\widetilde{\phi}_i=\widetilde{\phi}_i\vert_{\langle K\rangle}$ subject to a 
 rescaled  
  selection intensity $(1-\gamma^2)s=(1-{\rm var}(K(t))/ \langle K \rangle^2)s$.
Fast random switching therefore  reduces the selection intensity  proportionally 
 to the variance of $K$.
 Hence, under strong selection and fast switching,
 a  zero-one LOW law  appears  in the  switching-$K$ BDCLV only in a population whose 
 average size is $1/(1-\gamma^2)$ times
 greater  than in the constant-$K$ BDCLV. 
 This means that when $K$ has a large variance (large $\gamma$) 
 the onset of the zero-one LOW, with $\widetilde{\phi}_i \to 0,1/2,1$, 
  in the fast switching-$K$ BDCLV arises when $s\langle K \rangle \gg 1$ and $\langle K \rangle$ is at least
  one order of magnitude larger than in the constant-$K$ BDCLV ({\it e.g.}, $\langle K \rangle\gtrsim 10^4$ instead 
  of  $\langle K \rangle\gtrsim 10^3$ when $\gamma=0.8$), see also Fig.~\ref{fig:Fig8}.
\item {\it Random switching can yield new fixation scenarios}:
Which species is the most likely to fixate can vary with  $\nu$ and $\gamma$, at $s$ and $\langle K \rangle$ fixed,
and does not generally obey a simple law (neither LOW nor LOSO). When
the environmental variance is large enough ($\gamma\gtrsim\gamma^*$) 
the shaded areas of Fig.~\ref{fig:Fig7}  can overlap. This occurs
when the fixation probabilities of the two most likely species to prevail cross, see insets of Fig.~\ref{fig:Fig7}. This
yields
different fixation scenarios  below/above a critical switching rate $\nu^*(s)$:
one of these species  is the best off at low switching rate, while the other is the best to fare
under fast switching. These crossings therefore  signal a  stark departure from the LOW/LOSO laws.
For a crossing between $\widetilde{\phi}_{i}$ and $\widetilde{\phi}_{i+1}$ to be possible,
one, say $\widetilde{\phi}_{i}$, should decrease and the other  increase with $\nu$, i.e.
 $\widetilde{\phi}_{i}^{(\infty)}<\widetilde{\phi}_{i}^{(0)}$
 and  $\widetilde{\phi}_{i+1}^{(\infty)}>\widetilde{\phi}_{i+1}^{(0)}$
 Thus, if $\widetilde{\phi}_{i}^{(0)}>\widetilde{\phi}_{i+1}^{(0)}$ and 
 $\widetilde{\phi}_{i}^{(\infty)}<\widetilde{\phi}_{i+1}^{(\infty)}$, 
 there is a critical switching rate $\nu=\nu^*(s)$ where $\widetilde{\phi}_{i}=\widetilde{\phi}_{i+1}$.  The crossing conditions 
 can be determined  using (\ref{eq:phi_lim_inf}) and  
 (\ref{eq:phi_lim_zero}).
 A new fixation scenario emerges when the switching rate varies across $\nu^*$:
 $\widetilde{\phi}_{i+1}>\widetilde{\phi}_{i}$ when $\nu>\nu^*$,
 while $\widetilde{\phi}_{i+1}\leq \widetilde{\phi}_{i}$ when $\nu\leq\nu^*$. 
 Intuitively, crossings are possible when the variance of $K$ is large ($\gamma\gtrsim\gamma^*$),
 ensuring that Stage 1 ends up with comparable probabilities of hitting two edges of $S_3$, and  the two most likely species 
 to fixate have a different  $\nu$-dependence arising from Stage 2, see Fig.~\ref{fig:Fig6}~(c,d).
 In the inset of Fig.~\ref{fig:Fig7}~(a),
$\widetilde{\phi}_{1}$ decreases and $\widetilde{\phi}_{2}$  increases with $\nu$;
they intersect at $\nu=\nu^*\approx 0.01$ for $s=10^{-1/4}$: Species 1 is the most likely to fixate at
 $\nu<\nu^*$ and species 2 the most likely to prevail at $\nu>\nu^*$, and we have  
 $\widetilde{\phi}_{1}>\widetilde{\phi}_{2}\gg \widetilde{\phi}_{3}$  for 
$\nu<\nu^*$ and $\widetilde{\phi}_{2}\gg\widetilde{\phi}_{1}>\widetilde{\phi}_{3}$ when $\nu>\nu^*$. 
%; species 1 is thus the most likely to fixate when $\nu<\nu^*$, whereas 
%species $2$ is the most likely to prevail when $\nu>\nu^*$ 
This is to be contrasted with 
 Fig.~\ref{fig:Fig4}~(a), where the LOW yields $\widetilde{\phi}_{1}\vert_{\langle K\rangle}
 \gg \widetilde{\phi}_{2}\vert_{{\langle K\rangle}}\gg \widetilde{\phi}_{3}\vert_{{\langle K\rangle}}$.  
The inset of Fig.~\ref{fig:Fig7}~(b), shows another example of a fixation scenario that depends on $\nu$,
with 
$\widetilde{\phi}_{3}> \widetilde{\phi}_{1}>\widetilde{\phi}_{2}$ when $\nu<\nu^*\approx 0.03$
and  $\widetilde{\phi}_{1}\gtrsim \widetilde{\phi}_{3}>\widetilde{\phi}_{2}$ when $\nu>\nu^*$.
\end{itemize}

The main effect of the random switching of $K$
is therefore to  balance the influence of selection and to ``level the field'' of cyclic dominance
according to (\ref{eq:phi_lim_inf})-(\ref{eq:phi_fin}). 
This is particularly important under strong selection and large $K$ variability, when 
random switching  hinders the LOW by effectively promoting the fixation of the species 
that are less likely to prevail under constant $K=\langle K\rangle$. This can result in new fixation scenarios in which 
the most likely species to win varies with the variance and rate of change of the carrying capacity.
The CLV fixation scenarios are therefore richer and more complex when demographic and environmental noise are coupled than
when they are independent of each other as, {\it e.g.}, in Ref.~\cite{West18}.

 To rationalize further how environmental variability affects the fixation probabilities,
 we compute the ratio
 \begin{eqnarray}
  \label{eq:rhoi}
  \rho_i\equiv \frac{\widetilde{\phi}_{i}}{\widetilde{\phi}_{i}\vert_{\langle K \rangle}}.
 \end{eqnarray}
Using (\ref{eq:phi_lim_inf}) and (\ref{eq:phi_lim_zero}), we have 
 $\rho_i\to \rho_i^{(\infty)}\equiv  \widetilde{\phi}_{i}\vert_{(1-\gamma^2)\langle K 
\rangle}/\widetilde{\phi}_{i}\vert_{\langle K \rangle}$ 
and $\rho_i\to \rho_i^{(0)}\equiv  (\widetilde{\phi}_{i}\vert_{K_-}+ \widetilde{\phi}_{i}\vert_{K_+})/(2
 \widetilde{\phi}_{i}\vert_{\langle K \rangle})$ for fast and slow switching, respectively.
We  say that random switching enhances the fixation of species $i$ when $\rho_i>1$, whereas DMN hinders 
species $i$'s fixation when $\rho_i<1$ and environmental variability has no influence if  $\rho_i\approx 1$.
Simulation results of Fig.~\ref{fig:Fig8}
show that  $\rho_i$ varies non-monotonically across regime (i)-(iii), with a weak dependence on the 
switching rate $\nu$, and $\rho_i$ lying between $\rho_i^{(0)}$ and $\rho_i^{(\infty)}$ for intermediate $\nu$.

It is clear in Fig.~\ref{fig:Fig8} that, when there is enough environmental variance
(large $\gamma$),
the main effect of random switching arises  at the boundary of regimes (ii)-(iii) and in regime (iii):
In this case, the DMN balances the strong selection pressure yielding
$\widetilde{\phi}_{i}<1$ and  $\rho_i<1$ when
$\widetilde{\phi}_{i}\vert_{\langle K \rangle}\approx 1$  (for $r_i<r_{i\pm 1}$),
 and $\widetilde{\phi}_{i}>0$  and $\rho_i>1$ when 
$\widetilde{\phi}_{i}\vert_{\langle K \rangle}\approx 0$  (for $r_i>r_{i\pm 1}$).
This signals a systematic deviation from the asymptotic zero-one law predicted by the LOW
in the constant-$K$ BDCLV. The LOW and the zero-one LOW still arise in the switching-$K$
BDCLV with $s={\cal O}(1)$, but they set in for much larger values of $\langle K \rangle$
than in the constant-$K$
BDCLV
%: $\widetilde{\phi}_{i}\approx \widetilde{\phi}_{i}\vert_{\langle K \rangle}\to 0, 1/2$ or $1$, 
%with $\rho_i\to 1$
 (for $\langle K \rangle= 10^3- 10^4$), see insets of Fig.~\ref{fig:Fig8}. This 
demonstrates again that environmental variability acts to ``level the field''
of cyclic competition among the species by hindering the onset of the zero-one LOW.
%.

From Eq.~(\ref{eq:phitotsw1}), when $s \langle K \rangle \ll 1$, 
to leading order, we find 
 \begin{eqnarray}
  \label{eq:rhoi1}
  \rho_i=
1-s(\langle K\rangle-\kappa)\left(\frac{r_i -r_{i-1}}{12}\right),
 \end{eqnarray}
with  $\kappa=(1-\gamma^2)\langle K \rangle$ if $\nu/s \gg 1$ and $\kappa=\langle K \rangle$ if $\nu/s \ll 1$.
When $s \langle K \rangle \ll 1$ and $\nu/s \gg 1$, we thus have have 
$\rho_i\approx 1-s\gamma^2(r_i-r_{i-1})/12$ when $\nu/s \gg 1$ and $\rho_i=1 + {\cal O}(s^2)$ when $\nu/s \ll 1$.
This means that in regime (i), and at the boundary of regimes (i)-(ii),
when there is enough switching ($\nu \gg s$), $\rho_i>1$ if $r_i<r_{i-1}$ 
and $\rho_i<1$  if $r_i>r_{i-1}$, which is in agreement with the results of Fig.~\ref{fig:Fig8}. Accordingly, 
whether a fast switching environment promotes/hinders species $i$ under weak selection depends
only on its growth rate relative to  that  of its strong opponent. 
In Fig.~\ref{fig:Fig8}, we notice a non-monotonic dependence of $\rho_i$
on $s\langle K \rangle$ resulting from a different influence of environmental variability 
under weak and strong selection: In Fig.~\ref{fig:Fig8}, the fixation probability
of a species that is promoted/hindered under weak selection
is hindered/promoted under strong selection.

\subsection{Mean fixation time in the switching-$K$ BDCLV}
\label{Sec4.2}
In \ref{AppendixE2}, we analyze the effect of random switching
on the mean extinction and absorption times $T_1$ and $T_2$
characterizing respectively the stages 1 and 2 of the switching-$K$ BDCLV dynamics, see Fig.~\ref{fig:FigA5}(a,b).
We thus show 
  that, when $\vec{x}_0=\vec{x}_c$, the overall mean fixation time 
 $ T_F=T_1+T_2= {\cal O}(\langle N\rangle)={\cal O}(\langle K\rangle)$
 scales linearly with the average  population size, see Fig.~\ref{fig:FigA5}(c),
 similarly to $T_F$ in the constant-$K$ BDCLV. Hence, random switching makes
the cyclic competition more ``egalitarian'' but does not prolong species coexistence.
We also show that the average number 
of switches occurring in Stage 1 scales as $\nu\langle K \rangle$, see Fig.~\ref{fig:FigA6}~(a), while
the  average  number environmental switches along the edge $(i,i+1)$ in Stage 2
 scales as ${\cal O}(\nu/\alpha_i)$ 
when $s$ is neither vanishingly small nor too large.
\section{Fixation properties of  close-to-zero-sum rock-paper-scissors games in  fluctuating populations}
\label{genRPS}
The general, {\it non-zero-sum}, rock-paper-scissors
refers to the game
with payoff matrix (\ref{payoff}) where  $\epsilon\neq 0$
and non-zero 
average fitness $\bar{f}=1-\epsilon\sum_{i=1}^{3}\alpha_i x_i x_{i+1}$.
The mean-field description of the 
 general RPS game,
 formulated as the birth-death process (\ref{eq:birtheq})-(\ref{eq:ME}) with  $0\leq s\leq 1/(1+\epsilon)$,
 is given by (see Sec.~S1.1)
\begin{eqnarray}
\label{eq:xidot_eps}
\hspace{-3mm}
\dot{N} & = &  N\left(\bar{f}- \frac{N}{K}\right)  \nonumber\\
\hspace{-3mm}
\dot{x_i}&=&x_i [\alpha_i x_{i+1}-(1+\epsilon)\alpha_{i-1}x_{i-1}+1-\bar{f}].
\end{eqnarray}
In this model, the evolution of $N$ is 
coupled with  the $x_i$'s, whose mean-field dynamics is 
characterized by heteroclinic cycles when $\epsilon>0$
and a stable coexistence fixed point when $\epsilon<0$
~\cite{MayLeonard,Maynard,Hofbauer,Szabo07,Broom,SMR14,JPAtopical}

\begin{figure}[t!]
	\centering
	\includegraphics[width=0.98\linewidth]{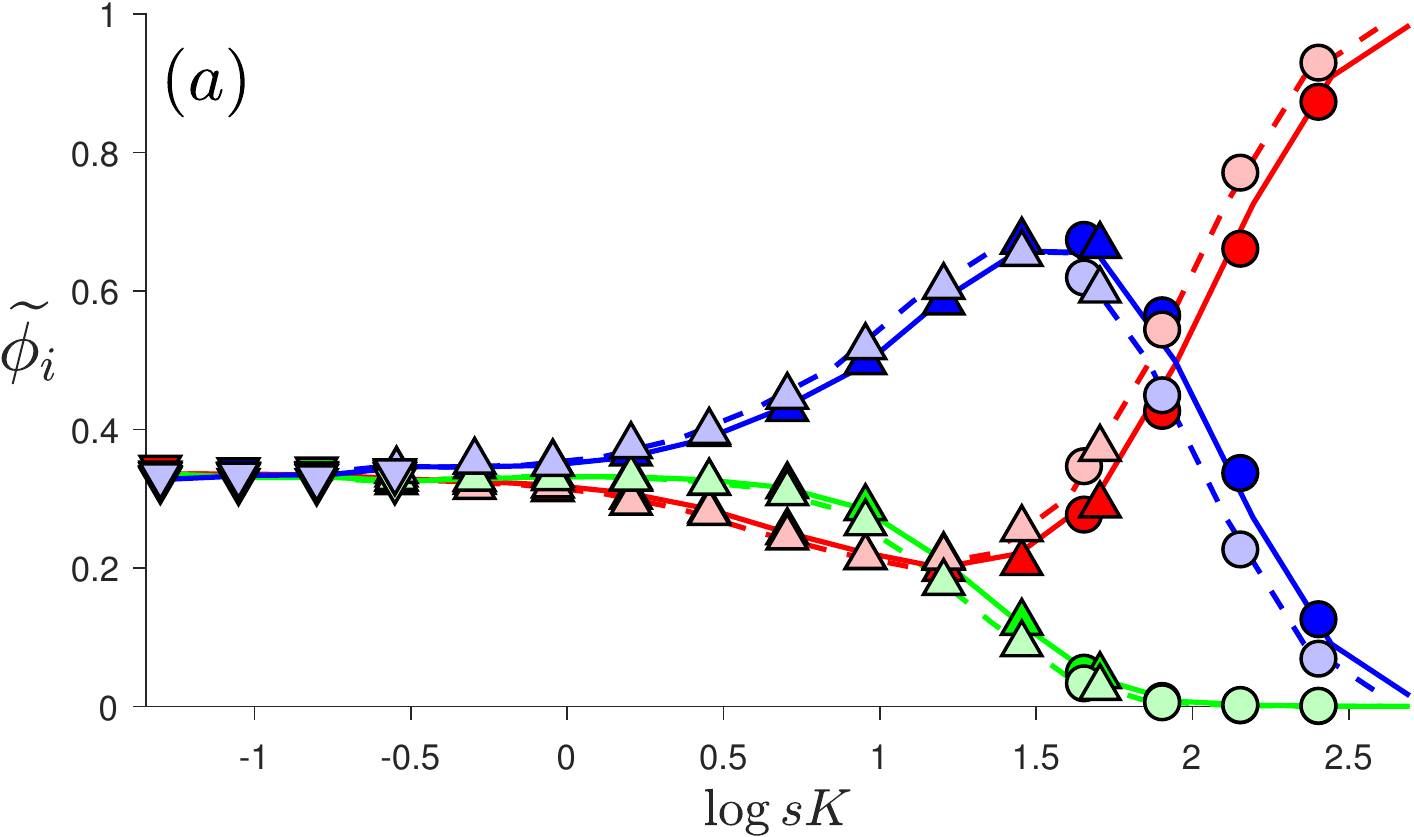}\\
	\includegraphics[width=0.98\linewidth]{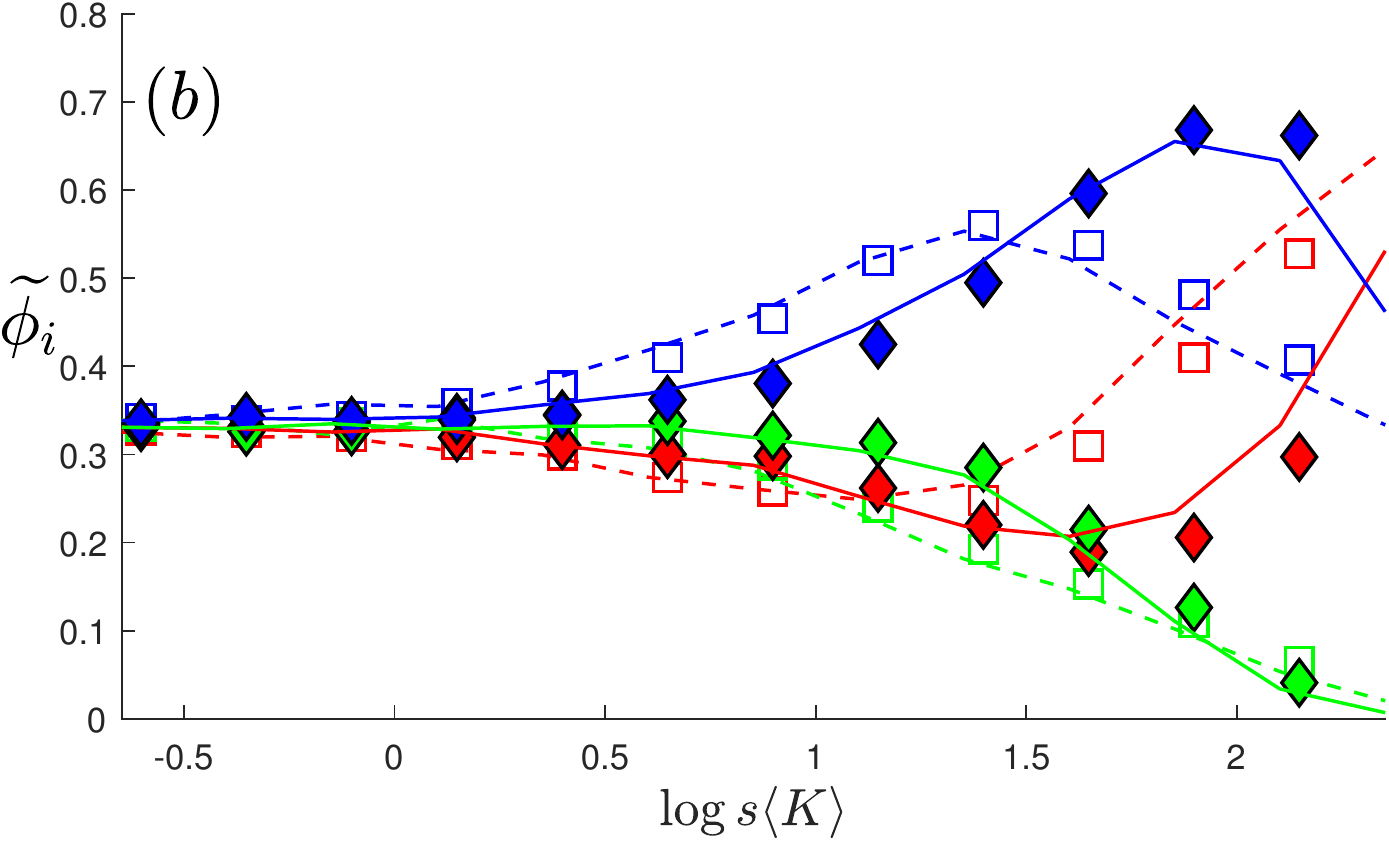}\\
	\includegraphics[width=0.98\linewidth]{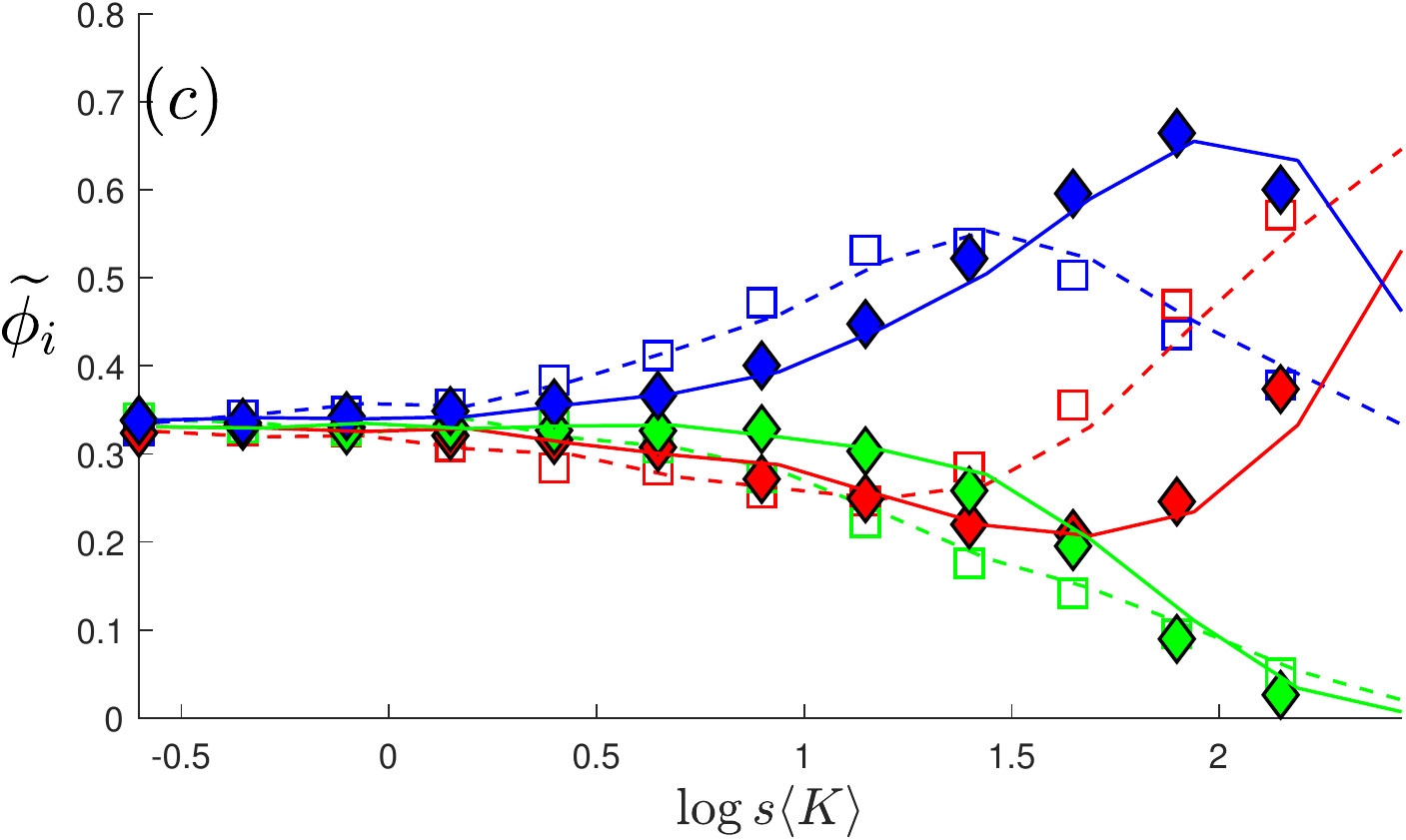}
	\caption{(a) ${\widetilde \phi}_i$ vs. $sK$ in the close-to-zero-sum RPS game
	with  constant  carrying capacity   
	$K = 450$~(circles), $90$~(upward triangles), $50$~(downward triangles),  $\epsilon = -0.2$ (light symbols)
	and $\epsilon = 0.2$ (dark symbols). 
	Lines show stochastic simulation results for the BDCLV
	 ($\epsilon=0$, see Fig.~\ref{fig:Fig4}) with rescaled selection intensity $s\to s(1+\epsilon/2)$
	 with $\epsilon = 0.2$ (solid) and $\epsilon = -0.2$ (dashed). Dark symbols / solid lines
	 and  light symbols / dashed lines collapse, demonstrating 
	 $\widetilde{\phi}_i^{\epsilon}(s)\simeq \widetilde{\phi}_i^{{\rm BDCLV}}(s(1+\epsilon/2))$, 
	see text.  
	(b) ${\widetilde \phi}_i$ vs. $s\langle K\rangle$ when $K$ switches between $K_- = 50$ and $K_+ = 450$~($\langle K 
	\rangle=250, \gamma=0.8$), with  $s\in (10^{-3},1)$.
	Symbols are stochastic simulation results for $\epsilon=-0.2$ and 
	and $\nu=10$ (filled diamonds) and $\nu=0.001$ (open squares).
	Lines are 
	stochastic simulation results from the BDCLV with same switching carrying capacity,
	$\nu=10$ (solid) and $\nu=0.001$ (dashed)
	and rescaled selection intensity $s\to s(1+\epsilon/2)$, see text and Fig.~\ref{fig:Fig7}.
	(c) Same as in panel (b) with $\epsilon>0$: Symbols are stochastic simulation results for $\epsilon=0.2$;
	 solid ($\nu=10$) and dashed ($\nu=0.001$) lines are results from the BDCLV with same switching carrying capacity and 
	selection intensity $s\to s(1+\epsilon/2)$.
	In all panels: red denotes species 1, blue  species 2, and green  species 3; $\vec{r}=\vec{r}^{(1)}$ and
	$\vec{x}_0=\vec{x}_c$.}
	\label{fig:Fig9}
\end{figure}

In this section, we briefly focus on the fixation probabilities of 
close-to-zero-sum rock-paper-scissors games when $|\epsilon| \ll 1$. 
We therefore approximate $\bar{f}\approx 1$ and  still 
assume that  there is a timescale separation 
between $N$ and $x_i$. This assumption is backed
up by simulations results which also show that
fixation properties are {\it qualitatively the same}
as in the BDCLV, see Fig.~\ref{fig:Fig9} (to be compared with Figs.~\ref{fig:Fig4} and \ref{fig:Fig7}). This suggests 
that the fixation probabilities of close-to-zero-sum  RPS games 
can be obtained from those of the BDCLV by rescaling the selection intensity
according to $s\to s(1+\sigma\epsilon +{\cal O}(\epsilon^2))$, see Fig.~\ref{fig:Fig9}. 
To determine the parameter $\sigma$, we consider the constant-$K$ RPS dynamics  with $|\epsilon|\ll 1$. 
Since the fixation properties of the BDCLV 
vary little with the selection intensity at quasi neutrality and under strong selection, we focus 
on the regime (ii) of weak selection where $s\ll 1$ and $sK={\cal O}(10)$,
and assume that $\phi_{ij}\approx 1/3$ and $P_{(i,j)}\approx 1$. As shown in \ref{AppendixC}, the  
absorption probability of species $i$  along  the edge $(i,i+1)$  in the realm of this approximation is
\begin{eqnarray*}
 \label{eq:phi_eps}
 \phi_i\simeq \frac{e^{-\alpha_i (1+\frac{\epsilon}{2})K}+\alpha_i (1+\frac{\epsilon}{2})K-1}{
 \alpha_i (1+\frac{\epsilon}{2})K(1-e^{-\alpha_i (1+\frac{\epsilon}{2})K})},
\end{eqnarray*}
which coincides with
(\ref{eq:condfixprob2}) upon rescaling the selection intensity according $s\to s(1+ (\epsilon/2))$.
Hence, if $\widetilde{\phi}_i^{\epsilon}(s)$ and $\widetilde{\phi}_i^{{\rm BDCLV}}(s)$ denote respectively
the fixation probability of species $i$ in close-to-zero-sum RPS game with $|\epsilon|\ll 1$
and in the BDCLV, we have $\widetilde{\phi}_i^{\epsilon}(s)\simeq \widetilde{\phi}_i^{{\rm BDCLV}}(s(1+\epsilon/2))$.
Since  $\widetilde{\phi}_i$ is related to $\phi_i\vert_{K}$, via
(\ref{eq:PhiTot1}), 
 the overall fixation probability 
is also obtained by rescaling the fixation probability
$\widetilde{\phi}_i^{\rm BDCLV}$ with the same carrying capacity $K$ according to
$s\to s(1+ (\epsilon/2))$. This is confirmed by the results of Fig.~\ref{fig:Fig9}~(a) where we find that
this scaling  holds across the regimes (i)-(iii).

This conclusion also holds when the carrying capacity $K(t)$
is randomly switching according to  (\ref{eq:K}) and $|\epsilon|\ll 1$, see Fig.~\ref{fig:Fig9}~(b,c).
In fact, proceeding as above and 
 focusing  on the weak selection regime where 
 $s\ll 1$ and $sK={\cal O}(10)$, we 
 can assume  
 $\phi_{ij}\approx 1/3$ and $P_{(i,j)}\approx 1$, and 
 find that $\phi_i$
 is given by (\ref{eq:phi3}) with the same carrying capacity $K(t)$ and a rescaled  selection intensity  $s\to s(1+ (\epsilon/2))$. 
 Along the same arguments as above, we  expect that also when the carrying capacity is switching,
 the overall fixation probabilities across the
regimes (i)-(iii) are approximately
 the same as in  the switching-$K$ BDCLV subject to a rescaled 
 selection intensity $s(1+ (\epsilon/2))$. This is confirmed by the results  of Fig.~\ref{fig:Fig9}~(b,c)
 where we have reported
$\widetilde{\phi}_i$  for  fast and slow switching rates. 
As in the BDCLV, 
values of $\widetilde{\phi}_i$  for intermediate  $\nu$ lie between the 
data shown in Fig.~\ref{fig:Fig9}~(b,c).

In Section Sec.~\ref{AppendixE6}, we show that the mean fixation time in the BDCLV with a rescaled  selection intensity 
$s\to s(1+ (\epsilon/2))$
allows us to obtain the mean fixation time of the close-to-zero-sum RPS game when $sK$
and $s\langle K\rangle$ are of order ${\cal O}(10)$ and
$|\epsilon|\ll 1$.

\section{Summary \& Conclusion}\label{disc}
Inspired by the evolution of microbial communities in volatile environments,
we have studied the evolution three species engaged in a
cyclic rock-paper-scissors competition when the environment varies randomly.
 In a static environment,  
the fixation probabilities in rock-paper-scissors games obey two different laws: The ``law of the weakest'' (LOW) prescribes that
the species with the lowest payoff  is the most likely to fixate in large populations, whereas 
a different rule (``law of stay out'', LOSO) arises in smaller populations~\cite{Berr09,Frean01,Ifti03,West18}. 
In this work, we have studied how this simple scenario changes 
when environmental and demographic noise are \emph{coupled}.  Environmental randomness is here introduced via a  randomly switching carrying capacity 
(dichotomous Markov noise) modeling  how the available resources 
switch continuously between states of scarcity and abundance.

We have  studied a birth-and-death process, 
in which a fluctuating population of three species competing cyclically
is subject to either a constant or  randomly switching carrying capacity. As demographic fluctuations (internal noise)
 depend on the population size which in turn varies with the switching carrying capacity, 
 internal and environmental noise are here coupled.  
The size of the fluctuating population can 
be subject to either the LOW (weak internal noise) or the LOSO (stronger internal noise),
or can switch between values subject to one and then the other law. This can greatly 
influence the fixation properties: It is not clear which species will be the most likely to prevail 
 when the population size fluctuates and how the outcome depends 
on the environmental variability. These questions have been studied in  detail for the 
zero-sum rock-paper-scissors game, equivalent to the
cyclic Lotka-Volterra model (CLV). 

The CLV dynamics consists of two stages: Species coexist in Stage 1
until one of them dies out initiating Stage 2 that consists of a two-species competition.
When the carrying capacity is constant, the CLV fixation probabilities under strong selection obey
the LOW and the LOSO holds under  weak selection.
When the CLV is subject to a randomly switching carrying capacity, the fixation probabilities %under fast and slow switching
can be expressed in terms of the fixation probabilities
 of the CLV subject to a suitable constant carrying capacity.
This has allowed us to analyze in detail how 
the variance and rate of change of the carrying capacity affect the fixation properties of the CLV. 
We have found that the general effect of random switching is to balance  selection, and to ``level the field'' of the cyclic 
competition: When the average carrying capacity is kept constant, the species that is the least likely to fixate has 
a higher probability to prevail under random switching than in a static environment. 
In particular, we have shown that when the rate of switching is large,
the effect of the environmental noise is to effectively reduce the selection strength by a factor increasing with the variance
of the carrying capacity. Hence, when the carrying capacity has a large variance,
the LOW becomes a zero-one-law only for much larger average population size than in the absence of switching. 
We have also found new fixation scenarios, not obeying neither the LOSO nor the LOW:
Under determined conditions, one of the  species surviving Stage 1 is best off below a critical
switching rate, whereas the other is most likely to win under faster switching.
Under random switching, fixation still occurs after a mean time that scales linearly with the average of the population size, with
the subleading prefactor affected by the switching rate. Hence, environmental variability
renders cyclic competition more ``egalitarian'' but does not prolong species coexistence.
Finally, we have considered close-to-zero-sum
rock-paper-scissors games  and have  shown that
the fixation probabilities can be obtained from those of the CLV by a suitable rescaling of the selection intensity.

\begin{center}
{\bf Acknowledgments}
\end{center}
We thank Alastair  Rucklidge for many helpful discussions.
The support of an EPSRC PhD studentship (Grant No.  EP/N509681/1) is gratefully acknowledged.

%%%%%%%%%%%%%%%%

\renewcommand{\thefootnote}{\fnsymbol{footnote}}
\renewcommand{\thefigure}{S\arabic{figure}}
\renewcommand{\theequation}{S\arabic{equation}}
\renewcommand{\thesection}{SM-\Roman{section}}
\setcounter{figure}{0}
\setcounter{equation}{0}
\setcounter{section}{0}

%\newpage

\begin{appendix}

\begin{onecolumn}
% Create the title.
\begin{center}
{{\itshape\Large Appendix: Supplementary Material for}\\~
\\\Large Fixation properties of rock-paper-scissors games in fluctuating populations}
\end{center}

In this Supplementary Material (SM), we provide additional information about
the relationships between various rock-paper-scissors models (Section A), 
and further technical details concerning 
the stages 1 and 2 dynamics (Section B and C). We also analyze the population composition at the inception of Stage 2
(Section D), as well as the mean extinction, absorption and fixation times (Section E) and discuss 
the average number of switches occurring in Stages 1 and 2. The notation in this SM is the same as in the main text; 
all  equations not given in this SM refer to those of the main text.

\section{Various cyclic Lotka-Volterra models (zero-sum rock-paper-scissors games): general properties, similarities and differences}
\label{AppendixA} 
In the literature, there are various
formulations of the zero-sum rock-paper-scissors games, here generically referred to as ``cyclic Lotka-Volterra'' models. 
Here, we consider  the birth-death cyclic Lotka Volterra model (BDCLV), defined in the main text by (2)-(7), 
the cyclic Lotka Volterra model formulated in terms of a Moran process (MCLV), and finally the so-called 
chemical cyclic Lotka volterra model (cCLV). These models are characterized by many similar features, but also some important differences.
Below, we outline some of the main properties of these models and discuss their similarities and differences.
\subsection{The birth-death cyclic Lotka-Volterra model (BDCLV): Mean-field equations and piecewise deterministic Markov process}
\label{AppendixA1} 
 The BDCLV is here defined in terms of the six reactions
\begin{eqnarray}
\label{eq:A1_BDCLV}
N_{i}  \xrightarrow{T_{i}^{+}}  N_{i}+ 1 \quad \text{and} \quad
N_{i}  \xrightarrow{T_{i}^{-}}  N_{i}- 1, \quad \text{with} \quad i\in \{1,2,3\},
 \end{eqnarray}
the first set of reactions corresponds to the birth of an individual of species $i$  and 
the other reaction is associated with the death of an $i$-individual. These reactions occur
with transition rates 
\begin{eqnarray}
\label{eq:A1_BDCLV_rates}
T_{i}^{+}= f_i N_i=(1+s\Pi_i)~N_i=(1+\{\alpha_i x_{i+1} -\alpha_{i-1} x_{i-1}\})~N_i \quad \text{and} \quad 
T_{i}^{-}=\frac{N}{K(t)}~N_{i}, \quad \text{where} \; N=\sum_{i=1}^{3}N_i
\end{eqnarray}
%
%where $N=\sum_{i=1}^{3}N_i$ 
is the total population size and $K(t)$ is the carrying capacity. In this work, we consider the case of a constant and randomly 
switching carrying capacity, namely
\begin{eqnarray*}
\label{eq:A1_K}
K(t)= \begin{cases}
K \quad \text{constant},  &\mbox{ see Section 3} \\
\frac{1}{2}\left[\left(K_+ + K_-\right) + \xi(t)\left(K_+ - K_-\right) \right],\; \text{with dichotomous  noise 
$\xi\in \{-1,+1\}$},  &\mbox{ see  Section 4 } 
\end{cases}.
\end{eqnarray*}
The formulation of the cyclic competition in terms
of the  BDCLV allows us to conveniently introduce the carrying capacity through the death rate $T_{i}^{-}$
and, the population size not being conserved, also enables us to aptly model the cyclic dynamics when 
the population size fluctuates and possibly varies greatly in time.

The BDCLV dynamics is fully described by the underpinning master equation (7) %(\ref{eq:ME})
from which the equation of motion of 
the average number of individual of species $i$ in the environmental state $\xi$
can be derived
as usual\footnote{In this section, for notational convenience $\langle X(\vec{N}) \rangle=\sum_{\vec{N}} X(\vec{N})P({\vec N}, \xi, t)$ denotes the average of 
the observable  $X(\vec{N})$ when the environment remains in the state $\xi$. 
This should not be confused with the notation used in the main text where the angular bracket refers to the average 
over the environmental noise $\xi$.}~\cite{Gardiner,VanKampen}
\begin{eqnarray*}
\hspace{-6mm}
 \label{eq:Ni}
 \frac{d}{dt}\langle N_i \rangle&=&  \frac{d}{dt} \sum_{{\vec N}}  N_i P(\vec{N},\xi,t)=
 \langle T_{i}^{+}\rangle-\langle T_{i}^{-}\rangle, \quad
 \text{where} \nonumber\\
 \langle T_{i}^{-}\rangle&\equiv&
 \begin{cases}
  \left\langle\frac{N}{K}~N_{i}\right\rangle
 & \mbox{, when $K$ is constant}\\
\left\langle T_{i}^{-}\vert_{+}\right\rangle=\left\langle \frac{N}{K_+}~N_{i}\right\rangle
 &\mbox{, in state $\xi=+1$ when $K$ is switching} \\
\left\langle T_{i}^{-}\vert_{-}\right\rangle=\left\langle \frac{N}{K_-}~N_{i}\right\rangle & \mbox{, in state $\xi=-1$ when $K$ is switching} 
\end{cases}.
\end{eqnarray*}
This readily leads to the following equations for the average population size
$\langle N \rangle$ in a static  environment (constant $K$):
\begin{eqnarray}
 \label{eq:A1_N_1}
 \frac{d}{dt}\langle N \rangle=  
 \sum_{i=1}^{3} \left(\langle T_{i}^{+}\rangle-\langle T_{i}^{-}\rangle\right), \quad \text{and in a varying enviroment with a randomly switching $K$:}
 \end{eqnarray}
%
%and in a varying enviroment with a randomly switching $K$:
%
\begin{eqnarray}
 \label{eq:A1_N_2}
 \frac{d}{dt}\langle N \rangle=  
 \begin{cases}
 \sum_{i=1}^{3} \left(\langle T_{i}^{+}\rangle-\langle T_{i}^{-}\vert_{+}\rangle\right) 
 &\mbox{if $\xi=+1$} \\
\sum_{i=1}^{3} \left( \langle T_{i}^{+}\rangle-\langle T_{i}^{-}\vert_{-}\rangle\right) &\mbox{if $\xi=-1$}
\end{cases}.
 \end{eqnarray}

For the population composition, we can proceed similarly to derive the equation motion for 
$\langle x_i \rangle \equiv \langle N_i/N \rangle=  \sum_{{\vec N}}(N_i/N)~P(\vec{N},\xi,t)$, 
paying due attention to the fact that now both $N_i$ and $N$ vary in time: 

\begin{eqnarray}
 \label{eq:xi1}
 %\vspace{-25mm}
&&\frac{d}{dt}\langle x_i \rangle= \sum_{\vec{N}} \frac{N_i}{N}\frac{d}{dt}P(\vec{N},\xi,t)\nonumber\\
&=& \sum_{\vec{N}} \frac{N_i}{N} \left\{
 T_i^+(\vec{N}-\vec{e}_i)P(\vec{N}-\vec{e}_i,\xi,t) + 
 T_i^-(\vec{N}+\vec{e}_i,K)P(\vec{N}+\vec{e}_i,\xi,t) -\left(T_i^+(\vec{N})+T_i^-(\vec{N},K)\right)P(\vec{N},\xi,t)
 \right\}\nonumber\\
&+& \sum_{j\in \{1,2,3\}\neq i}\sum_{\vec{N}} \frac{N_i}{N}
 \left\{
 T_j^+(\vec{N}-\vec{e}_j)P(\vec{N}-\vec{e}_j,\xi,t) + 
 T_j^-(\vec{N}+\vec{e}_j,K)P(\vec{N}+\vec{e}_j,\xi,t) -\left(T_j^+(\vec{N})+T_j^-(\vec{N},K)\right)P(\vec{N},\xi,t)
 \right\}\nonumber\\
 &=& \sum_{\vec{N}}  \left\{
\frac{N_i+1}{N+1} T_i^+(\vec{N})P(\vec{N},\xi,t) + 
 \frac{N_i-1}{N-1} T_i^-(\vec{N},K)P(\vec{N},\xi,t) -
 \frac{N_i}{N}\left(T_i^+(\vec{N})+T_i^-(\vec{N},K)\right)P(\vec{N},\xi,t)
 \right\}\\
&+& \sum_{j\in \{1,2,3\}\neq i}\sum_{\vec{N}} 
 \left\{\frac{N_i}{N+1}
 T_j^+(\vec{N})P(\vec{N},\xi,t) + \frac{N_i}{N-1}
 T_j^-(\vec{N},K)P(\vec{N},\xi,t) -\frac{N_i}{N+1}\left(T_j^+(\vec{N})+T_j^-(\vec{N},K)\right)P(\vec{N},\xi,t)
 \right\},\nonumber
\end{eqnarray}
where $\vec{e}_i$ is the unit vector such that $\vec{e}_1=(1,0,0)$, etc. By rearranging 
the right-hand-side of (\ref{eq:xi1}) and, for notational convenience, by writing $T^+\equiv T^+(\vec{N})$
and $T^-\equiv T^-(\vec{N},K)$,  we obtain
\begin{eqnarray}
 \label{eq:xi}
 \frac{d}{dt}\langle x_i \rangle&=&
 \left\langle \left( \frac{N_{i}+1}{N+1} -\frac{N_{i}}{N} \right)T_{i}^{+}\right\rangle +
  \left\langle \left( \frac{N_{i}-1}{N-1} -\frac{N_{i}}{N} \right)T_{i}^{-}\right\rangle +
   \sum_{j\in \{1,2,3\}\neq i}\left\{\left\langle \left( \frac{N_{i}}{N+1} -\frac{N_{i}}{N} \right)T_{j}^{+}
   \right\rangle +
  \left\langle \left( \frac{N_{i}}{N-1} -\frac{N_{i}}{N} \right)T_{j}^{-}\right\rangle\right\}\nonumber\\
  &=& \left\langle \frac{T_{i}^{+}-T_{i}^{-}}{N} \left(1+{\cal O}\left(\frac{1}{N}\right)\right)\right\rangle- 
  \left\langle \frac{x_i}{N}\left(1+{\cal O}\left(\frac{1}{N}\right)\right)~
  \sum_{j=1}^{3}\left(T_{j}^{+}-T_{j}^{-}\right)\right\rangle.
\end{eqnarray}
We can now derive the mean-field equations (constant $K$)
and the stochastic differential equation (SDE) defining the piecewise-deterministic Markov process (PDMP)
for the evolution of the population size. For this, as usual, 
we  ignore all demographic fluctuations 
and factorize all terms appearing on the right-hand-side
of (\ref{eq:Ni}) and (\ref{eq:xi}) in terms of $\langle x_i \rangle$ and $\langle N \rangle$,
respectively denoted by $x_i$ and $N$, {\it e.g.} $\langle x_i x_j \rangle \to  x_i x_j, \langle f_i x_j N \rangle 
\to  f_i x_j N$
and $\langle N_i N \rangle \to N_i N$. In the case of a constant carrying capacity, making the natural mean-field
assumption that  $N$ is always sufficiently large for contributions of order  ${\cal O}(x_i/N)$ to be negligible, using 
(\ref{eq:A1_N_1}),
we obtain:
\begin{eqnarray}
 \label{eq:BDCLV_MF1}
 \frac{d}{dt}N
&=&\sum_{i=1}^{3} \left(T_i^+ -T_i^-\right)=N\left(1-\frac{N}{K}\right),\nonumber\\
\frac{d}{dt}x_i
&=&\frac{T_i^+ -T_i^-}{N}-x_i\left(\frac{dN/dt}{N}\right)=s\Pi_ix_i=x_i\left(\alpha_i x_{i+1}-\alpha_{i-1}x_{i-1}\right),
\end{eqnarray}
where we have used $\bar{f}=1$ and $\alpha_i\equiv s r_i$.
These mean-field equations coincide with the decoupled 
REs (\ref{eq:Ndot}) and (\ref{eq:xidot}) 
% (8) and (9)
discussed in the main text.
 In the case of a randomly switching carrying capacity, 
the $x_i$'s still obey (\ref{eq:BDCLV_MF1}) while the population size evolves according to
\begin{eqnarray}
 \label{eq:N_MF2}
 \frac{d}{dt}N
=
 \begin{cases}
N\left(1-\frac{N}{K_+}\right) & \text{if $\xi=+1$}\\ 
   N\left(1-\frac{N}{K_-}\right) & \text{if $\xi=-1$,}
\end{cases}
\end{eqnarray}
which  can be rewritten as the SDE (\ref{eq:PDMP})
%(20)
defining the PDMP
governing the evolution of the population size $N(t)$ when demographic noise is ignored and whose stationary marginal probability density is given
(\ref{eq:pnustar}). 
%(21).

Similar derivations also hold in the 
general (non-zero-sum) rock-paper-scissors game, whose birth-death formulation 
is given by the rates $T_i^{\pm}$ of Eq.~(\ref{eq:transrates})
%  Eq.~(34)
and leads to the mean-field equations of Sec.~5.

\vspace{2mm}

A comment on our choice of the transition rates and of the model formulation is here in order: 
With (\ref{eq:birtheq}) and (\ref{eq:transrates}) 
% (5) and (6) 
we have arguably chosen the simplest formulation of the
RPS dynamics subject to a carrying capacity. It is however worth noting that other 
choices are of course also possible. Another natural possibility would be to use 
the transition rates
$T_i^+=f_i N_i/\bar{f}$ and $T_i^-=(N/K)N_i$~\cite{Moran62}.  Clearly, for the BDCLV
these transition rates coincide with (\ref{eq:transrates}) 
%  (6) 
since $\bar{f}=1$ when $\epsilon=0$. 
A difference  however arises when $\epsilon\neq 0$ and $\bar{f}=1-\epsilon\sum_{i=1}^{3}\alpha_i x_i x_{i+1}$. In fact, proceeding as above and using 
the rates $T_i^+=f_i N_i/\bar{f}$ and $T_i^-=(N/K)N_i$ in the master equation (\ref{eq:ME}),
%  (7),  
we obtain the following mean-field rate equations (MFREs):
 $\dot{N}=N(1-N/K)$ and $\dot{x}_i=x_i(f_i-\bar{f})/\bar{f}
=x_i [\alpha_i x_{i+1}-(1+\epsilon)\alpha_{i-1}x_{i-1}+1-\bar{f}]/\bar{f}$.
While these equations are decoupled, the  MFREs for 
the $x_i$'s do {\it not} coincide with the celebrated replicator equations (\ref{eq:xidot_eps}) 
% (34) 
of the general RPS game~\cite{Hofbauer,Broom}: The $x_i$'s MFREs obtained with the above
alternative transition rates~\cite{Maynard}, 
differ from  (\ref{eq:xidot_eps}) 
% (34) 
due to the nonlinear $\bar{f}$ term  appearing in the denominator on their right-hand-side.
The MFREs $\dot{x}_i=x_i(f_i-\bar{f})/\bar{f}$ and Eqs.~(\ref{eq:xidot_eps})
% (34) 
however coincide to leading order in $s\epsilon$.

\subsection{The Moran cyclic Lotka-Volterra model (MCLV)}
\label{AppendixA2} 
We now outline the main features of the Moran cyclic Lotka Volterra model (MCLV) in a static environment (no environmental noise).
The MCLV is defined by six pairwise reactions 
and is characterized by the {\it conservation} of the population size $N$~\cite{Moran62,Ewens,Blythe07,Antal,Nowak}.
Each of the six reactions corresponds to the {\it simultaneous} death of an individual of species $i$ 
and  the  birth of an individual of 
species $j\neq i\in \{1,2,3\}$~\cite{Moran62}. This occurs with a rate $T_{i\to j}$. If the
state of the system consisting of $N_{1}$ individuals of type $1$,
$N_2$ of species $2$, and $N_3=N-N_1-N_2$ of the third type is denoted by $[N_{1}, N_2]$,
the six reactions of the MCLV are~\cite{Claussen08,Mobilia10,Galla11}
\begin{eqnarray*}
[N_{1}, N_2]  \xrightarrow{T_{1\to 2}}  [N_{1} - 1,  N_{2}+  1]; \; [N_{1}, N_2]  \xrightarrow{T_{2\to 1}}  [N_{1} + 1,  N_{2}-  1]
\end{eqnarray*}
\begin{eqnarray*}
[N_{1}, N_2]  \xrightarrow{T_{1\to 3}}  [N_{1} - 1,  N_{2}]; \; [N_{1}, N_2]  \xrightarrow{T_{3\to 1}}  [N_{1} + 1,  N_{2}]
\end{eqnarray*}
\begin{eqnarray*}
[N_{1}, N_2]  \xrightarrow{T_{2\to 3}}  [N_{1},  N_{2}-1]; \; [N_{1}, N_2]  \xrightarrow{T_{3\to 2}}  [N_{1},  N_{2}+1],
\end{eqnarray*}
with the transition rates~\cite{Claussen08,Mobilia10}
\begin{eqnarray}
\label{eq:A1_MCLV_rates}
T_{j\to i}= f_i x_ix_j~N =(1+s\Pi_i)~x_ix_j~N=(1+\{\alpha_i x_{i+1} -\alpha_{i-1} x_{i-1}\})~x_ix_j~N, 
\end{eqnarray}
where $f_i$ and $\Pi_i$ are given by (\ref{eq:fi}) and (\ref{eq:Pi}). 
% (2) and (1).
Interestingly, the transition rates of the MCLV can be expressed in terms of those of the 
BDCLV for a population of constant size $N=K$. In fact, using (\ref{eq:A1_BDCLV_rates}) and  
$N=K$, we have  $T_{j\to i}=  T_i^+ T_j^-/K$. This means that the BDCLV coincides with
the MCLV in a population of constant size $N=K$, see  below. 
Proceeding as above, we can readily find the mean-field rate equations for the MCLV:
\begin{eqnarray*}
\label{eq:MF_MCLV}
\frac{d}{dt}x_i=\frac{1}{N}\sum_{j=1; j\neq i}^{3}\left(T_{j\to i}-T_{i \to j}\right)=s\Pi_ix_i=x_i\left(\alpha_i x_{i+1}-\alpha_{i-1}x_{i-1}\right),
 \end{eqnarray*}
which coincide with the mean-field (replicator) equations for the population composition in the BDCLV, see (\ref{eq:BDCLV_MF1}) 
and (\ref{eq:xidot}). 
% (9).
Clearly therefore, in the constant-$K$ BDCLV the dynamics of the population composition coincides
with that of the MCLV in the mean-field limit $K\to \infty$: both are characterized by the same 
neutrally stable fixed point $\vec{x}^*=(r_2,r_3,r_1)=(r_2,1-r_1-r_2,r_1)$ and constant of motion ${\cal R}=x_1^{r_2}x_2^{r_3}x_3^{r_1}$. 

Since in the  constant-$K$ BDCLV dynamics the population size
 obeys a logistic equation,
after a short transient $N(t)\approx K$, see Eq.~(\ref{eq:Ndot}) and Fig.~\ref{fig:Fig1}.
This establishes a useful relationship between the BDCLV and MCLV: 
Except for a short transient  (on  a timescale $t\sim {\cal O}(1)$), corresponding to the so-called 
exponential phase of the logistic equation,
the evolution of the constant-$K$ BDCLV is similar to the dynamics of the MCLV in 
a population of constant size $N=K$. The BDCLV and MCLV relation 
 is particularly useful to determine the absorption/fixation properties
of the former 
in terms of the well-studied fixation properties of latter, see Secs.~3.1.2 and \ref{AppendixC}.
%%%%%%%%%%
 In Fig.~\ref{fig:FigA1}, we show that the survival and absorption probabilities 
$\phi_{i,j}$ and $\phi_{i}$ in the constant-$K$  BDCLV 
are almost indistinguishable from those obtained in the MCLV (with $N=K$).
Since the overall fixation probabilities $\widetilde{\phi}_i=\phi_{i,i+1}\phi_{i}+\phi_{i-1,i}(1-\phi_{i})$, see 
Eq.~(\ref{eq:PhiTot}), we can consider that 
the absorption and total fixation probabilities in the constant-$K$ BDCLV  and  those of the MCLV with $N=K\gg 1$ coincide.
Similarly, the mean extinction and absorption times $T_1$ and $T_2$
in the  BDCLV with constant-$K$  and MCLV with $N=K\gg 1$ are indistinguishable, see the insets of Fig.~\ref{fig:FigA1} and 
below.

To study the absorption/fixation properties of the BDCLV and MCLV, it 
is useful to write down the two-dimensional forward Fokker-Planck equation (FPE) obeyed by the probability density
$P_{{\rm MCLV}}\equiv P_{{\rm MCLV}}(\vec{x},t)$ of the latter. Using  standard methods, see, {\it e.g.} 
Refs.~\cite{Gardiner,VanKampen,RMF06,Claussen08,Mobilia10}
we have the forward FPE 
\begin{eqnarray}
\hspace{-5mm}
\label{eq:FPE1_MCLV}
\left[\partial_t -{\cal G}_{{\rm fMCLV}}(\vec{x})\right]~P_{{\rm MCLV}}(\vec{x},t)=0, \quad 
\text{where} \quad{\cal G}_{{\rm fMCLV}}(\vec{x})\equiv -\sum_{i=1}^{2}\partial_i A_i^{{\rm MCLV}}(\vec{x}) +
\frac{1}{2}\sum_{i,j=1}^{2}\partial_i\partial_j
B_{ij}^{{\rm MCLV}}(\vec{x}), \; 
 \end{eqnarray}
is the forward FPE generator, with
$\partial_i\equiv \partial/\partial x_i$~\footnote{In Eq.~(\ref{eq:FPE1_MCLV}), the indices $i,j\in\{1,2\}$
since $x_3=1-x_1-x_2$ and, as usual in the diffusion theory, we have rescaled the time $t\to t/N$.}, defined by
\begin{eqnarray}
\label{eq:FPE2_MCLV}
\hspace{-5mm}
&&A_i^{{\rm MCLV}}(\vec{x})\equiv \sum_{{j=1}, j\neq i}^{3} \left(T_{j\to i}-T_{i\to j}\right); \nonumber\\
&&
B_{ii}^{{\rm MCLV}}(\vec{x})\equiv \sum_{{j=1}, j\neq i}^{3} \left(\frac{T_{j\to i}+T_{i\to j}}{N}\right) \quad
\text{and} \quad B_{12}^{{\rm MCLV}}(\vec{x})=B_{21}^{{\rm MCLV}}(\vec{x})\equiv -\left(\frac{T_{1\to 2}+T_{2\to 1}}{N}\right).
%T_{2\to 1}-T_{1\to 2}+ T_{3\to 1}-T_{1\to 3}
 \end{eqnarray}
Within the linear noise approximation~\cite{Gardiner,VanKampen}, upon linearising $A_i^{{\rm MCLV}}$
about the coexistence fixed point $\vec{x}^*$ and by evaluating $B_{ij}^{{\rm MCLV}}(\vec{x})$
at  $\vec{x}^*$, in the variables 
$\vec{y}={\bf S}\vec{x}=\frac{\sqrt{3}}{2}\begin{pmatrix}
\frac{(r_1+ r_2)\omega_0^{{\rm MCLV}}}{r_1 r_2} &\frac{\omega_0^{{\rm MCLV}}}{r_1}   \\                      
  0 & 1                      \end{pmatrix}~\vec{x}$, the forward FPE reads~\cite{RMF06,Mobilia10}
\begin{eqnarray}
\label{eq:FPE3_MCLV}
\partial_t~P_{{\rm MCLV}}(\vec{y},t)=-\omega_0^{{\rm MCLV}}\left[y_1 \partial_{y_1}-y_2 \partial_{y_2}\right]~P_{{\rm MCLV}}(\vec{y},t)
+D^{{\rm MCLV}}~[\partial_{y_1}^2+\partial_{y_2}^2]~P_{{\rm MCLV}}(\vec{y},t),
 \end{eqnarray}
where $\omega_0^{{\rm MCLV}}=s\sqrt{r_1r_2 (1-r_1-r_2)}$ and $D^{{\rm MCLV}}=3[r_1+r_2-4r_1r_2-(r_1-r_2)^2]/(4N)$. To study the 
fixation properties of the MCLV, the FPEs (\ref{eq:FPE2_MCLV}) and (\ref{eq:FPE3_MCLV}) have to be supplemented with absorbing 
boundaries at the corners  of $S_3$~\cite{RMF06,Berr09,West18}.

\begin{figure}[!htb]
	\centering
	\includegraphics[width=0.38\linewidth]{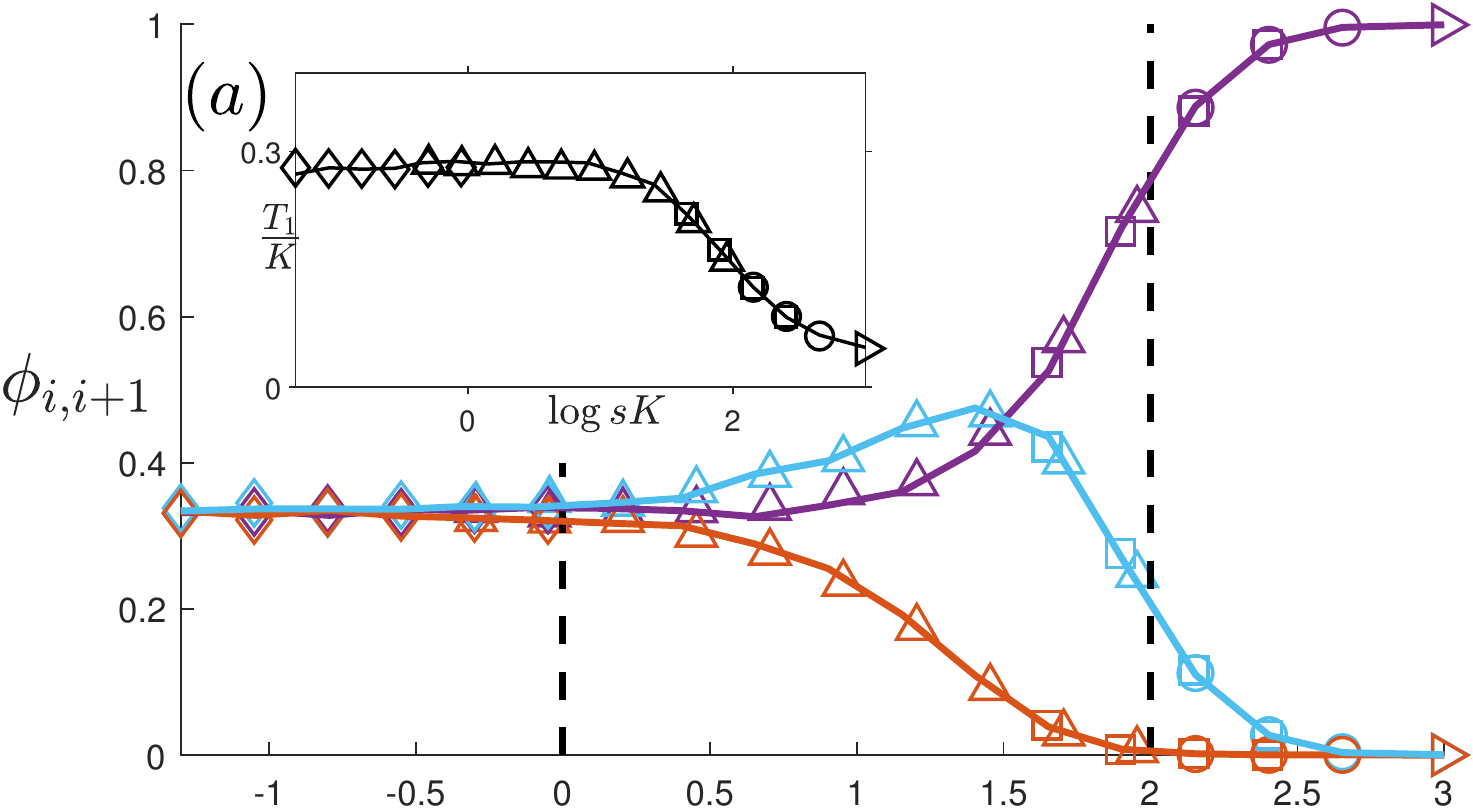}
	\includegraphics[width=0.38\linewidth]{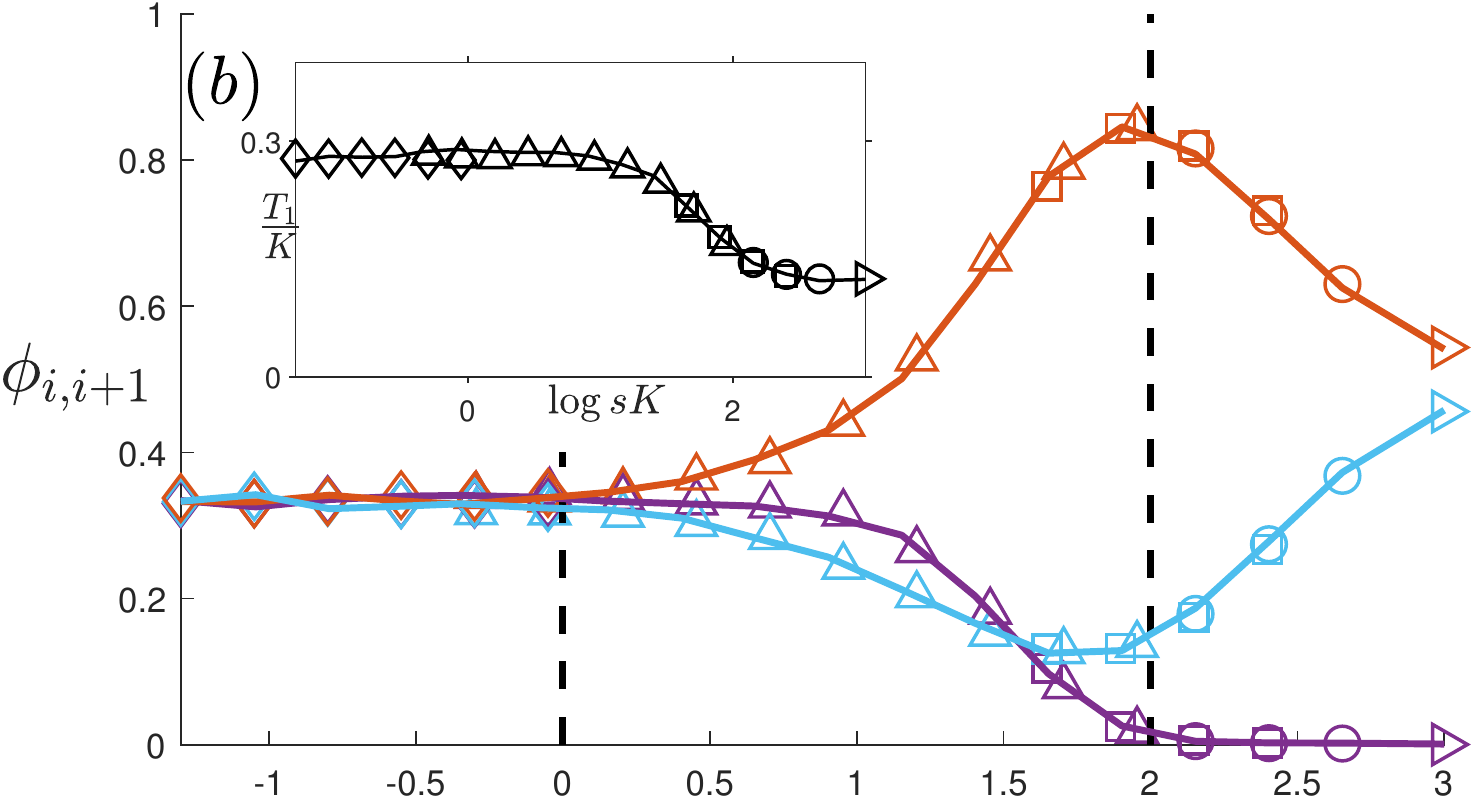}	
	\includegraphics[width=0.38\linewidth]{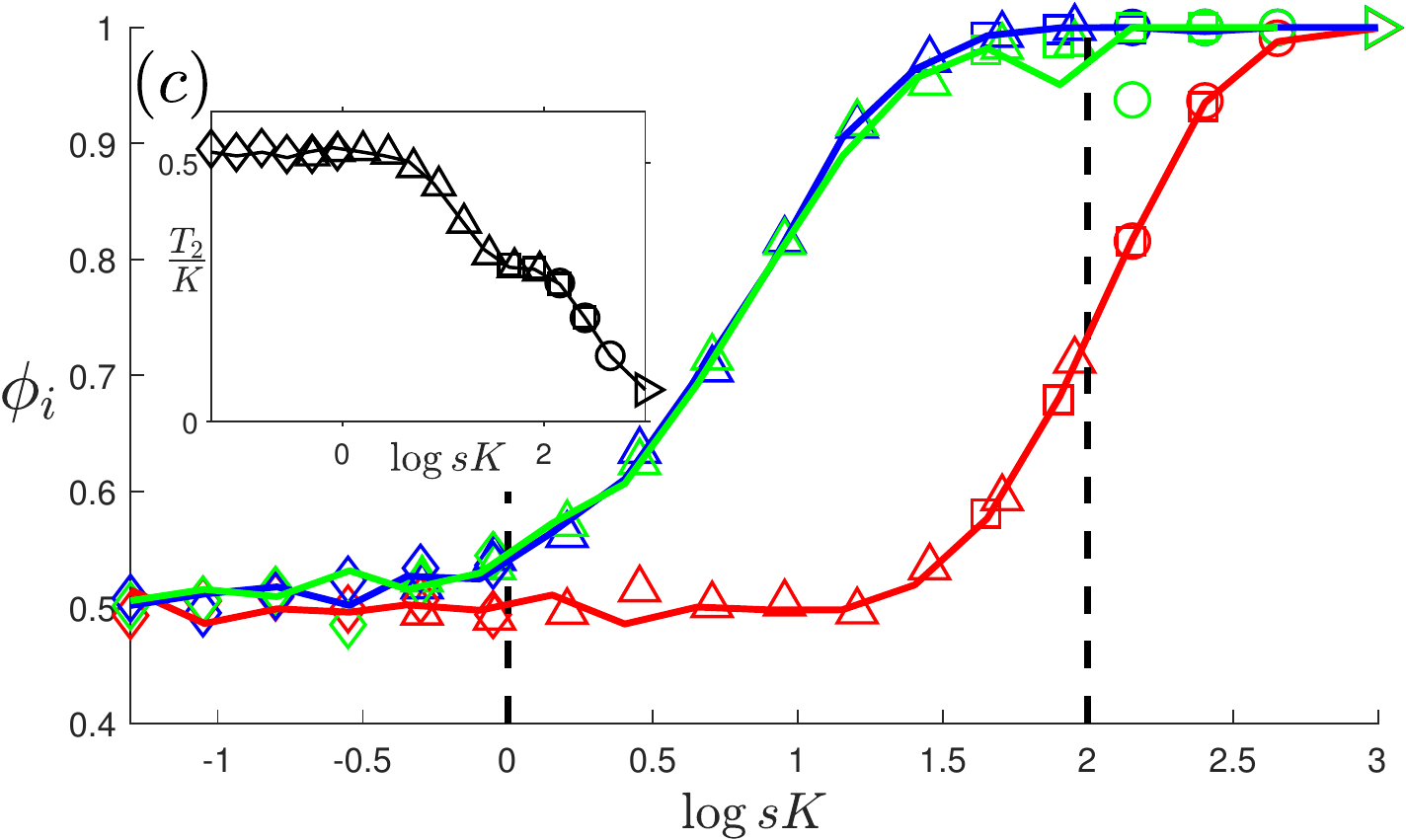}
	\includegraphics[width=0.38\linewidth]{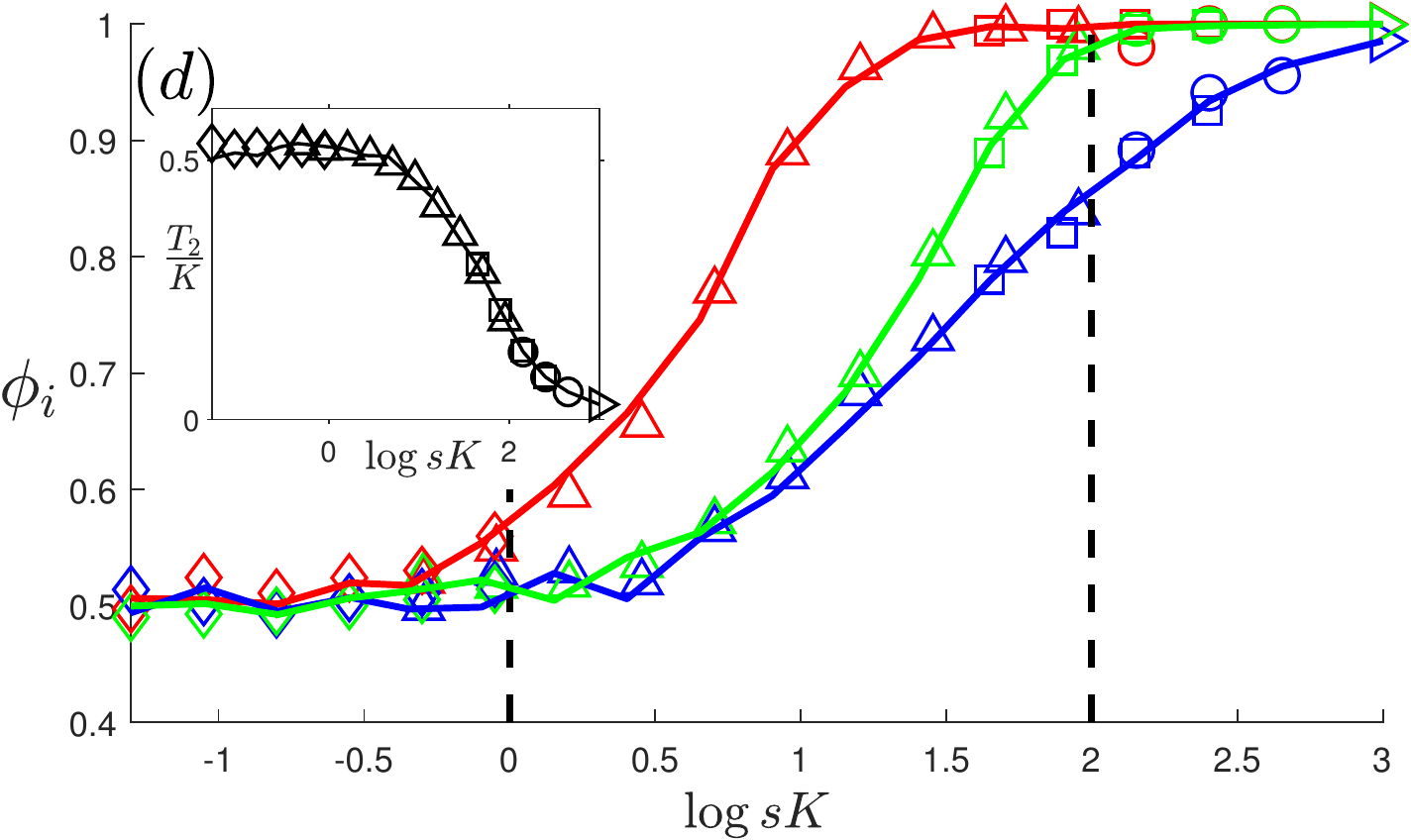}	
	\caption{Comparison of the fixation properties vs. $sK$ in the BDCLV (solid lines)
	with constant carrying capacity $K$
	and in the MCLV (symbols) with a constant population size $N=K\in \{1000~(\rhd), 450~(\circ), 250~(\diamond), 
	90~(\square), 50~(\triangle)\}$, with
	$\vec{r}=\vec{r}^{(1)}$ in (a,c) and $\vec{r}=\vec{r}^{(2)}$ in (b,d) and different values of selection intensity:
	$s\in \{10^{-j/4}, j\in J_{K}^{\rm MCLV}\}$ with $J_{1000}^{\rm MCLV}=\{0\}$,
	$J_{450}^{\rm MCLV}=\{0,\dots, 3\}, 
	J_{250}^{\rm MCLV}=
	\{0, \dots, 4\}, J_{90}^{\rm MCLV}=
	\{0, \dots, 10\}, J_{50}^{\rm MCLV}=\{7, \dots,12\}$  for the MCLV and 
	$s\in \{10^{-j/4}, j\in J_{K}^{\rm BDCLV}\}$ with $J_{1000}^{\rm BDCLV}=\{1\}$, $J_{450}^{\rm BDCLV}=
	\{0, \dots ,12\}, J_{90}^{\rm BDCLV}=
	\{10, 11, 12 \}, J_{50}^{\rm BDCLV}=\{12\}$  for the BDCLV.
	(a,b) Stage 1 survival probabilities  $\phi_{1,2}$~(purple), $\phi_{2,3}$~(light blue) and $\phi_{3,1}$~(orange)
	vs. $sK$: BCLV results (lines) match perfectly with those obtained for the MCLV (symbols).
	Insets: Rescaled mean extinction times $T_1/K$ vs. $sK$ for the  BDCLV (solid lines) and MCLV (symbols) virtually coincide, see text.
	(c,d) Stage 2 conditional fixation probabilities $\phi_{1}$~(red), $\phi_{2}$~(blue) and $\phi_{3}$~(green)
	vs. $sK$: BCLV results (lines) agree perfectly with those obtained for the MCLV (symbols).
	Insets: Rescaled mean absorption times $T_2/K$ vs. $sK$ for the  BDCLV (solid lines) and MCLV (symbols) almost coincide, see text.
	In all panels:
	$\vec{x}_0=\vec{x}_c$, $\epsilon=0$;  regimes (i)-(iii), from left to right, are indicatively separated by 
	dashed lines.
	Simulation results for the fixation probabilities of in the constant-$K$ BDCLV and MCLV with $N=K$ 
	are almost indistinguishable, see text.
	}
	\label{fig:FigA1}
\end{figure}

\subsection{The chemical cyclic Lotka-Volterra model (cCLV)}
\label{AppendixA3} 
The chemical cyclic Lotka Volterra model (cCLV) is defined by three  pairwise (``bimolecular'')
reactions involving the simultaneous death and birth of individuals of different species, therefore conserving the total population 
size $N$. Hence, in the cCLV, in contrast to the BDCLV and MCLV, species $i$ is the predator of species $i+1$ and the prey of
species $i-1$: an  $i$-individual kills and replaces an  $(i+1)$-individual with one of its offspring, while it is killed and replaced by 
individual of type $i-1$ according to the following ``bimolecular chemical reactions'', with $N_3=N-N_1-N_2$:
\begin{eqnarray}
\label{eq:A1_cCLV}
&&12 \xrightarrow{W_{2\to 1}} 11 \quad \text{(from state $[N_{1}, N_2]$ to state $[N_{1} + 1,  N_{2}-  1]$)} \nonumber\\
&&23 \xrightarrow{W_{3\to 2}} 22 \quad \text{(from state $[N_{1}, N_2]$ to state $[N_{1},  N_{2}+ 1]$)} \nonumber\\
&&31 \xrightarrow{W_{1\to 3}} 33 \quad \text{(from state $[N_{1}, N_2]$ to state $[N_{1} - 1,  N_{2}]$).} \nonumber\\
 \end{eqnarray}
These reactions occur with the transition rates~\cite{RMF06,Berr09,West18} 
\begin{eqnarray}
\label{eq:A1_cCLV_rates}
W_{i+1\to i}= k_i~\frac{N_i N_{i+1}}{N}=k_i x_i x_{i+1}~N, \quad \text{where $k_i\geq 0$.}
\end{eqnarray}
Clearly, the reactions (\ref{eq:A1_cCLV}) and transition rates (\ref{eq:A1_cCLV_rates}) differ from those of the BDCLV and MCLV.
Yet, as discussed below many of the features of the BDCLV, MCLV and cCLV are similar. The cCLV mean-field equations for the $x_i$'s are given by
\begin{eqnarray}
\hspace{-5mm}
\label{eq:MF_cCLV}
\frac{d}{dt}x_i=\frac{W_{i+1 \to i}-W_{i \to i-1}}{N}=x_i\left(k_i x_{i+1}-k_{i-1}x_{i-1}\right).
 \end{eqnarray}
We notice that upon rescaling the time as $t\to st/(k_1+k_2+k_3)$, the reaction rates become $k_i \to k_i/(k_1+k_2+k_3)=r_i$ and 
Eq.~(\ref{eq:MF_cCLV}) is identical to Eq.~(\ref{eq:BDCLV_MF1}).
Hence, upon time rescaling, the MCLV and cCLV are identical at mean-field level and their dynamics
coincide with the REs (\ref{eq:xidot}) 
%(9)
of the
BDCLV. Moreover, Eqs.~(\ref{eq:BDCLV_MF1}) and (\ref{eq:MF_MCLV})
admit the same marginally stable coexistence fixed point $\vec{x}^*=(k_2,k_3,k_1)/(k_1+k_2+k_3)=(r_2,r_3,r_1)$ and the 
same constant of motion ${\cal R}=\left(x_1^{k_2}x_2^{k_3}x_3^{k_1}\right)^{1/(k_1+k_2+k_3)}$.
The mean-field dynamics of the $x_i$'s is therefore identical for the BDCLV, MCLV and cCLV.

It is useful to proceed as above and consider the two-dimensional forward Fokker-Planck equation (FPE) obeyed by 
the cCLV probability density
$P_{{\rm cCLV}}\equiv P_{{\rm cCLV}}(\vec{x},t)$ (with $t\to t/N$):
\begin{eqnarray}
\hspace{-5mm}
\label{eq:FPE1_cCLV}
\left[\partial_t -{\cal G}_{{\rm cCLV}}(\vec{x})\right]~P_{{\rm cCLV}}(\vec{x},t)=0, \quad 
\text{where} \quad{\cal G}_{{\rm cCLV}}(\vec{x})\equiv -\sum_{i=1}^{2}\partial_i A_i^{{\rm cCLV}}(\vec{x}) +
\frac{1}{2}\sum_{i,j=1}^{2}\partial_i\partial_j
B_{ij}^{{\rm cCLV}}(\vec{x}), \; 
 \end{eqnarray}
with $A_i^{{\rm cCLV}}(\vec{x})\equiv W_{i+1\to i}-W_{i\to i-1}$, 
$B_{ii}^{{\rm cCLV}}(\vec{x})\equiv \left(W_{i+1\to i}+W_{i\to i-1}\right)/N$ where $i\in \{1,2\}$,
and $B_{12}^{{\rm cCLV}}(\vec{x})=B_{21}^{{\rm cCLV}}(\vec{x})\equiv -(W_{1\to 2}+W_{2\to 1})/N$. It is worth 
noting that the drift terms of the cCLV and MCLV are simply related by $A_i^{{\rm cCLV}}=sA_i^{{\rm MCLV}}/(k_1+k_2+k_3)$.
In the case of symmetric rates,
 $k_1=k_2=k_3=1$, within the linear noise approximation,   this forward FPE in the variables 
$\vec{y}={\bf S}\vec{x}$ reads:
\begin{eqnarray}
\label{eq:FPE2_cCLV}
\partial_t~P_{{\rm cCLV}}(\vec{y},t)=-\omega_0^{{\rm cCLV}}\left[y_1 \partial_{y_1}-y_2 \partial_{y_2}\right]~P_{{\rm cCLV}}(\vec{y},t)
+D^{{\rm cCLV}}~[\partial_{y_1}^2+\partial_{y_2}^2]~P_{{\rm cCLV}}(\vec{y},t),
 \end{eqnarray}
where $\omega_0^{{\rm cCLV}}=1/\sqrt{3}$ and $D^{{\rm cCLV}}=1/(12N)$~\cite{RMF06}.
This FPE is similar to Eq.~(\ref{eq:FPE2_MCLV}). The comparison with 
the MCLV with equal rates $r_i=1/3$ is particularly illuminating: 
 $\omega_0^{{\rm MCLV}}=s\omega_0^{{\rm cCLV}}/3$
and $D^{{\rm MCLV}}=2D^{{\rm cCLV}}$. Hence, upon a suitable rescaling of the timescale, the MCLV and cCLV 
deterministic drift and diffusive terms (about $\vec{x}^*$)  
can be mapped onto each other.
\subsubsection{Fixation probabilities in the cCLV: The law of the weakest and the law of stay out}
\label{AppendixA31} 
\begin{figure}
	\centering
	\includegraphics[width=0.25\linewidth]{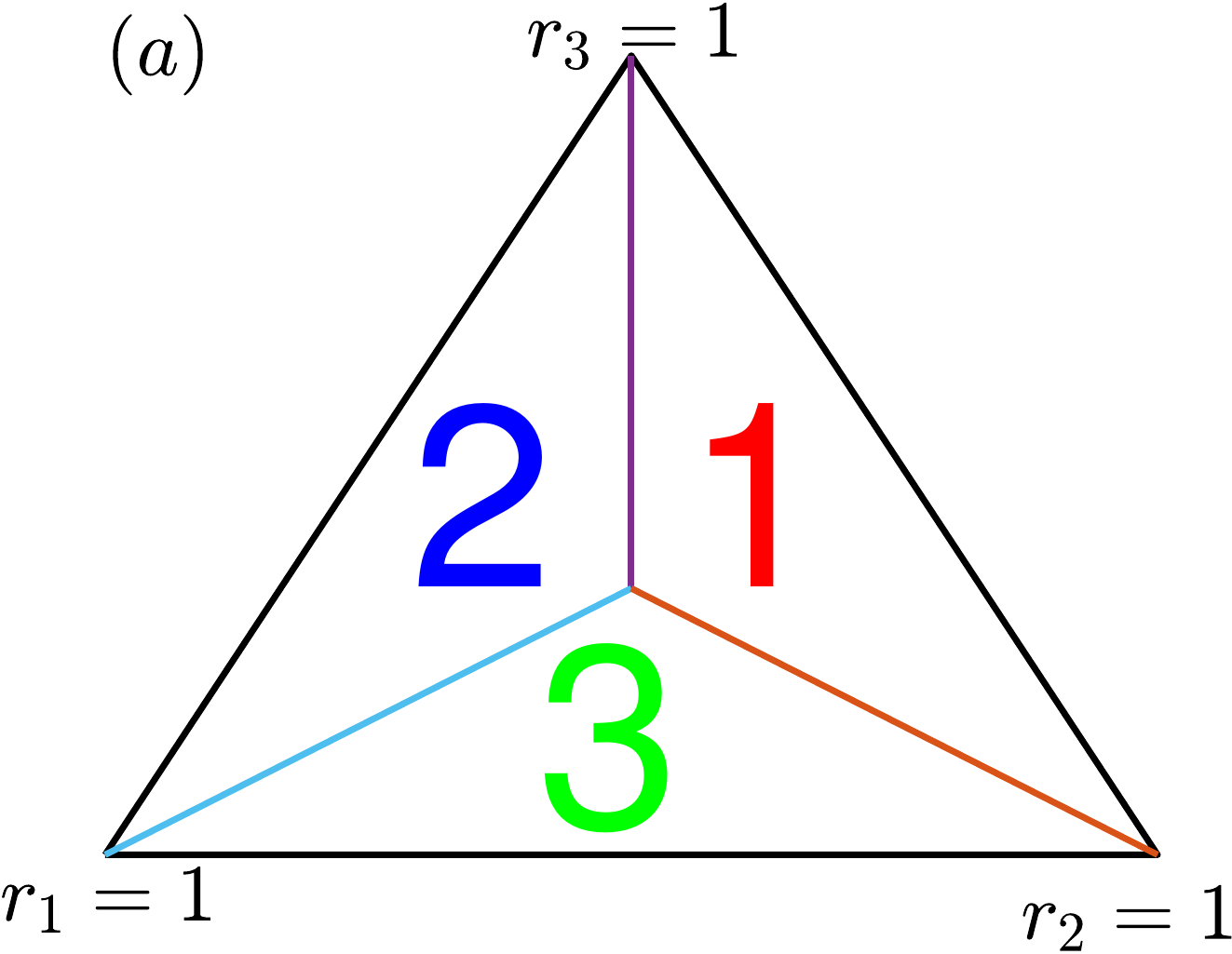}	
	\includegraphics[width=0.25\linewidth]{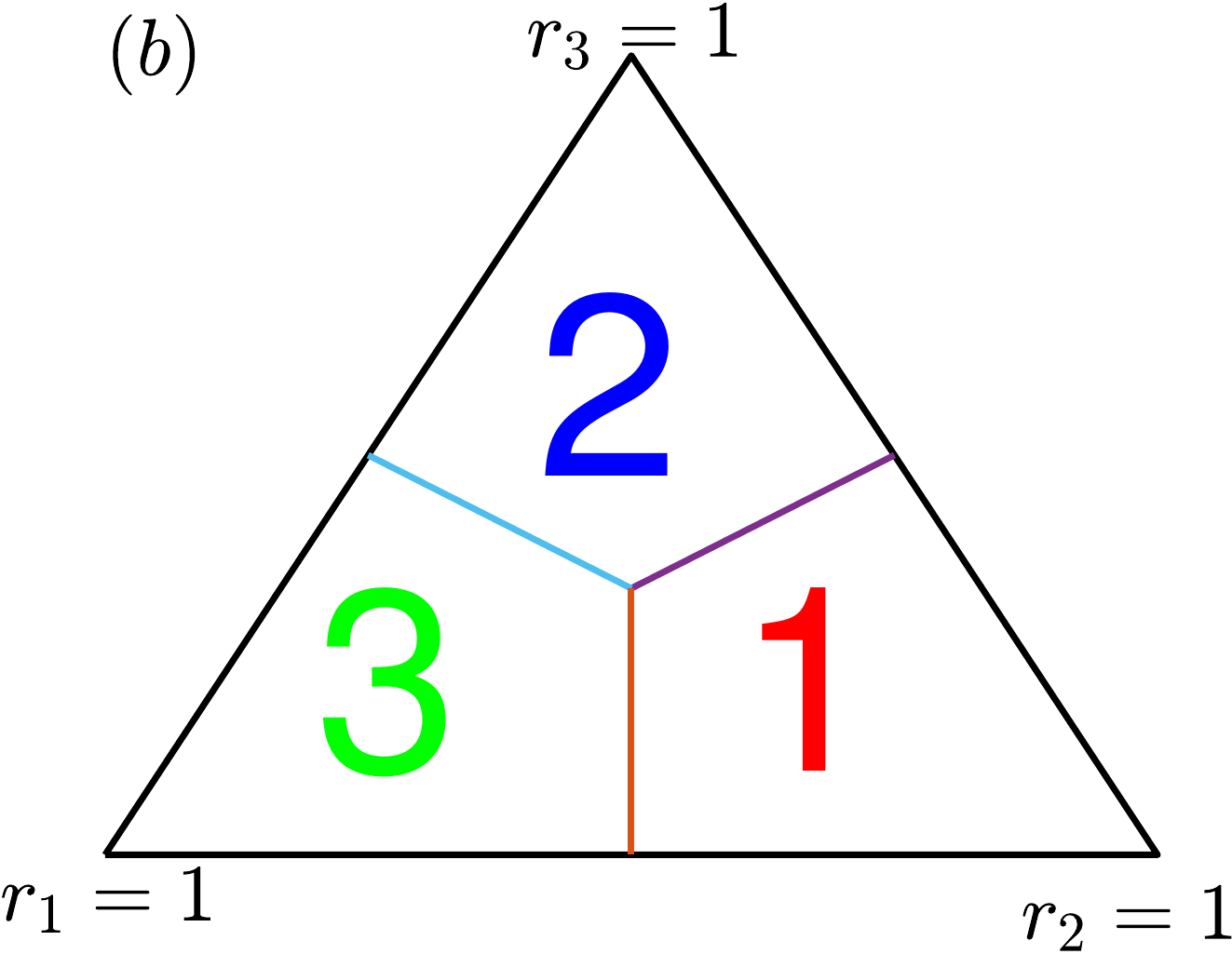}
	\caption{
	Law of the weakest (a) and law of stay out (b) in the simplex $S_3$ spanned 
	by $r_i$, divided into three regions where the most likely species to survive is labelled. 
	On the lines separating these regions, both adjacent species are equally likely to survive. 
	(a) Law of the weakest (LOW): In the cCLV, the most likely species to survive in a large population is that 
	with the lowest $r_i$. The LOW becomes asymptotically a zero-one law and
	also applies to the constant-$K$ BDCLV and MCLV when $N=K$
	and $sK \gg 1$ (regime (iii)), see text.
	(b) Law of stay out (LOSO) when all species initially coexist with the same density: 
	In the cCLV, no  species is guaranteed to survive
	is small populations, see text. The LOSO 
	also applies to the constant-$K$ BDCLV and MCLV when $N=K$
	and $sK ={\cal O}(10)$ (regime (ii)), see text.
	}
	\label{fig:FigA2}
\end{figure}
Due to the predator-prey interactions underpinning the cCLV,
its  fixation properties of the cCLV are entirely set by the stage 1 of its dynamics:
the probability $\phi_{i,i+1}^{{\rm cCLV}}$ that species $i$ and $i+1$ survive the 
 stage 1 coincides with the fixation probability $\phi_{i}^{{\rm cCLV}}$ of species $i$: 
 $\phi_{i,i+1}^{{\rm cCLV}}=\phi_{i}^{{\rm cCLV}}$.
The survival/fixation probability $\phi_{i}^{{\rm cCLV}}$ of the cCLV can be explained by two simple laws called the 
{\it law of the weakest} (LOW) and the {\it law of stay out} (LOSO)~\cite{Berr09,Frean01,Tainaka93,West18}, see Fig.\ref{fig:FigA2}.
The former applies to populations of large size 
and the latter to small populations.
The LOW says that in a sufficiently large population (when $N\gtrsim 10^2$)  evolving according to the cCLV,
the most likely species to survive is the one with the lowest rate $k_i$~\cite{Berr09} ($i\in \{1,2,3\}$), see Fig.\ref{fig:FigA2}~(a):
\begin{eqnarray}
\hspace{-4mm}
\label{eq:A1_LOW}
\phi_i^{{\rm cCLV}}> \phi_{i+ 1}^{{\rm cCLV}}, \phi_{i-1}^{{\rm cCLV}} \; \text{if \quad $k_i< k_{i\pm 1}, \;\;\;$ and \;\;\;
$\phi_{i}^{{\rm cCLV}}\approx \phi_{i+1}^{{\rm cCLV}}>\phi_{i-1}^{{\rm cCLV}} \;$  if $\quad k_i=k_{i+1}<k_{i-1},\; $ 
with $\; i\in \{1,2,3\}$.
%and 
%$\phi_{i-1}>\phi_{i},\phi_{i+1}$ if $k_i=k_{i+1}>k_{i-1}$
}
 \end{eqnarray}
The LOW becomes asymptotically a zero-one law (when $N\gtrsim 10^3$):
\begin{eqnarray}
\label{eq:A1_LOW2}
\phi_i^{{\rm cCLV}}\to 1, \phi_{i\pm 1} \to 0\quad \text{if $k_i< k_{i\pm 1}$, \quad while \quad $\phi_{i}^{{\rm cCLV}}= \phi_{i+1}^{{\rm cCLV}}\to 1/2$
and $\phi_{i-1}^{{\rm cCLV}}\to 0$ if $k_i=k_{i+1}<k_{i-1}$.}
 \end{eqnarray}
The LOW is independent of the initial condition and results from the fact that in large populations, due to the effect of weak demographic noise the cCLV trajectories
perform random walks between the  deterministic orbits until they reach the so-called ``outermost orbit''. This is obtained from 
the constant of motion ${\cal R}$ as
the deterministic orbit that lies at a distance $1/N$ from the closest edge of $S_3$~\cite{Berr09,West18}. 
In the cCLV, the extinction of a first species
occurs when a chance fluctuation pushes a trajectory along the edge of 
 $S_3$ from where the absorbing state  corresponding to the fixation of the ``weakest species'' (with lowest $k_i$)
 and death of its ``prey''
 is attained exponentially quickly.

The LOSO is a non-zero-one law prescribing which species is most likely to survive in small
populations ($3 \leq N\lesssim 50$). The LOSO
results from the interplay between the deterministic drift and demographic fluctuations
and its prescriptions depend on the initial condition. In the cCLV, when 
 initially all species have the same density, i.e.
$\vec{x}_0=\vec{x}_c$, the LOSO says that the most likely species to survive
is/are the one(s)  predating on the species with the highest $k_i$'s, see Fig.\ref{fig:FigA2}~(b)~\cite{Berr09,West18}\footnote{
In the cCLV, when the population size is $N=3$, we have $\phi_{i}^{{\rm cCLV}}=x_{i+1}^*$\cite{Berr09}.}:
\begin{eqnarray}
\label{eq:A1_LOSO}
\phi_{i-1}^{{\rm cCLV}}> \phi_{i}^{{\rm cCLV}},\phi_{i+1}^{{\rm cCLV}} \quad
\text{if $\quad k_i>k_{i+1},k_{i-1}$, \quad and \quad} \phi_i^{{\rm cCLV}} \approx \phi_{i+1}^{{\rm cCLV}}>\phi_{i-1}^{{\rm cCLV}} \quad 
\text{if $\quad k_{i+1}=k_{i-1}>k_i$}.
 \end{eqnarray}
The LOSO can be understood by estimating the 
 initial drift at $\vec{x}_0$ with the  Jacobian 
 $\bf{J}^*$ of (\ref{eq:MF_cCLV})
evaluated at $\vec{x}^*$. When, as here, $\vec{x}_0 \neq \vec{x}^*$,
the rate of the bias from $\vec{x}_0$
towards a corner $i$ of $S_3$ is $({\bf J}^*\vec{x}_0)_i=k_i x_{i+1}(0)-k_{i-1} x_{i-1}(0)$.
Hence,   $k_i^{-}\equiv k_i x_{i+1}(0)-k_{i-1} x_{i-1}(0)$,
gives the initial {\it deterministic rate} in the direction $i$.
The most likely species to die out first is therefore the one with the smallest
$k_i^-$ (edge $(i-1, i+1)$ as the most likely to be hit first). 
With this reasoning, and $k_i^{-}=(k_i-k_{i-1})/3$ when $\vec{x}_0=\vec{x}_c$,
we find that the species that is the least likely to survive/fixate 
in the cCLV satisfies (\ref{eq:A1_LOSO})
when all species initially coexist with the same density
$x_i(0)=1/3$.
\subsubsection{Mean extinction and fixation times in the cCLV}
\label{AppendixA32} 
The cCLV  dynamics is also characterized by two stages: in Stage 1, the three species coexist
until an edge of $S_3$ is hit and one of the species dies out (see Sec.~3.1.1) after a mean extinction time 
$T_1^{{\rm cCLV}}={\cal O}(N)$, see Sec.~E.1.1. While the stage 1 dynamics of the cCLV, MCLV and constant-$K$ BDCLV 
 are similar  
(when $s={\cal O}(1)$), a  major difference arises in Stage 2, when 
two species, say $i$ and its weak opponent $i+1$,
compete along the edge $(i,i+1)$ of $S_3$. 
According to the cCLV, the interaction between species $i$ (predator) and $i+1$ (prey)
is of predator-prey type, and the outcome of Stage 2 is certain: Contrarily to the MCLV and BDCLV, 
 species $i$  always wins against $i+1$ exponentially quickly in time. The overall cCLV mean 
fixation time  $T_F^{{\rm cCLV}}= T_1^{{\rm cCLV}}+T_2^{{\rm cCLV}}$ therefore
coincides with $T_1^{{\rm cCLV}}$ to leading order, yielding
$T_F^{{\rm cCLV}}\simeq T_1^{{\rm cCLV}}={\cal O}(N)$, when 
$N\gg 1$~\cite{RMF06,He10,West18}. 

It has been shown that the mean extinction/fixation time $T_1^{{\rm cCLV}}$ can be obtained from
the linear approximation about $\vec{x}^*$~\cite{RMF06}~(see also \cite{Mobilia10,Dob12}). For this,
it is useful to consider the FPE (\ref{eq:FPE2_cCLV}) in polar coordinates, via $y_1=r\cos{\theta}$
and $y_2=r\sin{\theta}$. Since there is no angular dependence when $\vec{x}_0=\vec{x}^*$,
one has
 $P_{{\rm cCLV}}(\vec{y},t) \to P_{{\rm cCLV}}(r,t)$ with
\begin{eqnarray}
\label{eq:FPE3_cCLV}
\partial_t~P_{{\rm cCLV}}(r,t)=
D^{{\rm cCLV}}~[r^{-1}\partial_{r}+\partial_{r}^2]~P_{{\rm cCLV}}(r,\theta,t),
 \end{eqnarray}
which is the two-dimensional diffusion equation
 in polar coordinates with only radial dependence  and diffusion constant $D^{{\rm cCLV}}=1/(12N)$. 
 By supplementing this FPE  with an absorbing boundary at $\partial S_3$, 
 approximated as a circle of radius $R$ in order to
exploit the symmetry about $\vec{x}^*$, the mean extinction time was found to scale with $N$:
\begin{eqnarray}
\label{eq:T1_cCLV}
T_1^{{\rm cCLV}}\simeq 3R^2~N, \quad \text{where} \quad R=\frac{1}{2\sqrt{3}}\left(1+\frac{1}{\sqrt{3}}\right).
 \end{eqnarray}
Hence, in the cCLV with equal rates ($k_i=1$), the  mean fixation and extinction time when the dynamics starts at $\vec{x}^*$
is $T_F^{{\rm cCLV}}\simeq T_1^{{\rm cCLV}}\simeq 0.62 N$. Qualitatively, the same conclusion $T_F^{{\rm cCLV}}\simeq T_1^{{\rm cCLV}}= {\cal O}(N)$ 
also holds when the rates $k_i$ are unequal~\cite{RMF06}.

\section{Stage 1 dynamics in the constant-$K$ BDCLV and MCLV: similarities with the cCLV}
\label{AppendixB} 

We have seen that the cCLV survival/fixation probabilities are set in Stage 1 by the outermost orbit and 
follow the LOW in large populations. The MCLV and cCLV obey the same 
 mean-field equations (up to time rescaling), with the same constant of motion ${\cal R}$
and fixed points, see Eqs. (\ref{eq:MF_cCLV}) and (\ref{eq:MF_MCLV}), and as such they admit the same outermost orbits.
Furthermore, with the same timescale, the diffusion constant in the 
MCLV is $1/(Ns)$ and $1/N$ in the cCLV.
The survival probabilities $\phi_{i,i+1}^{{\rm MCLV}}$ of a population evolving with the MCLV
are  therefore expected to correspond to those of 
the cCLV in a population of effective size ${\cal O}(Ns)$, with rates related according to $r_i=k_i/(k_1+k_2+k_3)$.
We have also seen that  in the constant-$K$ BDCLV  the population size rapidly fluctuates
about $K$, i.e. $N(t) \simeq K$, see Eq.~(\ref{eq:xidot}) and Fig.~\ref{fig:Fig1}, 
% see Eq.~(9) and Fig.~1,
and its survival probabilities are the same as in the MCLV with $N=K\gg 1$, see Fig.~\ref{fig:FigA1}.
The survival probabilities  $\phi_{i,i+1}$ in the constant-$K$ BDCLV are therefore 
the same as those, $\phi_{i,i+1}^{{\rm cCLV}}\vert_{Ks}$,  in the  cCLV with a population of 
size ${\cal O}(Ks)$: $\phi_{i,i+1}\approx \phi_{i,i+1}^{{\rm MCLV}}\vert_K\approx\phi_{i,i+1}^{{\rm cCLV}}\vert_{Ks}
=\phi_{i}^{{\rm cCLV}}\vert_{Ks}$.
We therefore expect that  the survival probabilities of the  constant-$K$ BDCLV  
 obey the LOW when $Ks\gtrsim 100$, whereas they obey the LOSO when $Ks= {\cal O}(10)$, see Fig.~\ref{fig:FigA2}. 
This is confirmed by the results discussed in Sec.~3.1.1, see Fig.~\ref{fig:Fig3}~(a,b).  
% see Fig.~3~(a,b).
%
We have also seen that the mean extinction time in the cCLV  scales with $N$ to leading order 
and can be obtained within a linear noise approximation about $\vec{x}^*$. We can proceed similarly with the 
MCLV, and since the  linear noise approximation about $\vec{x}^*$ of the cCLV and MCLV  
is similar, see Eqs.~(\ref{eq:FPE2_cCLV}) and (\ref{eq:FPE2_MCLV}), we can obtain the mean extinction time $T_1^{\rm MCLV}$
by solving the radial diffusion equation $\partial_t~P_{{\rm MCLV}}(r,t)=
D^{{\rm MCLV}}~[r^{-1}\partial_{r}+\partial_{r}^2]~P_{{\rm MCLV}}(r,\theta,t)$, with absorbing boundary on $\partial S_3$
and $D^{{\rm MCLV}}=2D^{{\rm cCLV}}$. This  yields $T_1^{\rm MCLV}\simeq \frac{3}{2}R^2 N\approx 0.3N$ when 
$r_i=r=1/3$ (symmetric rates). A similar relation, with a different expression of $R$, holds when the rates $r_i$ are asymmetric. 
Since $N(t)\simeq K$  in the constant-$K$ BDCLV  
(after a time $t={\cal O}(1)$), we readily obtain  its
mean extinction time: $T_1\simeq \frac{3}{2}R^2 K\approx 0.3K$ to leading order in $K\gg 1$, when 
$r_i=1/3$. The insets of Fig.~\ref{fig:FigA1} confirm that $T_1$ in constant-$K$ BDCLV is almost indistinguishable from 
$T_1^{\rm MCLV}$ obtained in the MCLV with $N=K\gg 1$. 
This result also holds when the dynamics towards 
extinction is driven by diffusion
(weak demographic noise). This is certainly the case when  $\vec{x}_0=\vec{x}^*$ and also when  $\vec{x}_0\neq \vec{x}^*$ and 
$s\ll 1$. In fact, 
under weak selection,
the deterministic drift arising when
$\vec{x}_0 \neq \vec{x}^*$ is weak and extinction is driven by weak demographic fluctuations when $s\ll 1$.
we therefore find
$T_1\simeq \frac{3}{2}R^2 N\approx 0.3N$ when 
$r_i=r=1/3$  and when $s \ll 1$ and $sK={\cal O}(1)$, as reported in Fig.~\ref{fig:FigA4}(a).

\section{Stage 2 dynamics in a population with constant carrying capacity}
\label{AppendixC} 
In stark contrast to the cCLV, the outcome of Stage 2 in  the MCLV/BDCLV is not certain.
This is because the interactions 
in the MCLV/BDCLV are {\it not}  of predator-prey type:
In Stage 2, the dynamics boils down to the competition between species $i$
and its ``weak opponent'', species $i+1$, that the latter has a non-zero chance to win it.

To study this two-species competition, we  focus on the stage 2 dynamics along the edge $(i,i+1)$. 
Since species $i-1$ has died out at the end of Stage 1, we have $x_i+x_{i+1}=1$ and $x_{i-1}=0$, and the 
constant-$K$ BDCLV
transition rates in Stage 2 are $T_j^+=(1+s\Pi_j)~Nx_j$
and $T_j^-=N^2 x_j/K$, with $j\in\{i,i+1\}$, see (\ref{eq:A1_BDCLV_rates}).
Similarly, the transition rates of the MCLV  along the  edge $(i,i+1)$  for a population of size $N=K$
are obtained from (\ref{eq:A1_MCLV_rates}) with $x_{i+1}=1-x_i$ and $x_{i-1}=0$:
\begin{eqnarray}
\label{eq:A3_MCLV_rates}
T_{i+1\to i}= \frac{T_i^+T_{i+1}^-}{K}=Kx_i(1-x_i)(1+\alpha_i(1-x_i))\quad \text{and} \quad
T_{i\to i+1}= \frac{T_i^-T_{i+1}^+}{K}=Kx_i(1-x_i)(1-\alpha_{i}x_i).
\end{eqnarray}
It is clear from these transition rates that, $x_i=0,1$
are the possible outcome of the stage 2 dynamics and correspond to
 either the absorption of species $i$ with probability $\phi_i\vert_K$
 ($x_i=1, x_{i+1}=0$), or the the absorption of $i+1$ ($x_i=0, x_{i+1}=1$) with  probability $1-\phi_i\vert_K$.

Clearly, (\ref{eq:A3_MCLV_rates})
define a one-dimensional Moran process whose fixation properties can be computed exactly~\cite{Ewens,Antal}. 
For our purposes, the diffusion theory allows us to obtain a concise and reliable characterization of $\phi_i|_{K}$.
In fact, the backward version of the FPE generator (\ref{eq:FPE1_MCLV}) for the MCLV (with $N=K$)
along the  edge $(i,i+1)$ 
is~\cite{Gardiner,VanKampen}
\begin{eqnarray*}
\hspace{-5mm}
\label{eq:A3_G_MCLV}
{\cal G}|_{K}(x_i)\equiv  \frac{x_i(1-x_i)}{K}\left[K\alpha_i\frac{\partial}{\partial x_i} +
\left\{1+\frac{\alpha_i}{2}(1-2x_i)\right\}\frac{\partial^2}{
\partial x_i^2}\right].
 \end{eqnarray*}
When Stage 2 starts with a fraction $\hat{x}_i$ of individuals of species $i$,
the fixation probability $\phi_i|_{K}$ of the underpinning MCLV 
is obtained in the realm of the diffusion theory by
solving ${\cal G}|_{K}(\hat{x}_i)~\phi_i|_{K}(\hat{x}_i)=0$ with $\phi_i|_{K}(0)=0$ and $\phi_i|_{K}(1)=1$. This yields  
\begin{eqnarray*}
\hspace{-5mm}
\label{eq:A3_phi1*}
\phi_i|_{K}(\hat{x}_i)
%\frac{1-\left(1-\frac{2\alpha_i x_0}{2+\alpha_i}\right)^{N+1}}{1-\left(1-\frac{2\alpha_i }{2+\alpha_i}\right)^{N+1}}
 = \frac{(2+\alpha_i)^{K+1}-\left\{2+\alpha_i(1-2\hat{x}_i)\right\}^{K+1}}{(2+\alpha_i)^{K+1}-(2-\alpha_i)^{K+1}}.
 \end{eqnarray*}
When $s\ll 1$, i.e. $\alpha_i \ll 1$, the backward FPE generator takes the classical form~\cite{Kimura,Ewens,Blythe07}
\begin{eqnarray}
\hspace{-5mm}
\label{eq:A3_phi1}
{\cal G}|_{K}(x_i)=\frac{x_i(1-x_i)}{K}\left[K\alpha_i\frac{\partial}{\partial x_i} +
\frac{\partial^2}{
\partial x_i^2}\right],
\quad\text{yielding the familiar expression}\quad
\phi_i|_{K}(\hat{x}_i)=
\frac{1-e^{-\alpha_iK~\hat{x}_i}}{1-e^{-\alpha_iK}}.
 \end{eqnarray}
In the realm of the diffusion theory, the MCLV mean absorption time $T_2^{{\rm MCLV}}$ (from the inception of 
Stage 2) with an initial fraction $\hat{x}_i$ of $i$ individuals, $T_2^{{\rm MCLV}}(\hat{x}_i)$,
is obtained from the FPE generator ${\cal G}_{(i,i+1)}|_{K}$
by solving ${\cal G}|_{K}(\hat{x}_i)T_2^{{\rm MCLV}}(\hat{x}_i)=-1$
with boundary conditions $T_2^{{\rm MCLV}}(\hat{x}_i=0)=T_2^{{\rm MCLV}}(\hat{x}_i=1)=
0$~\cite{Gardiner,VanKampen,Ewens,Kimura}. 
The insets of Fig.~\ref{fig:FigA1} confirm that the mean absorption time $T_2^{(i,i+1)}$ along the edge $(i,i+1)$ 
 in the constant-$K$ BDCLV virtually coincide
with  $T_2^{{\rm MCLV}}$ when $N=K\gg 1$: $T_2^{(i,i+1)}\simeq T_2^{{\rm MCLV}}$. 
The FPE for $T_2^{{\rm MCLV}}$ can be solved by standard methods 
and generally yields a cumbersome expression.
In the limit of weak selection, $s\ll 1$,
we can use the simpler form (\ref{eq:A3_phi1}) and find that 
$T_2^{(i,i+1)} \simeq T_2^{{\rm MCLV}}\sim (\log{K})/s$ when $s\ll 1$ and $sK \gg 1$. In this case,
$T_2^{(i,i+1)}$ scales as $1/s$ to leading order, with
a subleading dependence on the population size via the prefactor $\log{K}$.
When $s\ll 1$ and $sK \lesssim 1$, 
$T_2^{(i,i+1)}={\cal O}(K)$ at quasi-neutrality, while $T_2^{(i,i+1)}\sim \log{K}$ when  $sK \gg 1$
(strong selection), as shown in Fig.~\ref{fig:FigA4}~(b).
%
%We therefore conclude that in
% the constant-$K$ BDCLV, we have $T_2(\hat{x}_i)\sim (\log{K})/s$ when $s\ll 1$ and $sK \gg 1$
% and $T_2(\hat{x}_i)={\cal O}(K)$ when $s\ll 1$, and $sK \lesssim 1$, as shown in Fig.~\ref{fig:FigA4}~(b).

\vspace{0.3cm} 
A similar analysis can be carried out when $\epsilon>0$, see Section 5. In this case,
the non-zero-sum birth-death dynamics defined by (\ref{eq:transrates})
with constant carrying capacity $K\gg 1$ is similar to the dynamics of Moran model
defined by the transition rates 
\begin{eqnarray}
\label{eq:A3_esps-MCLV_rates}
T_{i+1\to i}= \frac{T_i^+T_{i+1}^-}{K}=Kx_i(1-x_i)(1+\alpha_i(1-x_i))\quad \text{and} \quad
T_{i\to i+1}= \frac{T_i^-T_{i+1}^+}{K}=Kx_i(1-x_i)(1-\alpha_{i}(1+\epsilon)x_i).
\end{eqnarray}
in a population of constant size $N=K$. 
In this case, the stage 2 dynamics along the edge $(i,i+1)$
is characterized by the backward FPE generator
${\cal G}^{\epsilon}|_{K}(x_i)\equiv  \frac{x_i(1-x_i)}{K}\left[K\alpha_i(1+\epsilon x_i)\partial_i +
\left\{1+\frac{\alpha_i}{2}(1-(2+\epsilon)x_i)\right\}
\partial_i^2\right]$. The absorption probability is thus obtained by solving
${\cal G}^{\epsilon}|_{K}(\hat{x}_i)\phi_i|_{K}(\hat{x}_i)=0$ with $\phi_i|_{K}(0)=1-\phi_i|_{K}(1)=0$ 
yielding
\begin{eqnarray}
\hspace{-5mm}
\label{eq:A3_phi2}
\phi_i|_{K}(\hat{x}_i)
 = \frac{(2+\alpha_i)^{Kh(\epsilon,\alpha_i)+1}-\left\{2+\alpha_i(1-(2+\epsilon)\hat{x}_i)\right\}^{Kh(\epsilon,\alpha_i)+1}}{(2+\alpha_i)^{Kh(\epsilon,\alpha_i)+1}-
 (2-\alpha_i(1+\epsilon))^{Kh(\epsilon,\alpha_i)+1}}, \quad \text{where} \quad h(\epsilon,\alpha_i)\equiv 
 \frac{1+\epsilon(1+1/\alpha_i)}{(1+\epsilon/2)^2}.
 \end{eqnarray}
When $|\epsilon| \ll 1$, this expression simplifies in the weak selection regime ($s\ll 1$)
where it takes the form 
\begin{eqnarray}
\hspace{-5mm}
\label{eq:A3_phi3}
\phi_i|_{K}(\hat{x}_i) \simeq
\frac{1-e^{-K\alpha_i(1+\epsilon/2)\hat{x}_i}}{1-e^{-K\alpha_i(1+\epsilon/2)}}.
 \end{eqnarray}
Hence, the stage 2 fixation probability in the weak selection regime when $|\epsilon| \ll 1$ 
is  the same as in the  constant-$K$ BDCLV with a selection intensity is rescaled by a factor $1+\epsilon/2$
($s\to s(1+\epsilon/2)$).
This
suggests to consider the following effective backward FPE generator
when $s\ll 1$ and $|\epsilon|\ll 1$:
${\cal G}^{\epsilon}|_{K}(x_i)\equiv  \frac{x_i(1-x_i)}{K}\left[K\alpha_i(1+\epsilon/2)~\partial_i +
\partial_i^2\right]$. In this case, the  mean absorption time
is given 
${\cal G}^{\epsilon}|_{K}(\hat{x}_i)T_2(\hat{x}_i)=-1$
with $T_2(\hat{x}_i=0)=T_2(\hat{x}_i=1)=
0$. Clearly, this implies that the mean absorption time 
is obtained from $T_2$  of the constant-$K$ BDCLV with a rescaled 
selection intensity $s\to s(1+\epsilon/2)$. We have checked in our simulations that this rescaling also applies to Stage 1
and therefore to the overall mean fixation time, see \ref{AppendixE6} and Fig.~\ref{fig:FigA6}~(b)-(d).

\begin{figure}[!b]
	\centering
	\includegraphics[width=0.48\linewidth]{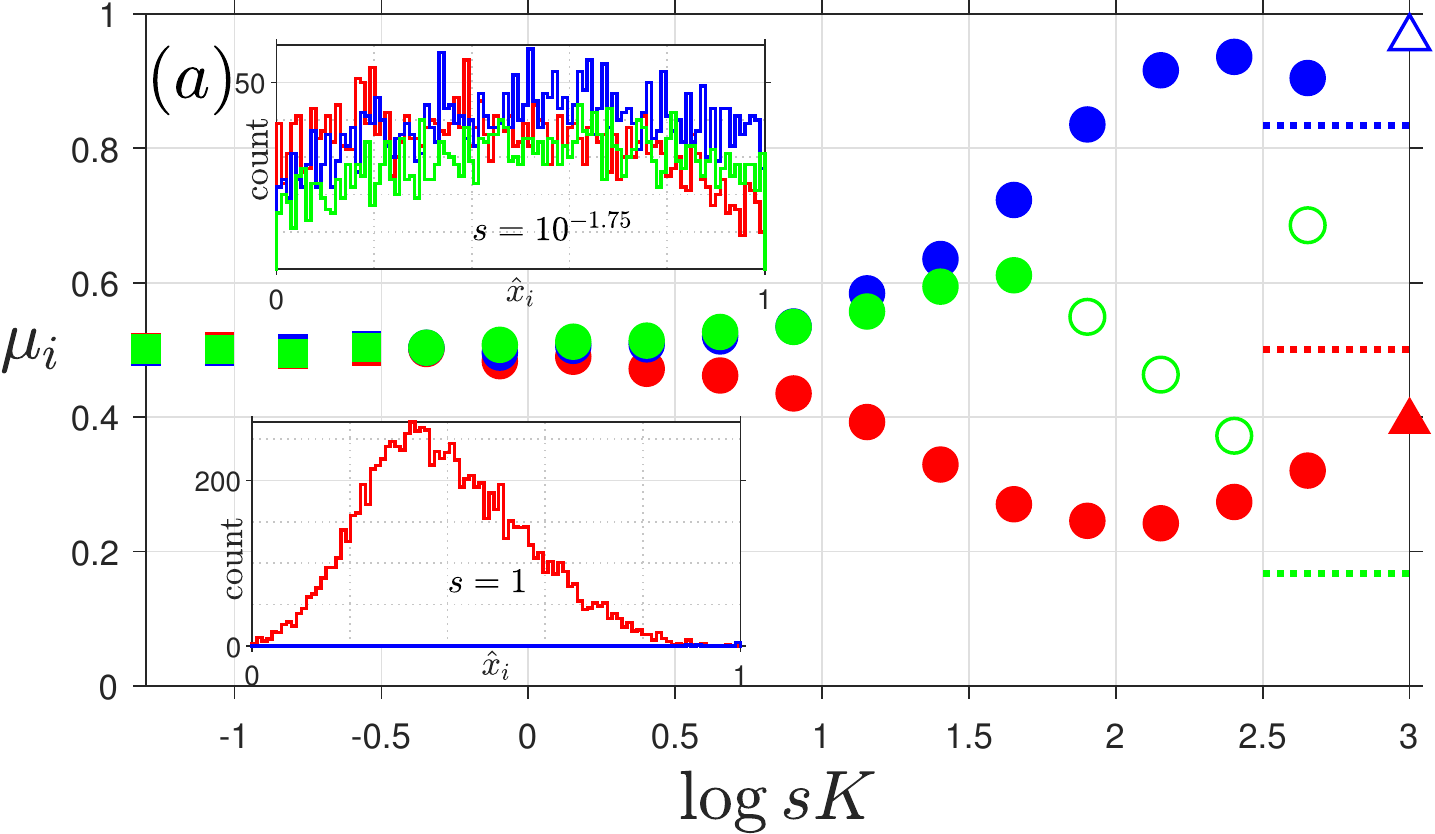}
	\includegraphics[width=0.48\linewidth]{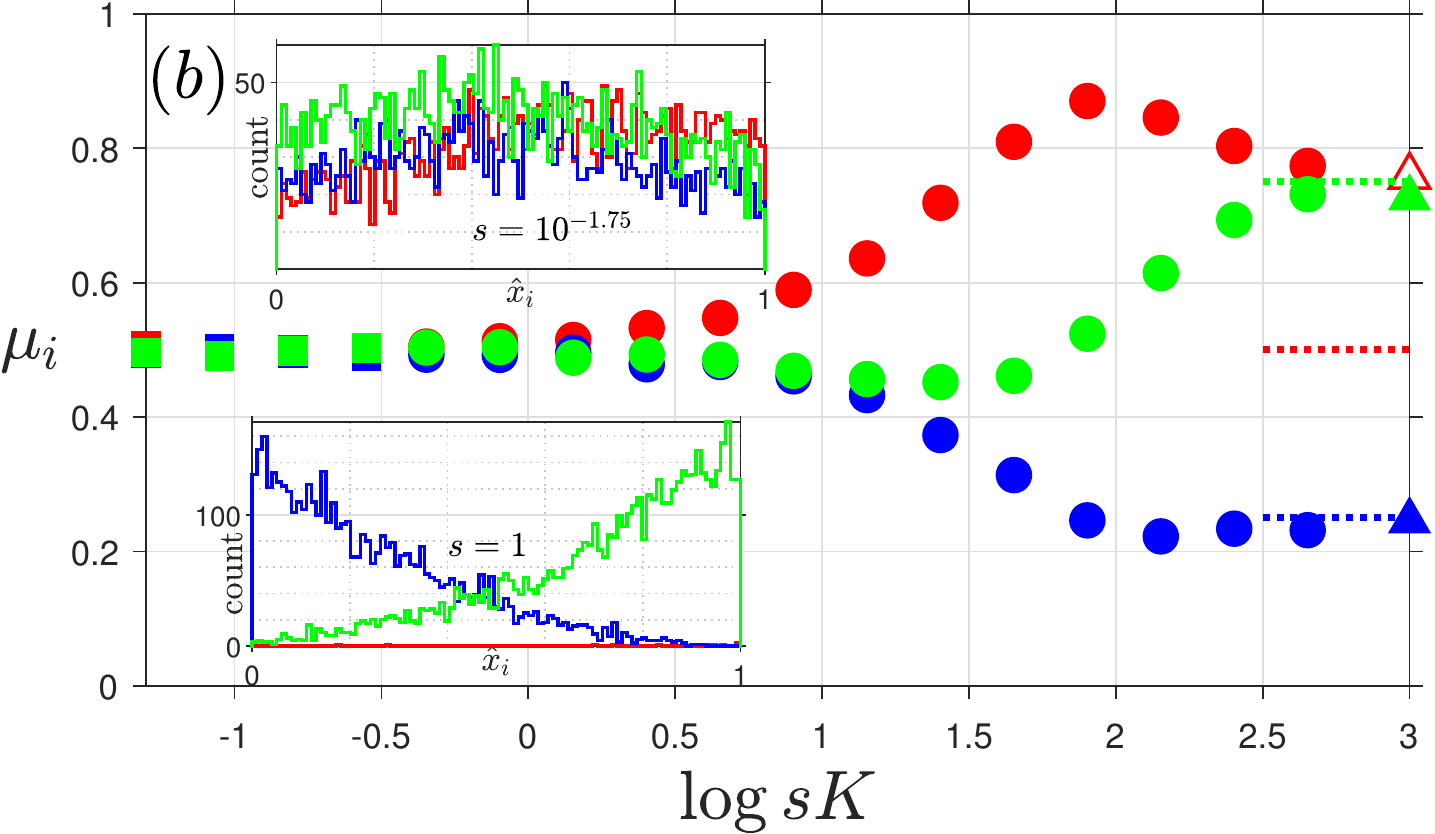}
	\includegraphics[width=0.48\linewidth]{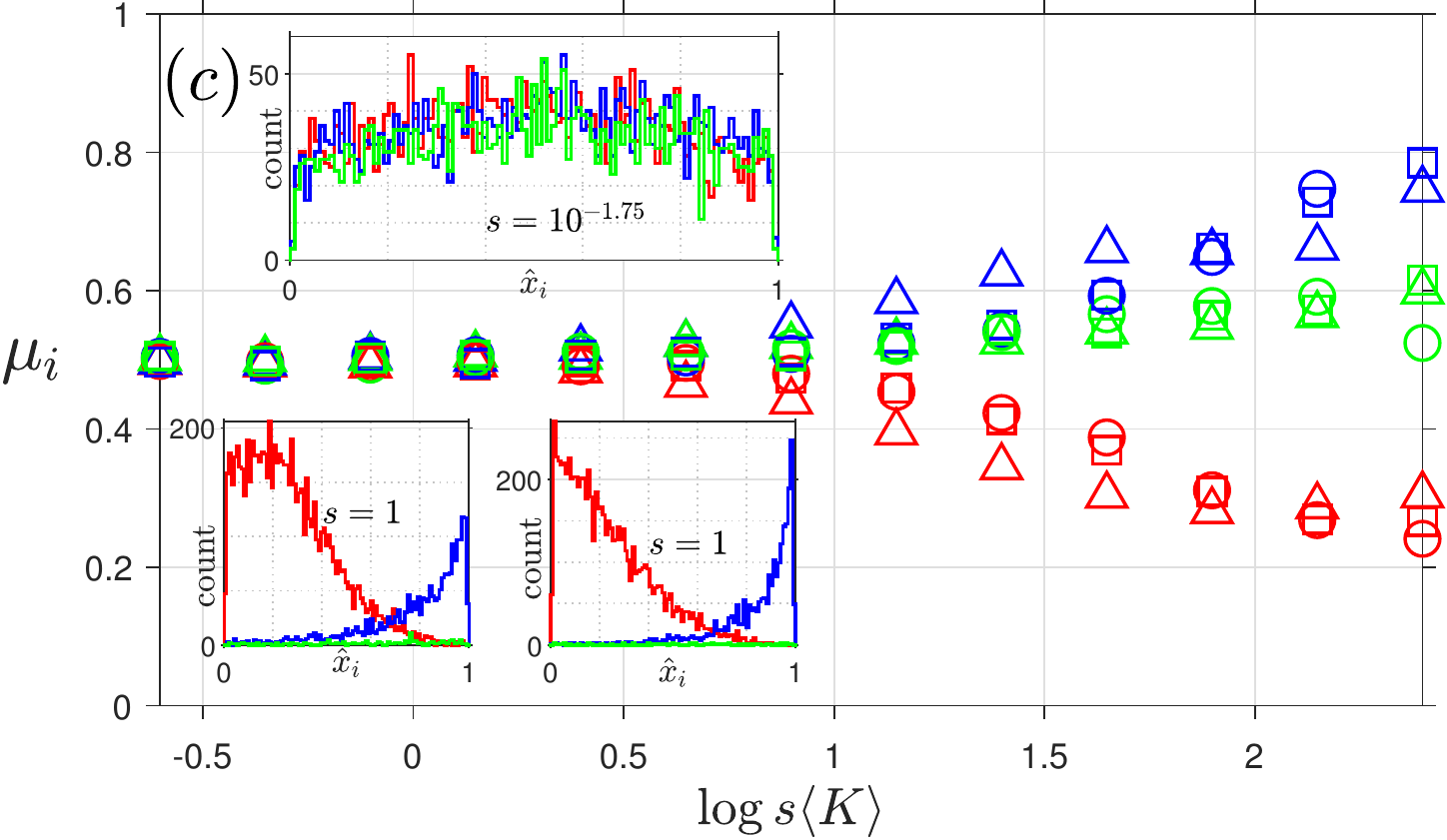}
	\includegraphics[width=0.48\linewidth]{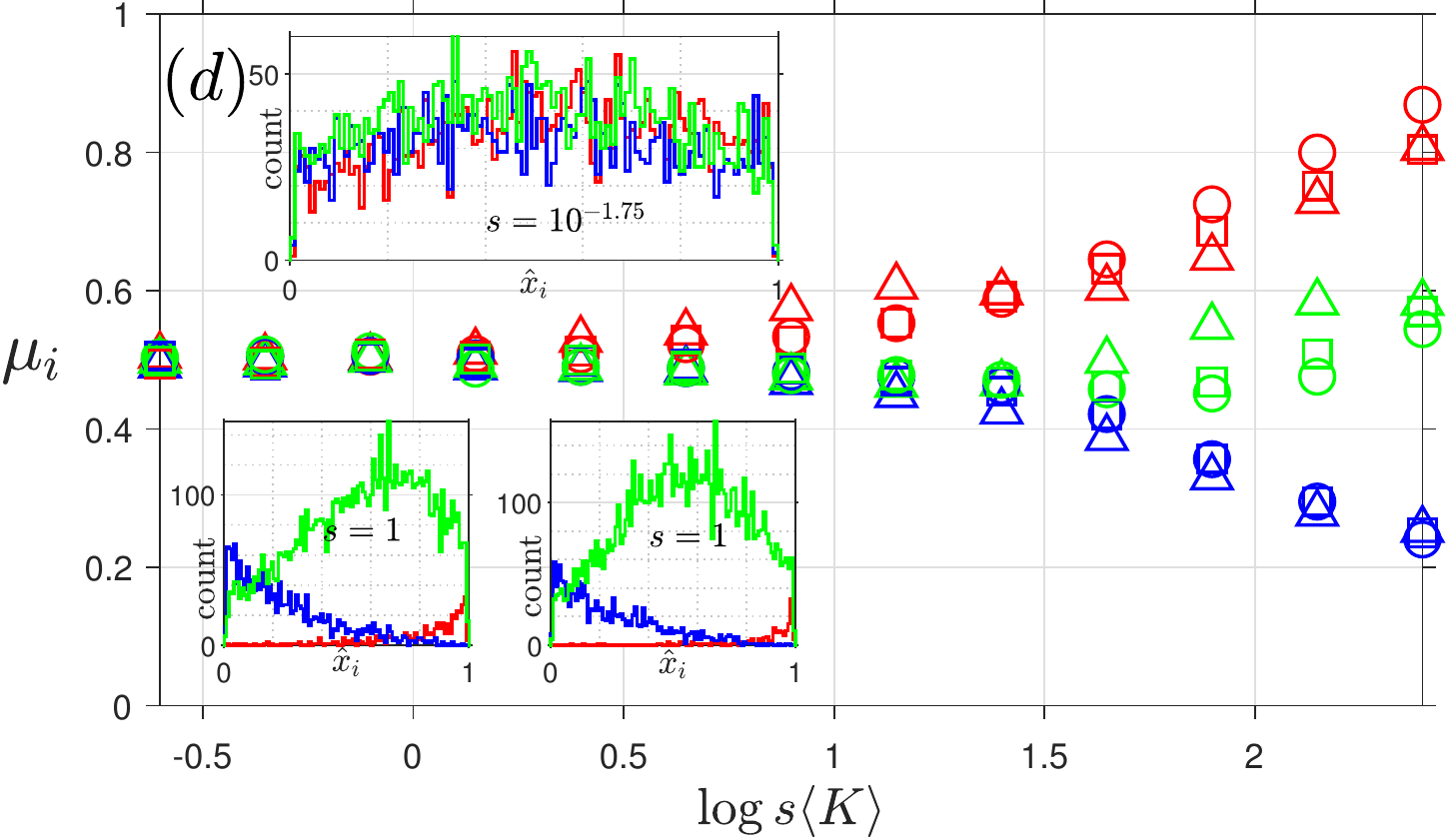}	
	\caption{
	Population composition at the inception of Stage 2 vs. $sK$ (a,b) and $s\langle K \rangle$ (c,d) 
	with $\vec{r}=\vec{r}^{(1)}$ in (a,c) and $\vec{r}=\vec{r}^{(2)}$ in (b,d).
	In all panels: $\mu_i=\int_{0}^{1}\hat{x}_iP_{(i,i+1)}(\hat{x}_i)~d\hat{x}_i$ is the 
	mean value of $\hat{x}_i$ for species $i=1$ (red), $2$ (blue),
	$3$ (green), with   $\vec{x}_0=\vec{x}_c$ and $\epsilon=0$.
	(a,b) $\mu_i$ vs. $sK$ in the  constant-$K$ BDCLV with $K = 1000~(\triangle), 450~(\circ), 50~(\square)$
	and $s\in (10^{-3},1)$. (Empty symbols denote data arising from small survival probability 
	$\phi_{i,i+1}<0.01$ that would require additional sampling).
	When $sK\lesssim 
	10$, $\mu_i\approx 1/2$ and $P_{(i,i+1)}\approx 1$ is approximately uniform.
	When $sK\gg 1$, the dynamics is dominated by the LOW and 
	$\mu_i \approx r_{i+1}/(r_{i+1}+r_{i-1})$ shown as dotted lines, see text.
	Upper insets:  Histograms corresponding to $P_{(i,i+1)}(\hat{x}_i)$ with $s=10^{-7/4}$
	and $K=250$,  is approximately uniform, corresponding to $P_{(i,i+1)}\approx 1$,
	along the three edges. Lower insets: Same with $s=1$ and $K=1000$, showing that 
	$P_{(i,i+1)}$ is no longer uniform when $sK\gg 1$
	and how it changes with $\nu=10$ (left) and $\nu=0.1$ (right).
	(c,d) $\mu_i$ vs. $s\langle K\rangle$ in the switching-$K$ BDCLV with $\langle K\rangle =
	250$ and $\gamma=0.8$ kept fixed and $s$ varies
 	with $\nu=10~(\square)$, $\nu=1~(\circ)$  and $\nu=0.001~(\triangle)$.
	Insets: (Upper)  Histograms corresponding to $P_{(i,i+1)}(\hat{x}_i)$ with $s=10^{-7/4}$, $\nu=0.1$ 
	and $\langle K\rangle=250, \gamma=0.8$ 
	for $i=1,2,3$. (Lower) Same with $s=1$,  $\langle K\rangle=250, \gamma=0.8$, $\nu=0.1$ (left) and $\nu=10$ (right).
	}
	\label{fig:FigD3}
\end{figure}

\section{Population composition at the inception of Stage 2}
\label{AppendixD} 
The stage 2 dynamics of the BDCLV and MCLV, as well as their fixation properties, depend on the 
population composition at the end of Stage 1 which coincides with the inception of Stage 2.
In the main text, we have seen that the initial fraction $\hat{x}_i$
of $i$ individuals along the edge $(i,i+1)$ of $S_3$ is given by the probability
density $P_{(i,i+1)}(\hat{x}_i)$ which can be approximated by a uniform distribution
$P_{(i,i+1)}(\hat{x}_i)\approx 1$ when $sK \lesssim 10$ (constant $K$) and $s\langle K \rangle \lesssim 10$  (switching $K$), 
yielding an average initial fraction 
$\mu_i=\int_{0}^{1}\hat{x}_iP_{(i,i+1)}(\hat{x}_i)~d\hat{x}_i\approx 1/2$
of $i$ individuals along $(i,i+1)$, see Fig.~\ref{fig:FigD3}. The same holds true also when
$0<\epsilon\ll 1$, see Sec.~5 and \ref{AppendixC}.

This is no longer the case under strong selection, when the $P_{(i,i+1)}$'s
are skewed and far from being uniform, see the lower insets of Fig.~\ref{fig:FigD3}.
When $K\gg 1$ is constant and the LOW holds, the extinction of the first species in Stage 1 occurs from the outermost orbit 
as in the cCLV~\cite{Berr09,West18}, see \ref{AppendixA3},  and $\mu_i$ can be estimated as follows: 
Along the outermost orbit that is closest ($x_{i-1}=1/K$)
to the edge $(i,i+1)$ in the constant-$K$ BDCLV, from the rate equations (\ref{eq:xidot}) we have $x_i/x_{i+1}=r_{i+1}/r_{i-1}$
yielding  $\mu_i=r_{i+1}/(r_{i+1}+r_{i-1})$. The results of  Fig.~\ref{fig:FigD3}~(a,b) for $sK\gg 1$
are in satisfying agreement with this prediction.

The results reported in Fig.~\ref{fig:FigD3}~(c,d) show that the averages $\mu_i$'s are closer to $1/2$
in regime (ii) than in the constant-$K$ BDCLV. This stems from the environmental
variability operating to balance the effect of selection and  implies that 
$P_{(i,i+1)}\approx 1$ is a better approximation in the regime (ii) when $K$
is randomly switching than when it is constant.
In the lower insets of Fig.~\ref{fig:FigD3}~(c,d), we find very similar  probability densities
$P_{(i,i+1)}$ for very different switching rates ($\mu=0.1$ and $\mu=10$),
showing that in the switching-$K$ BDCLV
$P_{(i,i+1)}$ varies little with $\nu$.

\section{Extinction, absorption and fixation times \&  number of switches}
\label{AppendixE} 
We study the overall mean fixation time $T_F$, which is the average
time after which one species takes over the entire population,
in the constant-$K$ and switching-$K$ BDCLV.
$T_F=T_1+T_2$ consists of the {\it mean extinction time} $T_1$
 and the  {\it mean absorption time}  $T_2$ arising from Stages 1 and 2, respectively.
 We also compute the average number of switches occurring 
in Stages 1 and 2 of the switching-$K$ BDCLV.
 
\subsection{Mean extinction, absorption and fixation times in the constant-$K$ BDCLV}
\label{AppendixE1}
We first consider the case of the  constant-$K$ BDCLV and show that the overall mean fixation time 
 $ T_F={\cal O}(K)$ across all regimes (i)-(iii), see Fig.~\ref{fig:FigA4}(a). 

\begin{figure}
	\centering
	\includegraphics[width=0.33\linewidth]{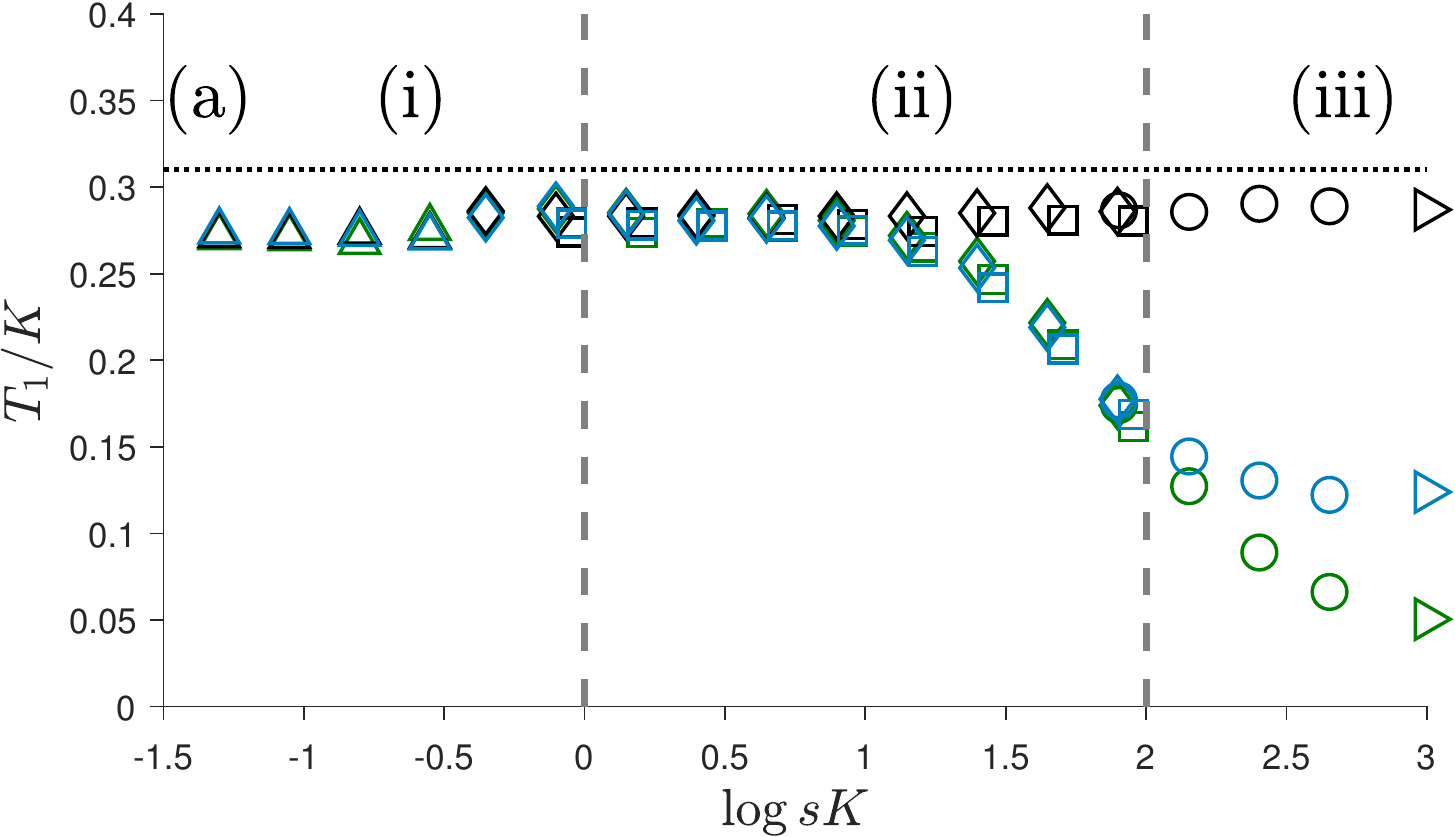}
	\includegraphics[width=0.33\linewidth]{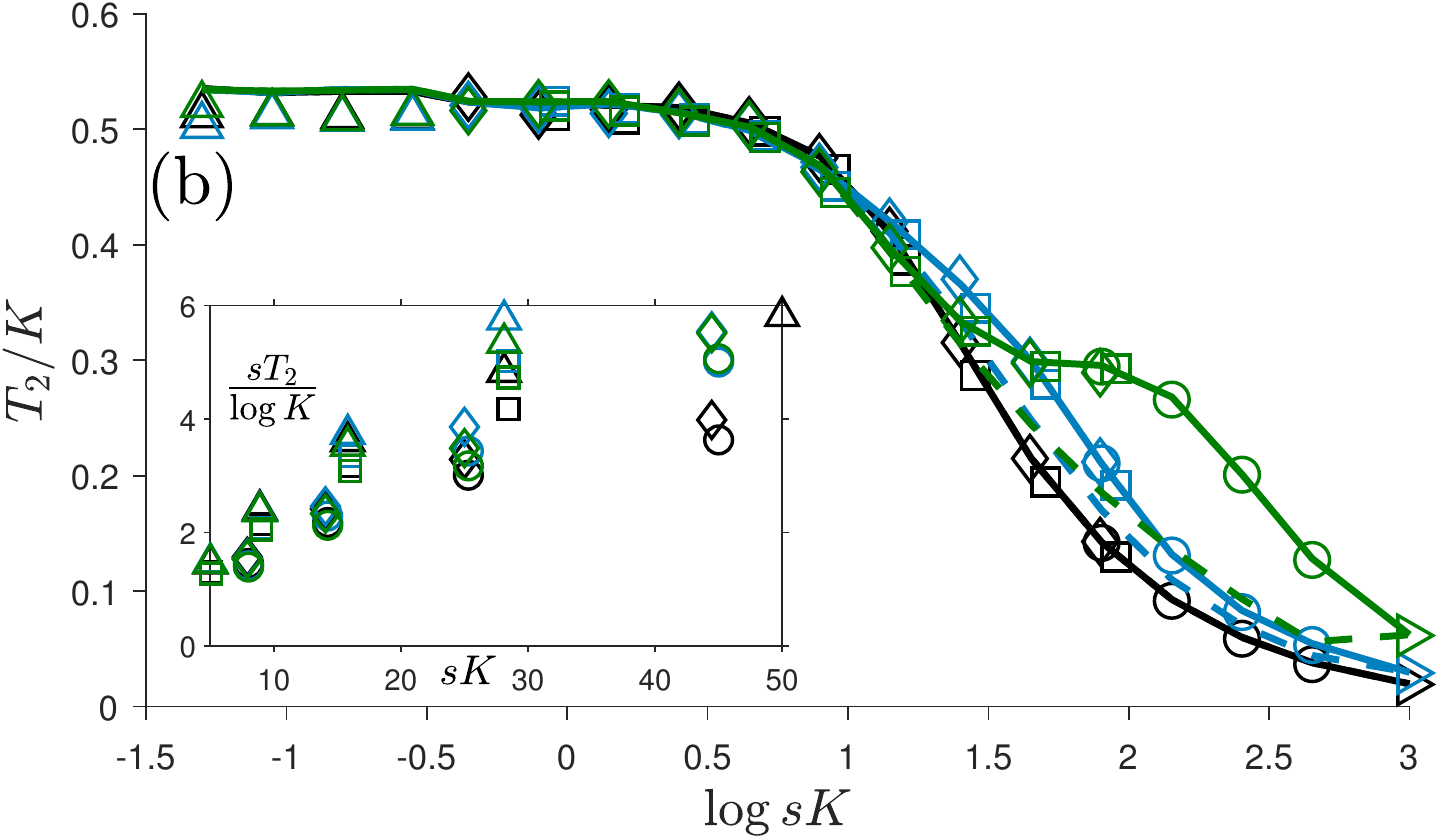}
	\includegraphics[width=0.33\linewidth]{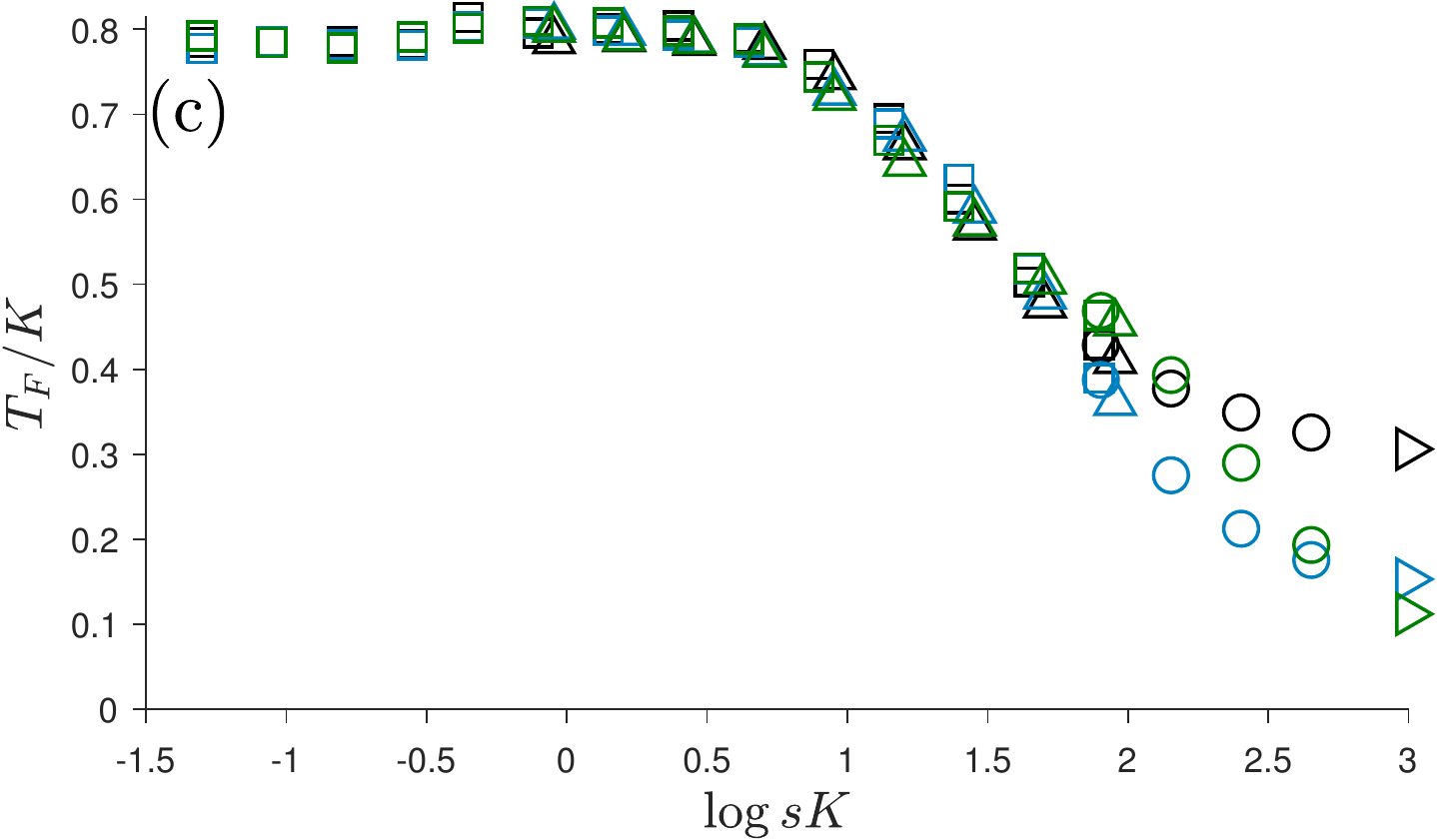}
	\caption{Mean
	extinction and absorption  times $T_1$ and $T_2$, and 
	mean fixation time $T_F$ in the
	constant-$K$ BDCLV for  
	$K\in \{1000~(\rhd), 450~(\circ), 250~(\diamond), 90~(\square), 50~(\triangle)\}$ and the same 
	values of $s$ as in Figs.~3 and 4: 
	(a)
	$T_{1}/K$ vs. $sK$; showing $T_{1}={\cal O}(K)$ when $K\gg 1$ and $T_{1}\approx 0.31K$ (dotted line)
	when $r_i=r$ and under weak selection ($sK\lesssim 10$) when  $\vec{x}_c \neq 
	\vec{x}^*$ (unequal $r_i$'s), see text. 
	(b) $T_{2}/K$ vs. $sK$; solid and dashed lines show the 
	respective predictions of $T_{2}\vert_{K}=\sum_i \phi_{i,i+1}T_{2}^{(i,i+1)}\vert_{K}$  and (\ref{eq:T2-bis}), see text.
	Inset: $sT_{2}/\log{K}={\cal O}(1)$ when $s\ll 1$ and $sK\gg 1$, see text.
	(c) $T_F/K$ vs. $sK$ showing that $T_{F}={\cal O}(K)$ across all regimes
	with subleading prefactors in regime (iii) shorter than in (i) and (ii).
	In all panels: symbols are from stochastic simulations, $\vec{x}_0=\vec{x}_c$, $\epsilon=0$ and 
	 $r_1=1/11$~(green), $1$~(black), $3/5$~(blue)  and  $r_2=r_3=1$.
	}
	\label{fig:FigA4}
\end{figure}
\subsubsection{Stage 1: Mean extinction time $T_1$ in the constant-$K$ BDCLV}
\label{AppendixE1.1}
The {\it mean extinction time $T_1$} is the average time for one of the species to go extinct at the end of  Stage 1. 
 As explained in \ref{AppendixB}, with the results obtained for the cCLV, we find 
 $T_1 \simeq  T_1^{{\rm cCLV}}/2\approx 0.3K$
 when $s\ll 1$ (regimes (i,ii)) and for arbitrary $s$ when all $r_i=1/3$, see 
 Fig.~\ref{fig:FigA4}~(a).
Deviations from  $T_1 \approx 0.3K$, and a weak dependence on $s$ and on the $r_i$'s,
are found near the boundary of regimes (ii)-(iii) and in regime (iii), where 
$T_1\simeq  \beta_c(s,\vec{r}) K$, where $\beta_c$ is a decreasing function of  $s$
when the $r_i$'s are unequal, see Fig.~\ref{fig:FigA4}~(a).

\subsubsection{Stage 2: Mean absorption time $T_2$ in the constant-$K$ BDCLV}
\label{AppendixE1.2}
The stage 2 mean absorption time $T_2$ is given by
\begin{eqnarray}
 \label{eq:T2}
 \hspace{-3mm}
 T_2=\sum_{i=1}^{3} \phi_{i,i+1}T_2^{(i,i+1)},
\end{eqnarray}
where the mean absorption time 
along the edge  $(i,i+1)$ of $S_3$, denoted by $T_2^{(i,i+1)}$,  is weighted by the probability $\phi_{i,i+1}$ that Stage 1 ends
on that edge.

The expression of $T_2^{(i,i+1)}$ is obtained from the mean fixation time of the MCLV with $N=K$, here denoted by 
$T_2^{(i,i+1)}\vert_{K}$ with $T_2^{(i,i+1)}\simeq T_2^{(i,i+1)}\vert_{K}$, 
see \ref{AppendixC}.  For a given initial fraction $\hat{x}_i$ of $i$'s 
at the start of Stage 2 is 
$(\hat{x}_i)$,  
$T_2^{(i,i+1)}(\hat{x}_i)\vert_{K}$ when $s\ll 1$
is obtained by solving 
$ {\cal G}_{(i,i+1)}\vert_{K}(\hat{x}_i)~T_2^{(i,i+1)}(\hat{x}_i)\vert_{K}=-1$,
with $T_2^{(i,i+1)}\vert_{K}(0)=T_2^{(i,i+1)}\vert_{K}(1)=0$, see (\ref{eq:A3_phi1}).
Since the exact population composition  along the edge $(i,i+1)$ at the inception of Stage 2 
is given by  $P_{(i,i+1)}(\hat{x}_i)$, we have:
\begin{eqnarray}
 \label{eq:T2cond2}
T_2\simeq \sum_{i=1}^{3}\phi_{i,i+1} T_2^{(i,i+1)}\vert_{K}= 
\sum_{i=1}^{3} \phi_{i,i+1}\int_0^1 P_{(i,i+1)}(\hat{x}_i)~T_2^{(i,i+1)}(\hat{x}_i)\vert_{K}~d\hat{x}_i.\nonumber,
\end{eqnarray}
with $T_2^{(i,i+1)}\vert_{K}\equiv \int_0^1 P_{(i,i+1)}(\hat{x}_i)~T_2^{(i,i+1)}(\hat{x}_i)\vert_{K}~d\hat{x}_i$.
A simpler expression for $T_2$ is obtained when $s\ll 1$ and $sK={\cal O}(10)$
upon substituting 
 $\phi_{i,i+1}\approx 1/3$ and $P_{(i,i+1)}\approx 1$ in (\ref{eq:T2cond2}):
\begin{eqnarray}
 \label{eq:T2-bis}
 \hspace{-3mm}
 T_2\simeq \frac{1}{3}\sum_{i=1}^{3} \int_0^1 T_2^{(i,i+1)}(\hat{x}_i)\vert_{K}~d\hat{x}_i.
\end{eqnarray}
While the expression of $T_2^{(i,i+1)}(\hat{x}_i)$ is not particularly illuminating,
its asymptotic behavior is simple
and allows us to determine the behavior of $T_2$:   In the weak-selection regime (ii) 
where $s\ll 1$ and $sK= {\cal O}(10)$, we obtain the classical result $T_2^{(i,i+1)}|_{K}={\cal O}((\log{K})/s)$
according to which $T_2$ scales as $1/s$ with  a subleading prefactor $\sim \log{K}$~\cite{Ewens,Blythe07},
which is confirmed by the results of  Fig.~\ref{fig:FigA4}~(c).

On the other hand, since the mean fixation time in the neutral Moran model scales linearly with the population size~\cite{Ewens,Kimura,Blythe07},
we readily find $T_2 ={\cal O}(K)$ in the quasi-neutral regime (i). 
The mean fixation time in the Moran model with strong selection favoring
species $i$ against $i+1$  scales logarithmically with the population size~\cite{Antal},
from which we infer that  $T_2
={\cal O}(\log{K})$ in  regime (iii).

Putting  the asymptotic behaviors
of $T_1$ and $T_2$ together, we find that to leading order in  $N\simeq K\gg 1$ the overall mean fixation
time $T_F=T_1+T_2={\cal O}(K)$ scales linearly with the population size across the regimes (i)-(iii), with different
subleading prefactors in each regime. We also notice that in regime (iii) $T_1\gg T_2$: The extinction of a
second species (Stage 2) occurs much faster than the death of a first species in Stage 1, see Fig.~\ref{fig:Fig1}(a).
In regime (i)  $T_1/T_2=  {\cal O}(1)$ and $T_1/ T_2 = {\cal O}(sK/\log{K})$ in regime (ii), see 
Fig.~\ref{fig:Fig1}(b).

\subsection{Mean extinction, absorption and fixation times in the switching-$K$ BDCLV}
\label{AppendixE2}

\begin{figure}
	\centering
\includegraphics[width=0.33\linewidth]{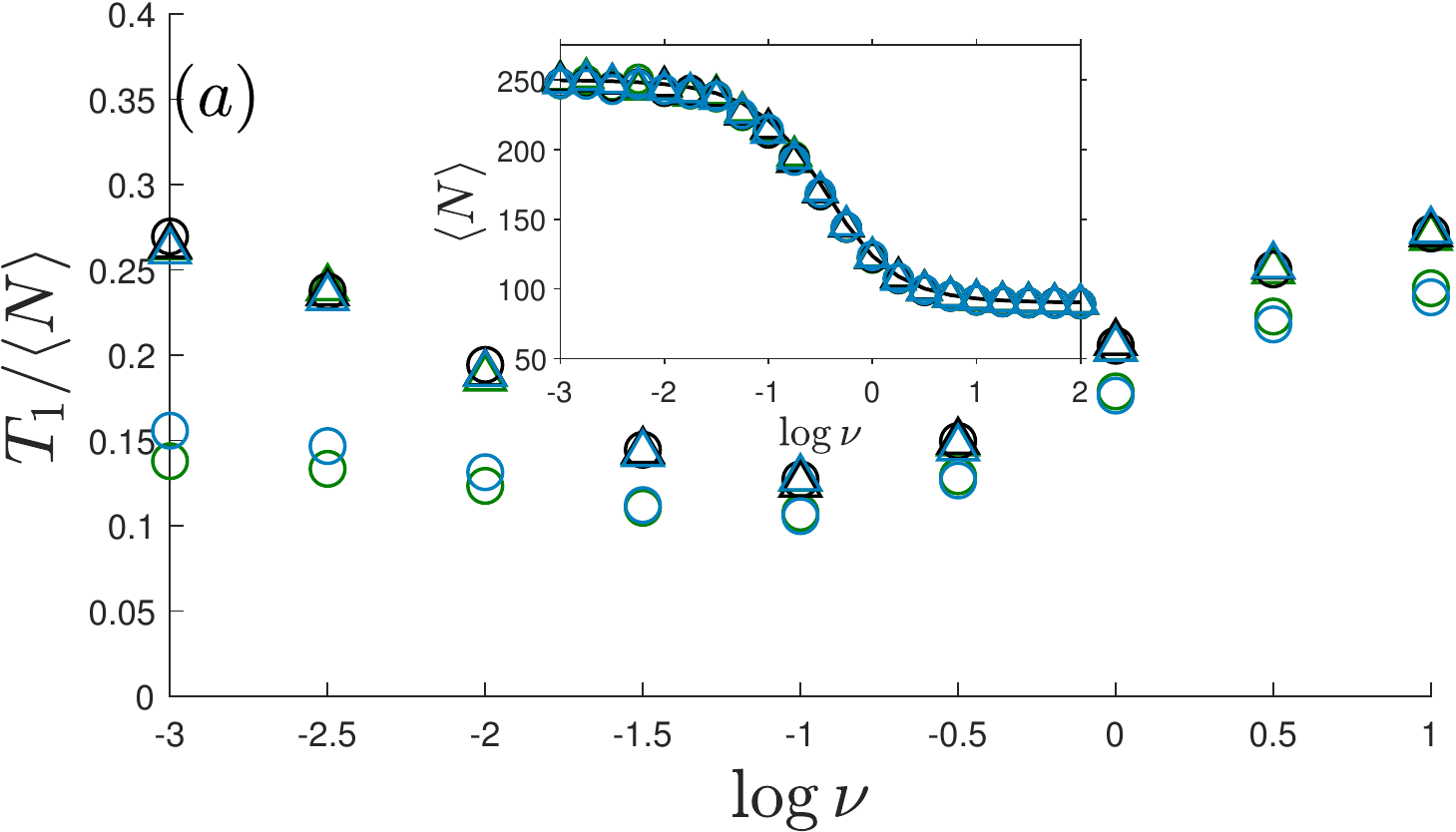}
	\includegraphics[width=0.33\linewidth]{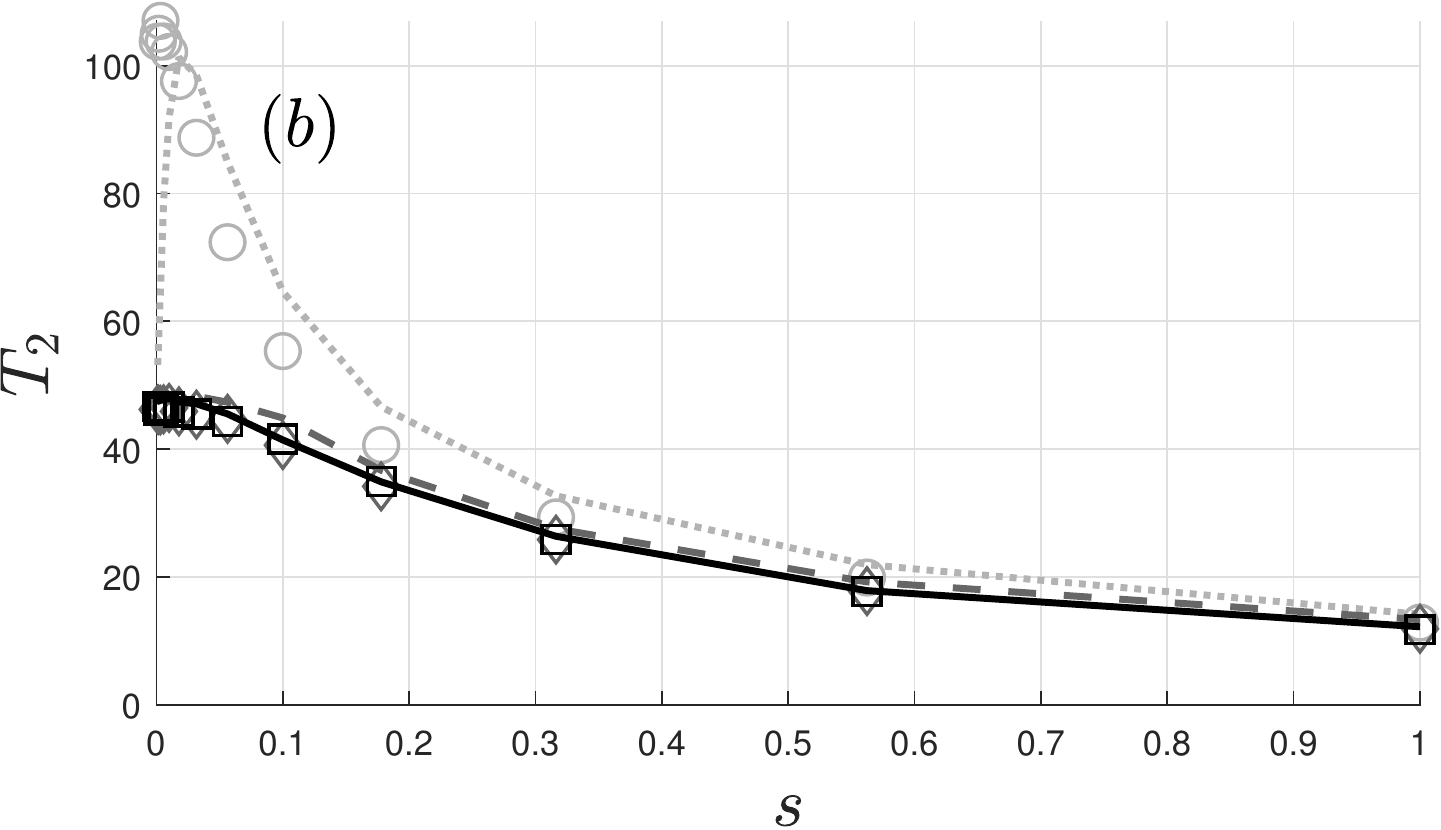}
\includegraphics[width=0.33\linewidth]{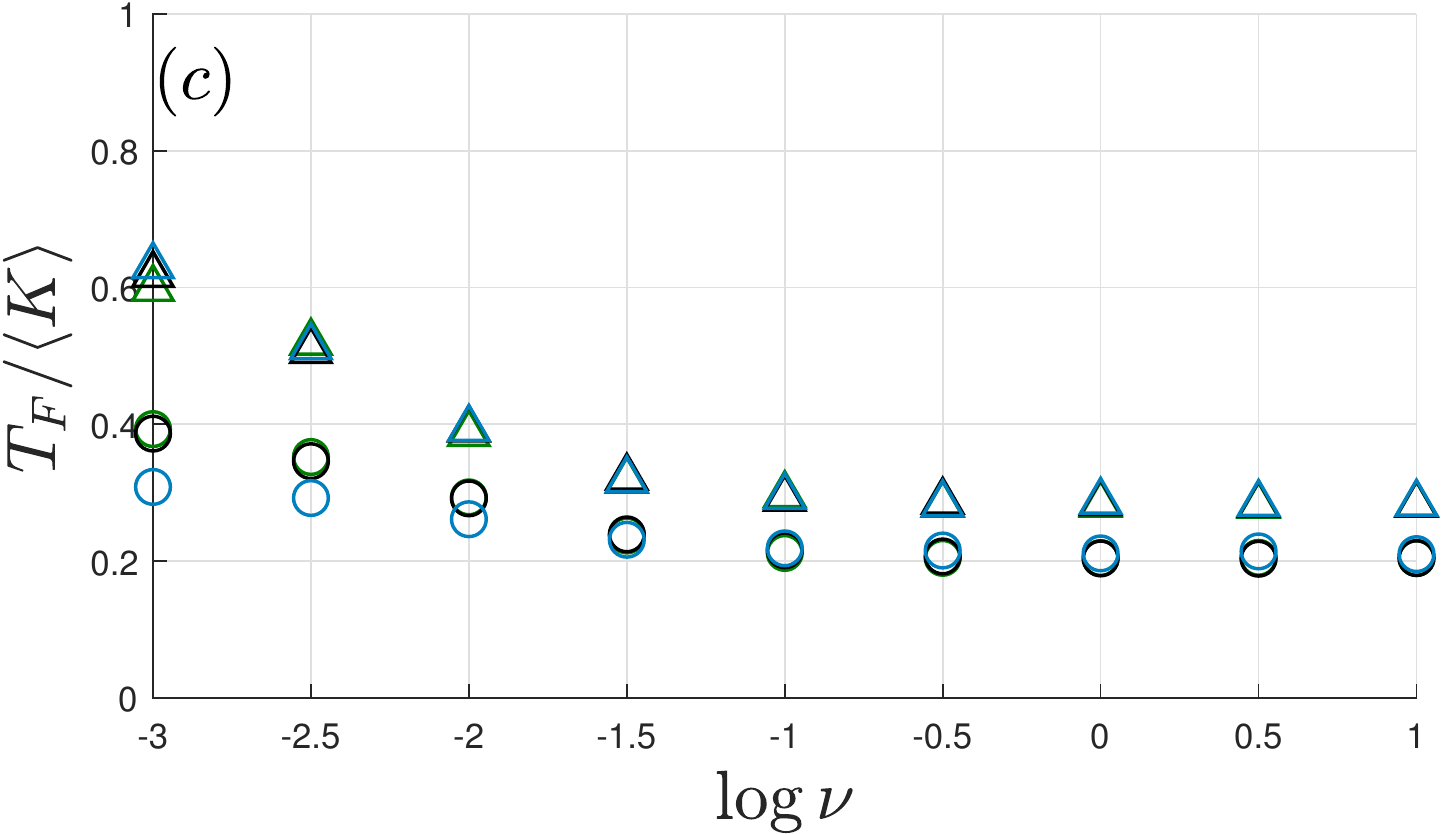}
	\caption{(a) $T_{1}/\langle N \rangle$ vs. $\nu$
	for $r_1 = 1/11 \text{(green)},1/3 \text{(black)}, 3/5 \text{(blue)}$ and $r_2=r_3=(1-r_1)/2$,
	with $s=10^{-1/2}$ (circles) and  $s=10^{-3/2}$ (triangles). In agreement with (\ref{eq:textp2th}),
	$T_{1}/\langle N \rangle=\beta_s={\cal O}(1)$ and slowly varies with $\nu$ and $s$.
	Inset: $\langle N \rangle$ vs $\nu$; solid lines are from the average over the marginal probability density (\ref{eq:pnustar}) of the process defined by \eqref{eq:N_MF2},
 and symbols
	are from stochastic simulations with $s=10^{-1/2}$ (circles) and  $s=10^{-3/2}$ (triangles),
	showing $\langle N \rangle={\cal O}(\langle K\rangle)$, see text.
	(b)   $T_{2}$ vs. $s$ for
	$\nu = 10^{-3}$~(circles, light  dotted gray), $10^{-1}$~(diamonds, dashed gray), $10$~(squares, solid black)
	and $\vec{r}=(1/3,1/3,1/3)$. Symbols are from stochastic simulations and lines are from 
	(\ref{eq:T_2_s}). $T_2$ scales as $1/s$ with 
	subleading prefactor $\sim \log{\langle K\rangle}$
	when $s\ll 1$ and $s\langle K\rangle= {\cal O}(10)$, see text.
	(c) Same as in (a) but for the overall mean fixation time: $T_{F}/\langle K \rangle$ vs. $\nu$ with
	$s=10^{-1/2}$ (circles) and  $s=10^{-3/2}$ (triangles), showing 
	$T_{F}= {\cal O}(\langle K\rangle)$ over a broad range of values $\nu$, see text.
	In all panels: $\langle K\rangle=250, \gamma =0.8$ $(K_-=50,K_+=450)$  and $\vec{x}_0=\vec{x}_c$; $\epsilon=0$.
	}
	\label{fig:FigA5}
\end{figure}
We  study the effect of random switching
on the mean extinction and absorption times, $T_1$ and $T_2$ characterizing Stages 1 and 2, respectively. This allows us
to show that the mean  fixation time $T_F=T_1+T_2={\cal O}(\langle N \rangle)={\cal O}(\langle K \rangle)$
scales linearly with the average population size, and to compute the 
average number of switches occurring in Stages 1 and 2.
\subsubsection{Stage 1: Mean extinction time in the switching-$K$ BDCLV}
\label{AppendixE2.1}
Guided by the results of the constant-$K$ BDCLV, where $T_1$ scales linearly
with $N\approx K$ to leading order in $\langle K \rangle\gg 1$,
we expect
\begin{eqnarray}
\label{eq:textp2th}
T_{1}  = \beta_s \langle N \rangle
\quad \text{with} \quad \beta_s = \beta_s (s,\vec{r},\nu),
\end{eqnarray}
 where  $\langle N \rangle={\cal O}(\langle K \rangle)$ is  the long-time average population size
 that is in principle obtained by averaging $N$ over the $N$-QSD. In the inset of Fig.~\ref{fig:FigA5},
 this quantity is accurately computed in the realm of the PDMP approximation as
 $\langle N \rangle=\int_{K_-}^{K_+} N p_{\nu}^*(N) dN$, see the inset of Fig.~\ref{fig:FigA5}~(a),
 and is shown to be  independent of $s$ and a decreasing function of $\nu$.
For fast/slow switching, we have
$\langle N \rangle=(1-\gamma^2)\langle K\rangle$ when $\nu \to \infty$
and $\langle N \rangle=\langle K\rangle$ when $\nu \to 0$~\cite{KEM1,KEM2}.
Comparison with simulation results of Fig.~\ref{fig:FigA5} confirm that 
$T_{1}/\langle N \rangle=\beta_s={\cal O}(1)$  is a slowly varying function of  $\nu$
 and a weakly decreasing function of $s$. Since  $\langle N \rangle= {\cal O}(\langle K \rangle)$
when $\gamma={\cal O}(1)$, we obtain 
$T_{1} ={\cal O}(\langle N \rangle)={\cal O}(\langle K \rangle)$
to leading order in $\langle K \rangle$. 
\subsubsection{Stage 2 mean absorption time and overall mean fixation time in the switching-$K$ BDCLV}
\label{AppendixE2.2}
Proceeding as in  Sec.~5.1.2, the Stage 2 mean absorption time is given  by $T_2 = \sum_{i=1}^3 \phi_{i,i+1} T_{2}^{(i,i+1)}$.
In the realm of the PDMP approximation, when $s\ll 1$ and $s\langle K \rangle \gg 1$,
$T_{2}^{(i,i+1)}$ is obtained by averaging
the constant-$\langle K \rangle$ mean absorption time $T_{2}^{(i,i+1)}\vert_{\langle K \rangle}$ along 
the edge $(i,i+1)$ over the probability density function (\ref{eq:pnustar})
% (21)
~\cite{KEM1,KEM2}:
\begin{equation*}
\hspace{-4mm}
\label{eq:T_i_fix_x}
T_{2}^{(i,i+1)} \simeq \int_{0}^{1} \int_{K_-}^{K_+}  P_{(i,i+1)}(\hat{x}_i)~T_{2}^{(i,i+1)}(\hat{x}_i)\vert_{\langle K \rangle}~
p_{\nu/\alpha_i}^*(N)~d\hat{x}_i~dN.
\end{equation*}
As in Sec.~4.1.2, the switching rate is rescaled $\nu \to \nu/\alpha_i$
due to the average number ${\cal O}(\nu/\alpha_i)$ 
of switches  occurring in Stage 2 along the edge $(i,i+1)$ when $s\ll 1$ and $s\langle K \rangle \gg 1$~\cite{KEM1,KEM2}.
The above equation can be simplified using  $\phi_{i,i+1}\approx 1/3$ and $P_{(i,i+1)}(\hat{x}_i)\approx 1$
when $s\ll 1$ and $s\langle K \rangle \lesssim 10$ (\ref{AppendixD}): 
\begin{equation}
\hspace{-2mm}
\label{eq:T_2_s}
T_2\approx \frac{1}{3} \sum_{i=1}^{3}
T_{2}^{(i,i+1)}
\simeq \frac{1}{3} \sum_{i=1}^{3}
\int_{K_-}^{K_+} T_{2}^{(i,i+1)}(\hat{x}_i)\vert_{\langle K \rangle}~
p_{\nu/\alpha_i}^*(N)~dN,
\end{equation}
where $T_{2}^{(i,i+1)} \sim T_{2}^{(i,i+1)}\vert_{\langle K \rangle}(\hat{x}_i)$ which 
 scales as $1/\alpha_i$  with a prefactor $\sim \log{\langle K \rangle}$
 and  a weak dependence on $\nu$ when $s\ll 1$ and $s\langle K \rangle\gg 1$~\cite{KEM1}. This yields   $T_{2}^{(i,i+1)}={\cal O}((\log{\langle K \rangle})
/s)$ in regime (ii): In agreement with the results of 
Fig.~\ref{fig:FigA5}~(b),
 $T_{2}={\cal O}(1
/s)$ with a subleading prefactor $\sim \log{\langle K \rangle}$ when $s\ll 1$ and $s\langle K \rangle \lesssim 10$. 
As in the constant-$K$ BDCLV, the quasi-neutral regime (i), where $s\langle K \rangle\ll 1$, 
$T_2={\cal O}(\langle K \rangle)$,
whereas under strong selection, $s\langle K \rangle\gg 1$,
$T_2={\cal O}(\log{\langle K \rangle})$, see Fig.~\ref{fig:FigA5}~(b).
\vspace{0.2cm}
\\
Putting together the results for $T_1$ and $T_2$, we obtain
the overall mean fixation time $T_F=T_1+T_2 \sim \langle N \rangle$. Since 
$\langle N \rangle={\cal O}(\langle K \rangle)$, we have 
$T_F={\cal O}(\langle K \rangle)$ which,  with subleading prefactors that vary slowly with $\nu$ and $s$,
as illustrated by Fig.~\ref{fig:FigA5}(c).

\subsection{Average number of switches in Stages 1 and 2 of the switching-$K$ BDCLV}
\label{AppendixE5}
Since the average duration of Stage 1 in the the switching-$K$ BDCLV is $T_1=\beta_s \langle N \rangle={\cal O}(\langle K \rangle)$, see 
Eq.~(\ref{eq:textp2th}),
the average number of switches occurring prior one of the species die out scales as ${\cal O}(\nu \langle K \rangle)$ as shown in 
Fig.~\ref{fig:FigA6}~(a), i.e. the average number of switches increases as $\nu \langle K \rangle$, with a prefactor that depends on $s$
via $\beta_s$ which is a weakly decreasing function of $s$ (i.e. the number of switches is greater for smaller values of $s$).
Hence, for any non-vanishingly small switching rate $\nu\gg 1/\langle K \rangle$ and $\langle K \rangle \gg 1$, a 
large number of switches occur
during Stage 1 prior to the extinction of the first species and the DMN self averages, see Sec.~4.1.1.

\begin{figure}[!t]
	\centering
	\includegraphics[width=0.49\linewidth]{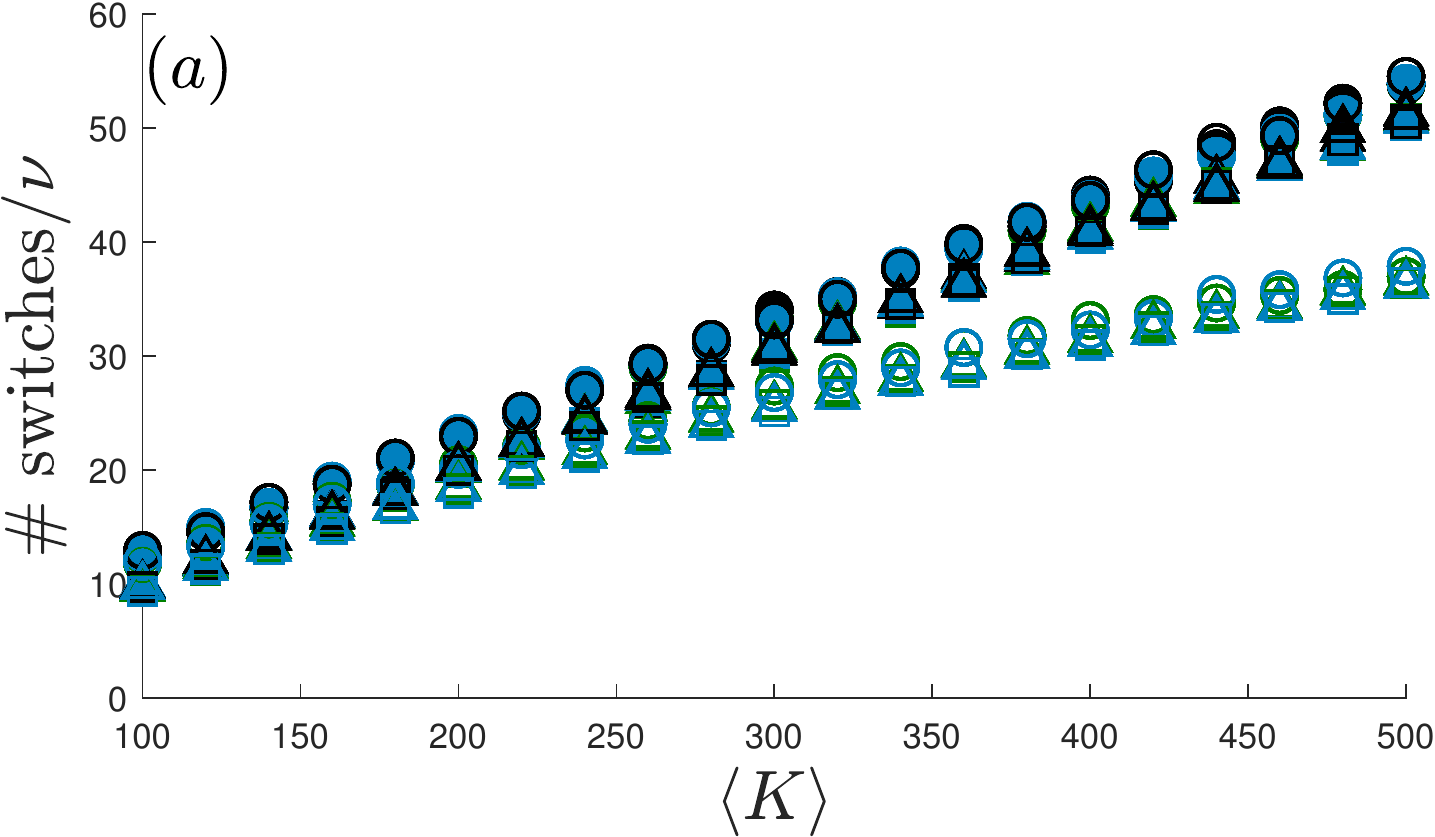}
	\includegraphics[width=0.49\linewidth]{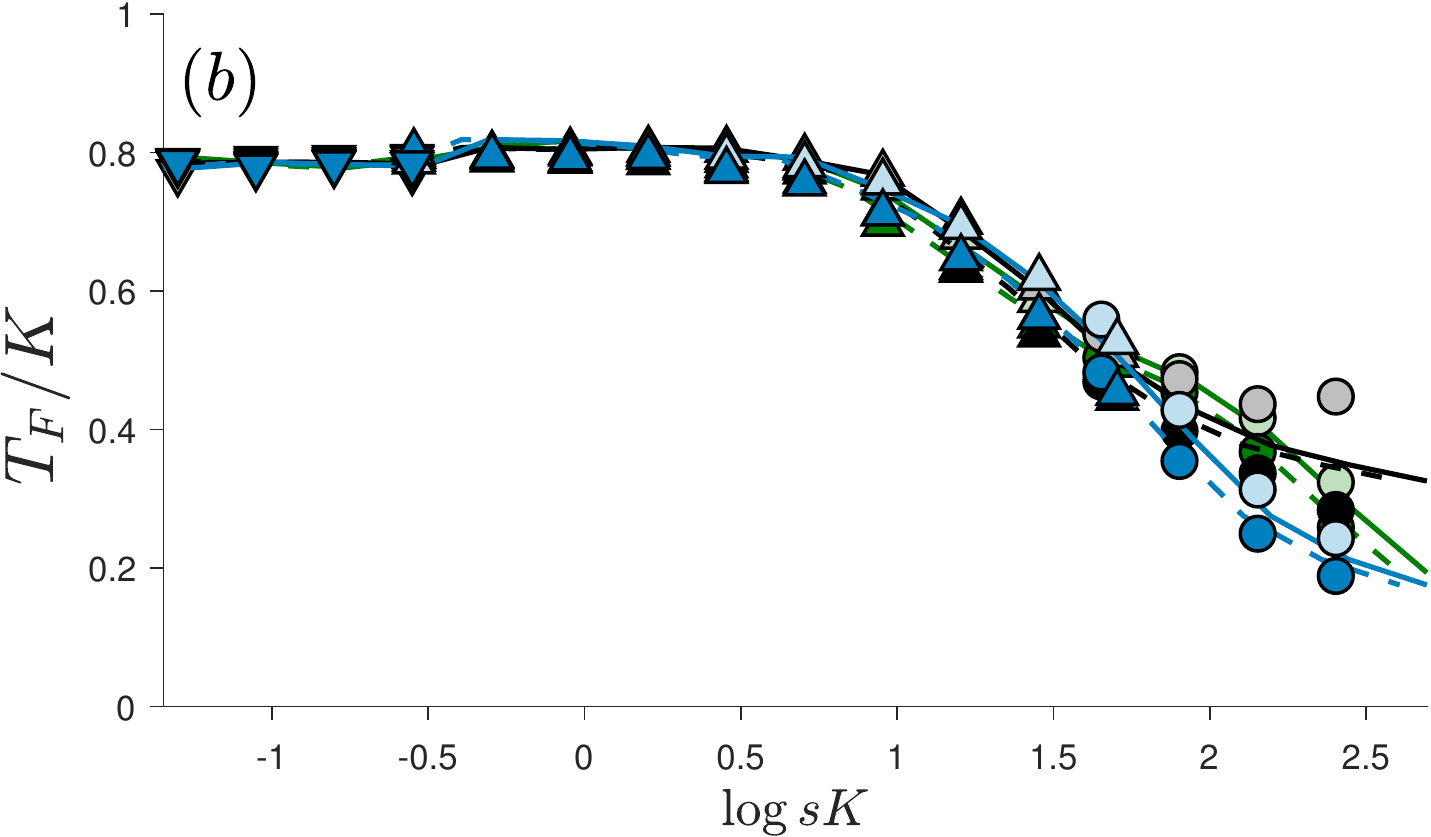}
	\includegraphics[width=0.49\linewidth]{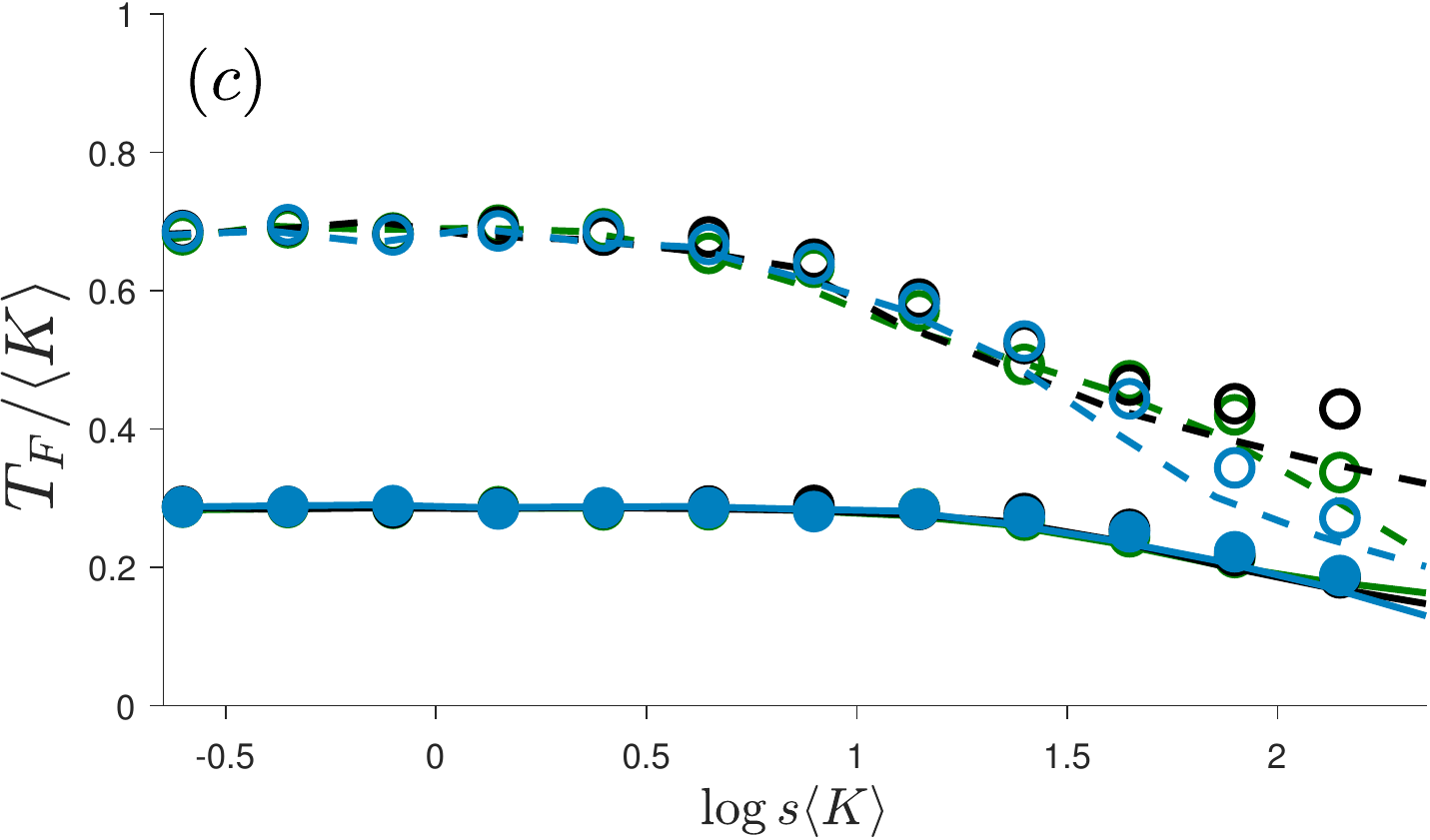}
	\includegraphics[width=0.49\linewidth]{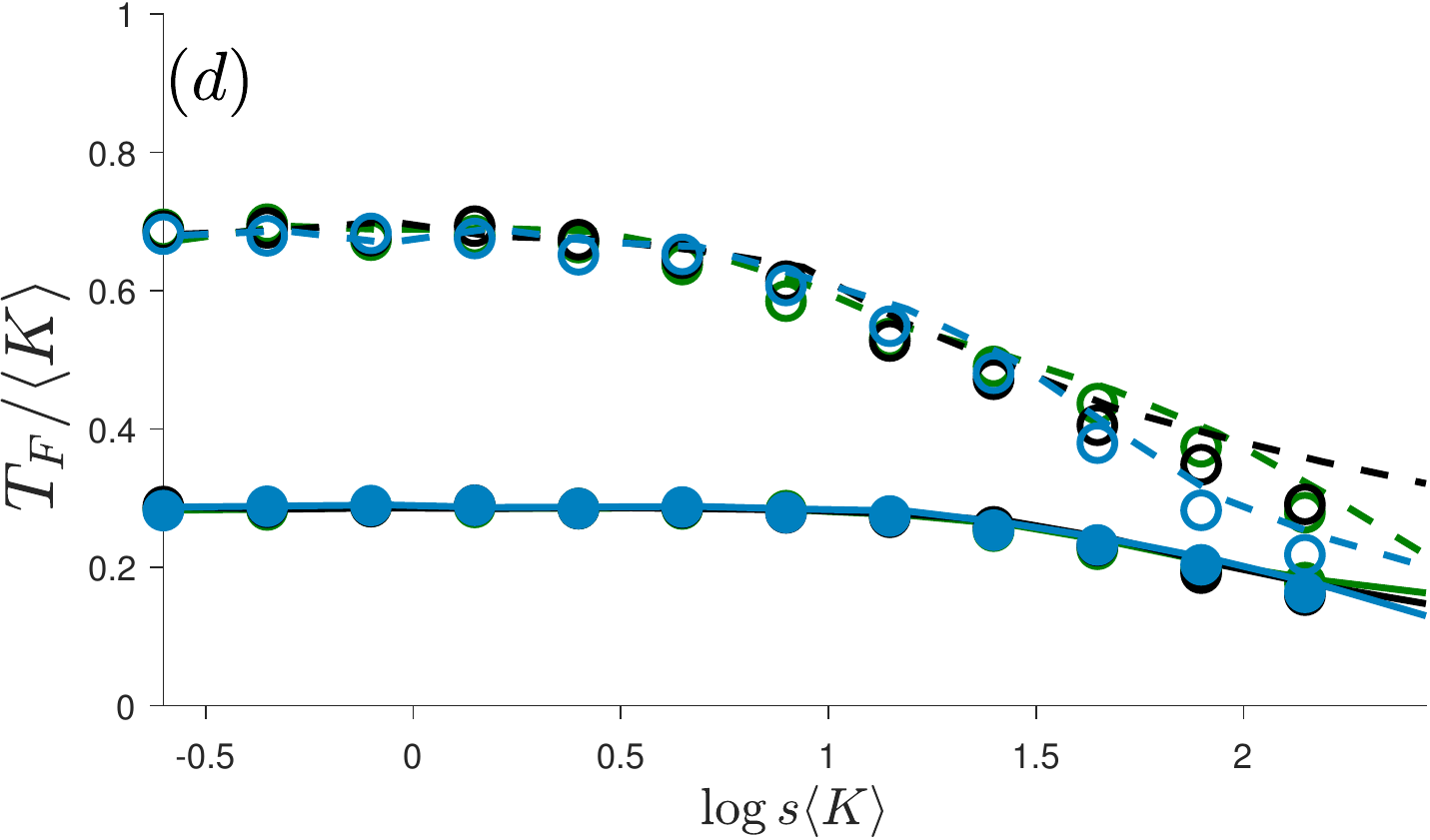}
	\caption{(a) Average number of switches in Stage 1 of the BDCLV
	for $\nu = 0.1$ (circles), $1$ (triangles), $10$ (squares).
	Selection intensity is $s=10^{-3/2}$ (filled symbols) and $s=10^{-1/2}$ (open symbols).
	Data for (average number of switches in Stage 1)/$\nu$ vs $\langle K\rangle$ and different values of $\nu$ and $\vec{r}$
       essentially collapse onto a curve (almost a line). 
%       , showing that the average number of switches in Stage 1 
%	scaled as $\nu\langle K\rangle\gg 1$ with the number of switches that is greater for smaller values of $s$, see text.
	(b)  Rescaled mean fixation time $T_F/K$  vs. $sK$ in the close-to-zero-sum game ($|\epsilon|\ll 1$)
	and constant $K$  for  values of  $s\in (10^{-3},1)$  and 
	$K = 450$~(circles), $90$~(upward triangles), $50$~(downward triangles). Symbols are from stochastic simulations for
	$\epsilon = -0.2$ (light)
	and $\epsilon = 0.2$ (dark). 
	%Same color code as in Fig.~\ref{fig:FigA4}.
	Lines are from the constant-$K$ BDCLV obtained
	with the same carrying capacity but a rescaled selection intensity $s(1+\epsilon/2)$.
	%Dark symbols and solid lines
	%are for $\epsilon=0.2$, light  symbols and dashed lines are for $\epsilon=-0.2$.
%
	(c)  $T_F/\langle K \rangle$ vs.
	$s\langle K\rangle$ when $K$ switches between $K_- = 50$ and $K_+ = 450$ with  $s\in (10^{-3},10^{-1/4})$, 
	and $\nu = 10$ (closed symbols) and  $\nu = 0.001$ (open symbols).
	Symbols are from stochastic simulations obtained for $\epsilon=-0.2$; solid ($\nu=10$) and dashed ($\nu=0.001$) lines  are from the switching-$K$ BDCLV obtained
	with the same $K(t)$ but  selection intensity $s(1+\epsilon/2)=0.9s$.
	(d) Same as in panel (c) with $\epsilon>0$: Symbols are stochastic simulation results for $\epsilon=0.2$;
	 solid ($\nu=10$) and dashed ($\nu=0.001$) lines are results from the BDCLV with same switching carrying capacity and 
	selection intensity $s\to s(1+\epsilon/2)=1.1s$.
	In all panels: $\vec{r}=(1,1,1)/3$ (black), $\vec{r}=(1,5,5)/11$ (green),  $\vec{r}=(3,2,2)/5$ (blue); $\vec{r}=\vec{r}^{(1)}$ and
	$\vec{x}_0=\vec{x}_c$. 
	In panels (a) and (b): dark symbols and solid lines
	are for $\epsilon=0.2$, light  symbols and dashed lines are for $\epsilon=-0.2$.}
	\label{fig:FigA6}
\end{figure}

In Refs.~\cite{KEM1,KEM2}, it has been shown that that under weak selection the 
population experiences, on average, ${\cal O}(\nu/\alpha_i)$
 switches during the two-species competition characterizing the stage 2 dynamics
 along the edge $(i,i+1)$. This supports the rescaling 
 $\nu \to \nu/\alpha_i$ in formula (\ref{eq:phi3})
 which has been  found to be actually valid
 when the selection intensity $s$ is neither vanishingly small nor too large~\cite{KEM2}.

\subsection{Mean fixation time of a close-to-zero-sum rock-paper-scissors game in fluctuating populations}
\label{AppendixE6}
The mean fixation time of the close-to-zero-sum rock-paper-scissors game ($|\epsilon|\ll 1$) under weak selection 
can be obtained with a similar argument usd in Sec.~5 for the fixation probabilities. 
In fact,  the mean absorption time $T_2$ and 
the mean fixation time $T_F=T_1+T_2$ ($T_1$ varies little with $s$ in regime (ii), see Fig.~\ref{fig:FigA4}~(a)) 
under weak selection  can be obtained 
from their values in the BDCLV with a rescaled 
selection intensity $s\to s(1+ (\epsilon/2))$, as shown in Fig.~\ref{fig:FigA6}~(b)-(d). This is valid both for the case of a constant
$K$, see Fig.~\ref{fig:FigA6}~(b), and a randomly switching carrying capacity, see Fig.~\ref{fig:FigA6}~(c,d).
This confirms  that the effect of $0<\epsilon\ll 1$ on the fixation properties simply boils down to increasing the selection intensity
by a factor $1+ (\epsilon/2)$ with respect to the BDCLV when $sK$ and $s\langle K \rangle$ are in regimes (i) and (ii). 
When $sK\gg 1$ and $s\langle K \rangle\gg 1$ (regime (iii)), the above argument breaks down and rescaling 
the selection intensity of the BDCLV's mean fixation time is no longer a good approximation: Under strong selection, 
the actual $T_F$ is systematically overestimated and underestimated by the  $s\to s(1+ (\epsilon/2))$ rescaling when $\epsilon>0$ 
and $\epsilon<0$. Deviations from the rescaled BDCLV results are particularly pronounced under strong selection  in the case
 $\epsilon<0$ and $\vec{x}_c= \vec{x}^*$ (with $r_i=1/3$).
 
\end{onecolumn}
\end{appendix}

%%%%%%%%%%%%%%%%

%
\end{document}